\documentclass[12pt,english]{article}
\usepackage[T1]{fontenc}
\usepackage[latin9]{inputenc}
\usepackage{geometry}
\usepackage[active]{srcltx}
\usepackage{color}
\usepackage{verbatim}
\usepackage{float}
\usepackage{mathtools}
\usepackage{amsmath}
\usepackage{amsthm}
\usepackage{amssymb}
\usepackage{stackrel}
\usepackage{graphicx}
\usepackage[onehalfspacing]{setspace}
\usepackage[labelformat=simple]{subcaption}

\usepackage{apacite}
\usepackage[dvipsnames]{xcolor}

\definecolor{darkcerulean}{rgb}{0.05, 0.2, 0.7}

\usepackage{microtype}
\makeatletter
\usepackage{natbib}
\providecommand{\tabularnewline}{\\}

\theoremstyle{definition}
\newtheorem*{example*}{\protect\examplename}
\theoremstyle{plain}

\theoremstyle{plain}
\theoremstyle{plain}

\theoremstyle{plain}
\newtheorem{prop}{\protect\propositionname}

\newtheorem{definition}{Definition}

\newtheorem{lemma}{Lemma}

\newtheorem{assumption}{Assumption}

\usepackage{amsthm,amssymb}
\usepackage{array}
\usepackage{bbm}
\usepackage{multirow,hhline}
\usepackage{threeparttable}
\usepackage{color, colortbl}
\usepackage{booktabs}
\usepackage{bookmark}
\usepackage{mathtools}
\usepackage{verbatim}
\usepackage{bm}
\usepackage{tikz,siunitx}
\usetikzlibrary{arrows,shapes,matrix,positioning,calc}
\usepackage{ragged2e}
\usepackage{lipsum}
\usepackage{dsfont}
\usepackage{pdflscape}
\usepackage{capt-of}
\usepackage[figurewithin=none]{caption}
\usepackage{setspace}
\usepackage{footmisc}
\usepackage{enumitem}

\newif\ifold
\oldfalse 

\usepackage{babel}
\providecommand{\propositionname}{Proposition}
\providecommand{\theoremname}{Theorem}

\definecolor{Gray}{gray}{0.85}
\newcolumntype{C}{@{\extracolsep{2cm}}{c}@{\extracolsep{6pt}}}%

\renewcommand*{\@fnsymbol}[1]{\ifcase#1\or*\else\@arabic{#1}\fi}

\makeatletter
\newcommand{\thickhline}{%
    \noalign {\ifnum 0=`}\fi \hrule height 2 pt
    \futurelet \reserved@a \@xhline
}

\newcolumntype{"}{!{\vrule width 2pt}}

\makeatother

\makeatletter
\renewcommand*{\@fnsymbol}[1]{\ensuremath{\ifcase#1\or *\or \dagger\or \ddagger\or
    \mathsection\or \mathparagraph\or \|\or **\or \dagger\dagger
    \or \ddagger\ddagger \else\@ctrerr\fi}}
\makeatother

\makeatother
\usepackage{babel}
\providecommand{\corollaryname}{Corollary}
\providecommand{\examplename}{Example}
\begin{document}

\title{\Large{Firm Heterogeneity, Market Power and Macroeconomic Fragility}\thanks{\fontsize{8pt}{15pt}\selectfont This is a substantially revised version of a previously circulated paper under the title "Low Competition Traps".
We thank \'{A}rp\'{a}d \'{A}brah\'{a}m, Pol Antr{\`a}s, Fernando Broner, Jes\'{u}s Bueren, Giacomo Calzolari, Vasco Carvalho, Andrea Colciago, Russell Cooper, David Dorn, Jan Eeckhout, Matteo Escud\'e, Luca Fornaro, Manuel Garc\'{i}a-Santana, Matteo Gatti, Basile Grassi, David Hemous, Giammario Impullitti, Nir Jaimovich, Tullio Jappelli, Chad Jones, Philipp Kircher, Omar Licandro, Ramon Marimon, Isabelle Mejean, Morten Olsen, Marco Pagano, Giacomo Ponzetto, Edouard Schaal, Florian Scheuer, Armin Schmutzler, Jaume Ventura, Nic Vincent, Xavier Vives and seminar participants at the EUI, UPF, CSEF, the University of Konstanz, EIEF, University of Nottingham, University of Bonn, Cornell University, Bank of Portugal, Bank of England, the EEA, SMYE, T2M, Swiss Macro Workshop and the Barcelona Summer Forum. All errors are our own.} }
\author{\color{blue}\large{Alessandro Ferrari}\color{black}
\\ \small University of Zurich \& CEPR \and \color{blue}\large{Francisco Queir\'os}\color{black}
\\ \small  University of Naples Federico II and CSEF}

\date{\small{First Version: March 2019\\
This Version: April 2024 }\\\vspace{5pt}
    \href{https://www.dropbox.com/s/vwrcqttqktqc8g8/Paper_complete.pdf?dl=0}{\textsc{{\small Latest Version Here}}}
}
\maketitle
\vspace{-1cm}

\begin{abstract}
We study how firm heterogeneity and market power affect macroeconomic fragility, defined as the probability of long slumps. We propose a theory in which the positive interaction between firm entry, competition and factor supply can give rise to multiple steady-states. We show that when firms are highly heterogeneous in terms of productivities, even small temporary shocks can trigger firm exit and make the economy spiral in a competition-driven poverty trap. We calibrate our model to incorporate the well-documented trends on rising firm heterogeneity in the US economy, and show that they significantly increase the likelihood and length of slow recoveries. We use our framework to study the 2008\textendash{}09 recession and show that the model can rationalize the persistent deviation of output and most macroeconomic aggregates from trend, including the behavior of net entry, markups and the labor share. Post-crisis cross-industry data corroborates our proposed mechanism. We conclude by showing that firm subsidies can be powerful in preventing long slumps and can lead to welfare gains between 10\% and 50\%.
\end{abstract}
\vfill{}


\thispagestyle{empty}
\newpage
\onehalfspacing

\section{Introduction}

\begin{figure}[H]
\centering{}%
\begin{tabular}{ccc}
     \hspace{-1cm} \includegraphics*[scale=0.4]{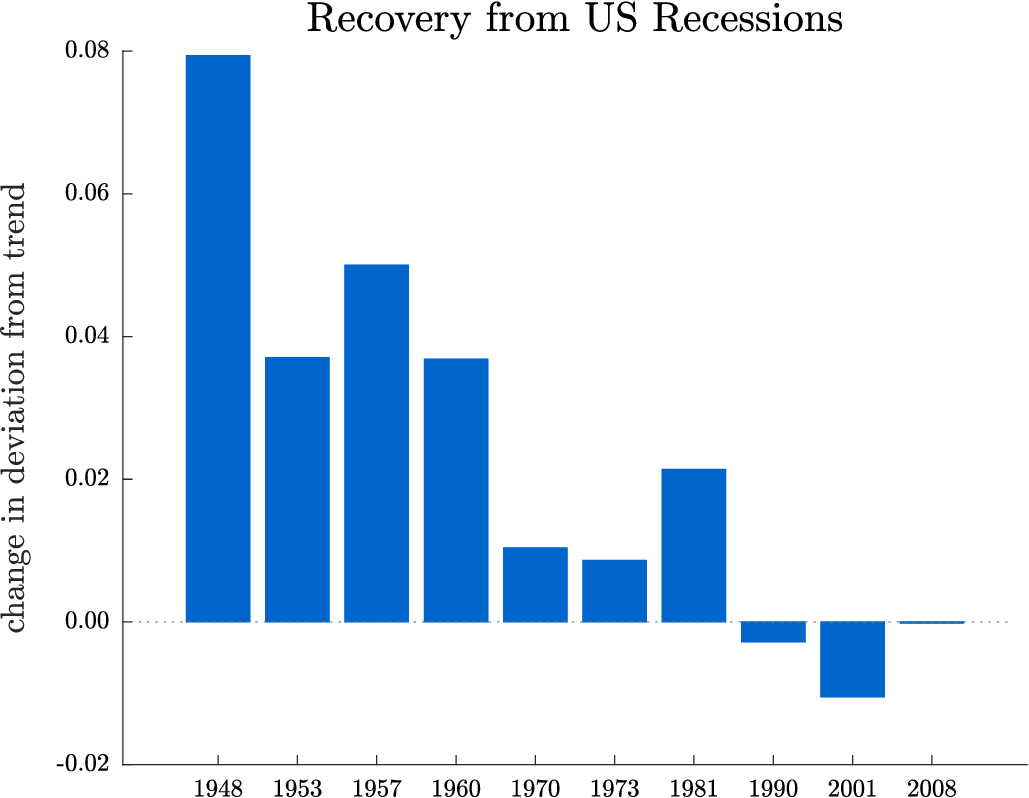} &
     \hspace{0.75cm}
     \includegraphics*[scale=0.4]{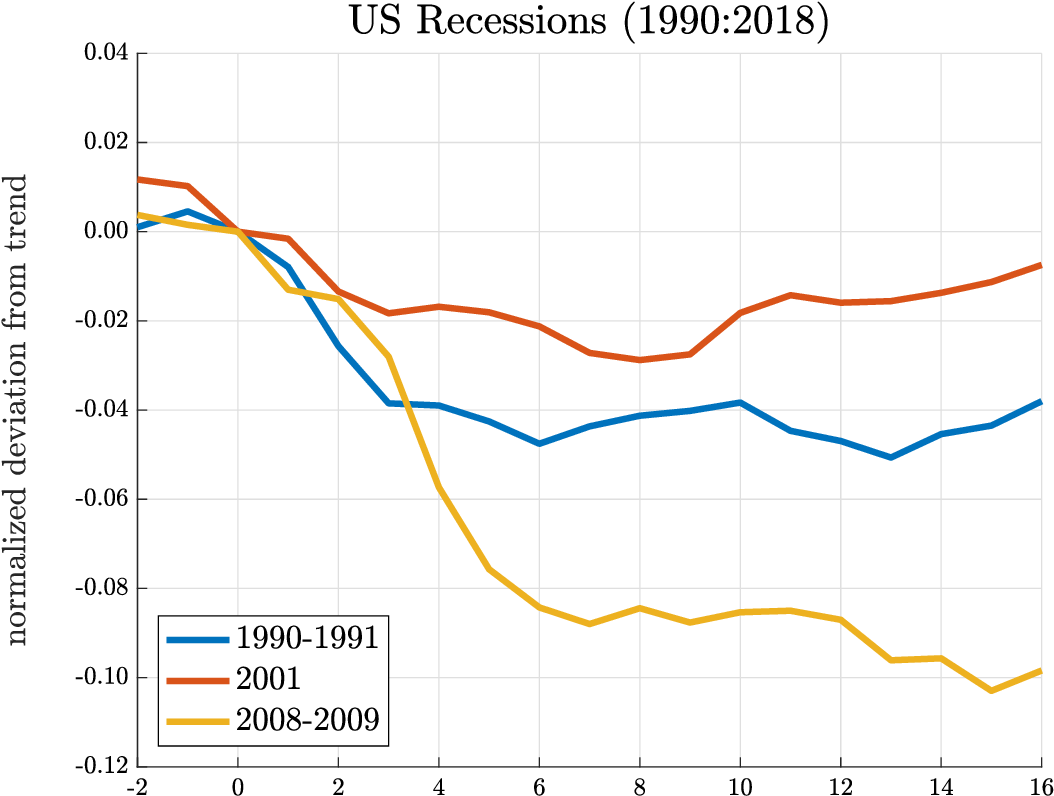}
\end{tabular}
\caption{\small{} Recovery from US Recessions\protect \\ {Note: \footnotesize{} The left panel shows, for each US postwar recession, the change in detrended output in the first two years after the trough. We use quarterly real GDP per capita (in logs, from BEA) and compute a linear trend for the period 1947-2019. We consider all recessions that lasted longer than 6 months, according to the NBER. The right panel shows the evolution of detrended output for the 1990-1991, 2001, and 2008-2009 recessions.}} 
\label{intro_figure}
\end{figure}

\begin{figure}[htb]
\centering{}%
\includegraphics*[scale=0.6]{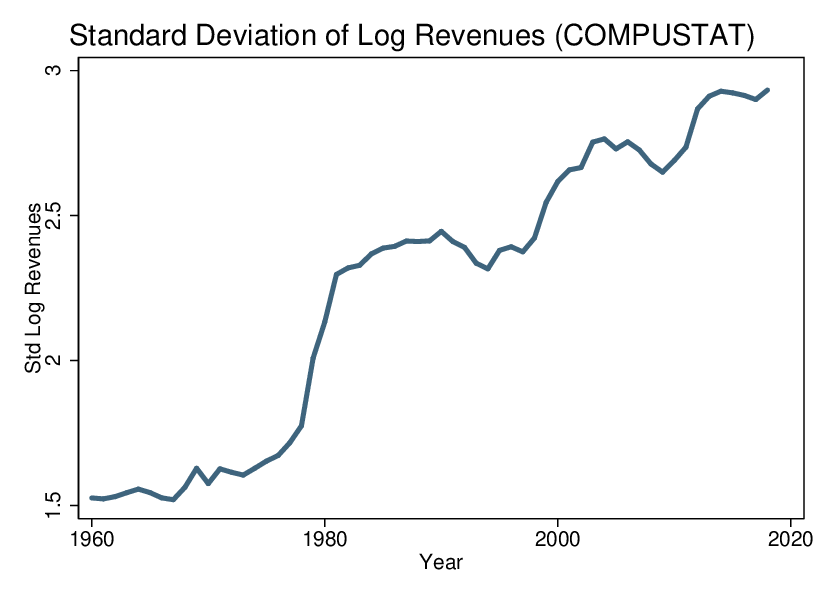} 
\hspace{0.4cm}
\includegraphics*[scale=0.6]{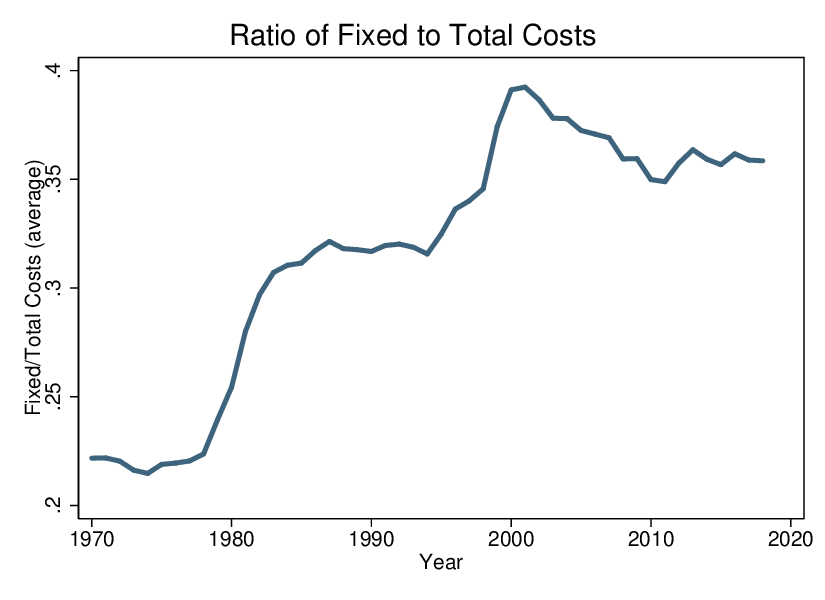}
\caption{\small {}Long-Run Trends for US Public Firms\protect \\ {Note: \footnotesize{} Data is from COMPUSTAT. The left panel shows the standard deviation of log sales for US public firms. The right panel shows the average ratio of fixed to total costs. Fixed costs are equal to `Selling, General and Administrative Expenses' (COMPUSTAT item XSGA). Total costs are the sum of fixed costs and variable costs, where the latter corresponds to the `Cost of Goods Sold' (Compustat item COGS).}} 
\label{fig:intro_figure_2}
\end{figure}

The US economy appears to be recovering more slowly from its recessions. The left panel of Figure \ref{intro_figure} shows, for different recessions, the change in detrended output in the first two years after the trough. Over the postwar period the pace of recoveries has significantly slowed down. 
This has been especially clear for the last three recessions \citep[][]{gali2012slow}. As shown in the right panel of Figure \ref{intro_figure}, four years after the beginning of each recession, detrended output was still below its pre-crisis value. For the 2008-2009 crisis, the gap is substantial (about 10 log points below trend) and has been referred to as the \textit{long slump} or \textit{great deviation} \citep[][]{H_slump}. During the same period, the US experienced a significant increase in firm heterogeneity along several dimensions, such as productivity, size, and markups. This is illustrated in the left panel of Figure \ref{fig:intro_figure_2}, which shows the evolution of the standard deviation of sales for US public firms.\footnote{The figure focuses on listed firms, but increasing firm-level dispersion has been documented using census data, and focusing on narrowly defined industries. Several studies have documented rising firm differences in terms of i) revenue TFP \citep{ACG,K, DHJM2}, ii) size \citep{BCG, ADKPR}, and iii) markups \citep{LEU,CCM,DLT}. See \cite{VR} for a summary of the recent findings.} In this paper, we argue that the rise of firm heterogeneity can partially explain the slowdown in the pace of US recoveries.

The main contribution of this paper is to show that rising firm heterogeneity can have a significant impact on business cycle fluctuations, and result in a higher probability of long slumps. We refer to such a probability as \textit{macroeconomic fragility}. We show this in the context of an RBC model with oligopolistic competition, endogenous firm entry and elastic capital and labor supply. We use a quantitative version of our framework to study the effects of rising firm heterogeneity. We find that these forces are quantitatively important. They can rationalize episodes such as the 2008-2009 recession and its aftermath and are consistent with cross-industry empirical evidence from the \textit{great deviation}.

At the heart of our model is a complementarity between competition and factor supply. First, economies with intense competition in product markets feature low-profit shares and high factor shares and factor prices, which induce high factor supply. Second, a high factor supply allows more firms to enter, which results in greater competition. This complementarity can give rise to multiple competition regimes or (stochastic) steady-states. A key contribution of our theory is to show that rising heterogeneity in idiosyncratic TFP can make transitions from high to low steady-states more likely to occur. When firm heterogeneity increases, large/productive firms expand, while small/unproductive firms contract. As a consequence, smaller negative shocks can be enough to trigger firm exit and a transition to a steady-state featuring lower competition, capital, and output. An identical result is obtained when fixed costs of production increase. In fact, there is also evidence that US firms have been facing increasingly larger fixed costs, as shown in the right panel of Figure \ref{fig:intro_figure_2}.  We characterize these results formally by showing that the minimum size (negative) shock required to trigger a transition from a high to a low steady-state decreases when firm TFP heterogeneity rises or when fixed costs increase. 


To quantify these economic forces, we provide three calibrations of our model, where we target US firm-level moments in 1975, 1990, and 2007. These different calibrations differ in the fixed cost level and the degree of TFP dispersion (which both increase over time). We first find that, while the 1975 economy is characterized by an unimodal ergodic distribution of output, the 1990 and 2007 feature bimodal distributions (an indication of two stochastic steady-states). Second, when we subject the three economies to the same shocks, the 2007 economy exhibits significantly greater amplification and propagation. For example, we find that the 2007 economy experiences a recession greater than 10\% every 70 years, while the 1990 and 1975 economies experiences one every 95 and 380 years, respectively. This suggests that rising firm heterogeneity and fixed costs made the US economy significantly more fragile and prone to long-lasting slumps.

We also use our model economy as a laboratory to study the Great Recession and its aftermath. The 2008 crisis was marked by a large and persistent deviation of output and other aggregates from the trend, something unusual in the entire postwar period. For example, in 2019, output per capita was 14\% below its pre\textendash2007 trend, a deviation far larger and more persistent than in previous recessions (Figure \ref{fig:great_deviation_figure}). We ask whether our model can replicate such a quasi-permanent drop in output and other variables. To this end, we feed our 2007 economy a sequence of shocks calibrated to match the behavior of aggregate TFP in 2008\textendash9 and study the economy's response. The model generates the observed persistent deviation from trend of GDP as well as of investment, hours, and aggregate TFP. It also rationalizes the sharp and persistent drop in the labor share after 2008, and the acceleration in markup growth. Importantly, when we subject the 1990 and 1975 economies to the same shocks, the model does not predict such a persistent deviation from trend.


We also provide empirical evidence in favor of our proposed mechanism. Our theory provides predictions in terms of cross-industry responses to the business cycle. In particular, for any two markets with the same number of firms, the one with more heterogeneity/concentration reacts more to a negative shock. 
Consistent with our model predictions, we show that industries that were more concentrated in 2007 experienced greater cumulative declines in the number of firms, the labor share, and economic activity over the 2008-2016 period.

In terms of policy lessons from our theory, we show that firm subsidies can be effective in preventing deep recessions, and lead to large welfare gains. An entry subsidy can be enough to prevent the economy from experiencing persistent slumps and can lead to a welfare gain of 10\% (in consumption-equivalent terms). These gains can be even larger under a revenue subsidy that reduces the markup distortion at all states of the world, increasing welfare by approximately 50\%, a number that is in line with \cite{EMX}. 


\paragraph{Related Literature}

Our paper speaks to three different strands of the literature. First, it is related to the macroeconomic literature studying models of coordination failures \citep{CJ1,Mat,BenhabibFarmer,FarmerGuo,herrendorf2000ruling,big_push_distorted_economies}. While we are not the first to show how multiple equilibria and/or steady-states can arise in a context of imperfect competition and variable markups \citep{MP,CCR,RW,GZ,J}, 
we contribute to this literature by studying the role of firm-level heterogeneity in shaping macroeconomic \textit{fragility}. As we discuss, this concept is distinct from the \textit{existence} of steady-state multiplicity. We also provide a quantification of our mechanism and link it to the 2008 crisis.\footnote{Our paper also speaks to the literature studying the cyclicality of markups, which includes \cite{RS}, \cite{JF}, \cite{EC}, \cite{BGM}, \cite{BKM}, \cite{nekarda2020cyclical} and \cite{AVB}.}

Second, this paper relates to a large and growing literature documenting long-term trends in firm heterogeneity and market power. There are several signs that indicate rising market power in the US and other advanced economies. For example, \cite{ADKPR} use data from the US census to document rising sales and employment concentration, while \cite{AA2} document a rise in patenting concentration. Other studies have documented a secular rise in price-cost markups. Using data from national accounts, \cite{H} finds that the average sectoral markup
increased from 1.12 in 1988 to 1.38 in 2015. \cite{LEU} document a steady increase in sales-weighted average markups for US public firms between 1980 and 2016.\footnote{\cite{EMX} show that a cost-weighted average markup displays a less pronounced trend. See also \cite{T}, \cite{KN2}, and \cite{BHKZ} on trends in markups.} This was driven by both an increasing share of large firms and by rising dispersion in the markup distribution.

Close to our approach, \cite{de2021quantifying} argue that declining dynamism and rising market power can be explained, among other channels, by increasing productivity dispersion and fixed costs. \cite{EMX} estimate that the welfare cost of markups can be large and represent a loss of up to 50\% in consumption-equivalent terms.
We contribute to this literature by investigating the business cycle implications of these trends, particularly their impact on the 2008 crisis and the subsequent \textit{great deviation}. 

Lastly, this paper relates to the literature studying slow recoveries and the persistent impact of the 2008 crisis. A large part of this literature has focused on the secular decline in interest rates and/or the possibility of liquidity traps, which constrain monetary policy \citep{Christiano_Eichenbaum_Trabandt,BF,guerrieri_lorenzoni,eggertsson_et_al}, or on the long-run consequences of R\&D decisions \citep{ACGM,Bianchi_Kung_Morales,queralto2020model,fasil2021heterogeneous}. \cite{clementi_palazzo} argue that the decline in firm entry observed after 2008 is crucial to understanding the slow recovery. Close to this paper, \cite{ST} build an RBC model with complementarities in capacity utilization choices among monopolistic firms to generate multiple steady-states. We also use coordination to obtain multiplicity and interpret the post\textendash{}2008 deviation as a transition to a low steady-state. 
We complement their analysis and the aforementioned literature by arguing that rising firm-level heterogeneity has increased the likelihood of slumps.\footnote{In related work, \cite{Baley_Blanco_ECMT} show that heterogeneity in capital-to-productivity ratios may indicate large capital adjustment frictions, which may result in slow transitional dynamics of investment.} Our theory can also account for a number of trends observed after 2008, such as the acceleration in the labor share decline, the acceleration of markup growth, and the decline in the number of firms.

The rest of the paper is organized as follows. Section \ref{sec:Model} contains the model and the main theoretical results. Section \ref{sec:Quantitative-Evaluation} discusses the calibration and presents the quantitative results. Section \ref{sec:aggregatefacts} provides an extended application to the US Great Recession and its aftermath and presents the cross-industry empirical evidence. In Section \ref{policy}, we study the welfare effects of fiscal policy in our model. Finally, section \ref{sec:Conclusion} concludes.
\section{A Model with Firm Heterogeneity and Variable Markups \label{sec:Model}}
The model builds upon the neoclassical growth model, with a representative household that supplies labor and capital. The technology side is comprised of a large number of product markets, where firms compete oligopolistically.
We start by describing the demand side and the technology structure. Then, we analyze the equilibrium of a particular product market and the general equilibrium.
\subsection{Preferences}
Time is discrete and indexed by $t=0,1,2,\ldots$.
There is a representative, infinitely-lived household with lifetime utility
\begin{align}
U_{t}=\mathbb{E}\sum_{t=0}^{\infty}\:\beta^{t}\:U\left(C_{t},L_{t}\right),
\end{align}
where $0 < \beta < 1$ is the discount factor, $C_{t} \geq 0$ is consumption of the final good and $L_{t} \geq 0$ is labor.
We adopt the period utility function as in \cite{GHH}
\begin{align}
U\left(C_{t},L_{t}\right)=\dfrac{1}{1-\psi}\left(C_{t}-\dfrac{L_{t}^{1+\nu}}{1+\nu}\right)^{1-\psi},
\label{eq:HH_utility}
\end{align}
where $\psi >0$ is the inverse of the intertemporal elasticity of substitution and $\nu>0$ is the inverse of the Frisch elasticity of labor supply. 

The representative household owns all firms in the economy, but runs them in a non-cooperative way \textendash{} i.e. they compete against each other and do not collude. Nevertheless, all individuals pool together the profits they make. Hence, the budget constraint is given by
\begin{align}
K_{t+1}=\left[R_{t}+\left(1-\delta\right)\right]K_{t}+W_{t}L_{t}+\Pi^{N}_{t}-C_{t},
\label{eq:budget_constraint}
\end{align}
where $K_t$ is capital, $R_t$ is the rental rate, $W_{t}$ is the wage rate and $\Pi^{N}_{t}=\sum_{j} \Pi_{jt}^{N}$ are the aggregate profits net of fixed costs. Capital depreciates at rate $0\leq\delta\leq1$ and factor prices $R_{t}$ and $W_{t}$ are taken as given. The representative household maximizes (\ref{eq:HH_utility}) subject to (\ref{eq:budget_constraint}). Our choice of GHH preferences implies that the aggregate labor supply is given by $L^{S}_{t} = W_{t}^{1/\nu}$.

\subsection{Technology}

There is a final good (the \textit{numeraire}), which is a CES aggregate of $I$ different product markets 
\[
Y_{t}=\left(\sum_{i=1}^{I}{\vphantom{\Pi^P}y}_{it}^{\rho}\right)^{1/\rho}, \]
where $y_{it}$ is the quantity of product market $i\in\left\{1,\ldots,I\right\}$, and $\sigma_{I} = {1}/{(1-\rho)}>1$ is the elasticity of substitution across markets. $I$ is assumed to be large, so that each individual market has a negligible size in the economy. The output of each market $i$ is itself a CES
composite of differentiated goods
\[
y_{it}=\left(\sum_{j=1}^{n_{it}}{\vphantom{\Pi^P}y}_{jit}^{\eta}\right) ^{1/\eta},
\]
where $n_{it}$ is the number of active firms in product market $i$ at time $t$ (to be determined endogenously) and $\sigma_{G} = 1/(1-\eta) > 1$ is the within market
elasticity of substitution. Following \cite{AB},
we assume that goods are more easily substitutable within product market than across product markets: 
$0<\rho<\eta\leq1$.
The inverse demand for variety $j$ in market $i$ is given by
\begin{align}
p_{ijt}=\left(\dfrac{Y_{t}}{y_{it}}\right)^{1-\rho}\left(\dfrac{y_{it}}{y_{ijt}}\right)^{1-\eta}.\label{eq:inv_demand_good}
\end{align}
In every product market $i$, there is a maximum number of firms $M \in \mathbb{N}$, so that $n_{it} \leq M$. Firm $j$ can produce its variety
by combining capital $k_{ijt}$ and labor $l_{ijt}$ through a CRS
technology
\begin{align}
y_{ijt}=\underbrace{{A_t}\;\gamma_{ij}}_{\tau_{ijt}}\left(k_{ijt}\right)^{\alpha}\left(l_{ijt}\right)^{1-\alpha}.\label{eq:CD_pf}
\end{align}
Note that the productivity of each firm $\tau_{ijt}$ is the product of two terms (i) a time-varying aggregate component $A_t$
(common to all firms) and (ii) a time-invariant idiosyncratic
term $\gamma_{ij}$. We refer to $A_{t}$ as aggregate productivity and to $\gamma_{ij}$ as $j$'s idiosyncratic productivity. Aggregate productivity follows an auto-regressive process
\[
\log A_{t} \; = \; \phi_{A}\log A_{t-1}+\varepsilon_{t},
\]
with $\varepsilon_{t}\sim N \left(0,\sigma_{\varepsilon}^{2}\right)$. Without loss of generality,
we order idiosyncratic productivities according to $\gamma_{i1}\geq\gamma_{i2}\geq\hdots.$ 
Labor is hired at the competitive wage $W_{t}$ and capital at the
rental rate $R_{t}$. Firm $j$ can thus produce
its variety at a marginal cost $\Theta_{t} / \tau_{ijt}$,
where 
\begin{align}
\Theta_{t}\coloneqq\left(\frac{R_{t}}{\alpha}\right)^{\alpha}\left(\frac{W_{t}}{1-\alpha}\right)^{1-\alpha}
\label{eq:factor_price_index}
\end{align}
is the marginal cost function for a Cobb-Douglas technology with unit
productivity. We refer to $\Theta_{t}$ as the \textit{factor
price index}. In addition to all variable costs, the production of each variety entails a fixed production cost $c_{i} \geq 0$ per period (which may differ across product markets). Such a cost is in units of the \textit{numeraire}.\footnote{In earlier versions, we have also allowed fixed costs to change with factor prices, see  \cite{FQ_working_paper}.}

\subsection{Market Structure \label{subsec:Market-Structure}}
We assume that all firms that enter pay a fixed cost $c_i$ and live for one period. We follow \cite{AB} and assume that firms enter sequentially in decreasing order of TFP. Active firms within a market play a static Cournot game, announcing quantities simultaneously while taking the output of the other competitors as given.\footnote{Since firms only live for one period and, therefore, do not make any dynamic choice, we assume, without loss of generality, that the matrix of productivities of potential entrants is constant over time. This assumption substantially simplifies the quantitative application in Section \ref{sec:Quantitative-Evaluation}.}
Therefore, each firm $j$ solves 
\begin{align}
\underset{y_{ijt}}{\textrm{max}}\quad\left(p_{ijt}-\dfrac{\Theta_{t}}{\tau_{ijt}}\right)y_{ijt}\qquad\text{s.t.}\;  p_{ijt}=\left(\dfrac{Y_{t}}{y_{it}}\right)^{1-\rho}\left(\dfrac{y_{it}}{y_{ijt}}\right)^{1-\eta} \text{ and }
\quad y_{it}=\left(\sum\limits_{k=1}^{n_{it}}y_{kit}^{\eta}\right)^{\frac{1}{\eta}}.
\label{eq:Cournot_problem}
\end{align}
The solution to (\ref{eq:Cournot_problem}) yields a system of $n_{it}$ non-linear equations in $\left\{ p_{ijt}\right\}_{j=1}^{n_{it}}$
\begin{align}
p_{ijt} = \underbrace{\dfrac{1}{\eta-\left(\eta-\rho\right)s_{ijt}}}_{\mu_{ijt}} \,  \dfrac{\Theta_{t}}{\tau_{ijt}},\label{eq:firm_FOC}
\end{align}
where $s_{ijt}$ is the market share of firm $j=1,\ldots,n_{it}$ and $\mu_{ijt}$ is the markup. Eq. (\ref{eq:firm_FOC}) establishes a positive relationship between market shares and markups. This follows from firms internalizing the impact of their size on the price they charge; large firms end up restricting output relative to productivity, thereby charging a high markup. 
Rearranging (\ref{eq:firm_FOC}), one can also see that market shares are a positive function of revenue TFP $\left(p_{ijt} \: \tau_{ijt}\right)$. Our model thus features a positive association between revenue productivity, size, and markups. A shock that increases dispersion in revenue TFP is also associated with greater dispersion in market shares and markups.

To conclude the description of the product market equilibrium, we need to determine the number of active firms $n_{it}$. To this end, let $ \Pi\left(j ,n_{it},\Gamma_{it},X_{t}\right) \coloneqq \left(p_{ijt}-{\Theta_{t}}/{\tau_{ijt}}\right)y_{ijt}$ denote the gross profits of firm $j\leq n_{it}$ in product market $i$, when there are $n_{it}$
active firms, given a productivity distribution $\Gamma_{i} \coloneqq \left\{ \gamma_{i1} \, , \, \gamma_{i2} \, , \, \ldots \right\}$ and a vector of aggregate variables $X_{t} \coloneqq \left[ A_{t} \, , \, Y_{t} \, , \, \Theta_{t} \right]$. The equilibrium number of firms must be such that (i) the profits of each active firm are not lower than the fixed cost $c_{i}$ and (ii) if an additional firm were to enter, its profits would be lower than the fixed cost. Formally, an interior solution $n^{*}_{it} < M$ to the equilibrium number of firms must satisfy
\begin{align}
\left[\Pi\left({\color{blue}n_{it}^{*}},{\color{blue}n_{it}^{*}},\Gamma_{i},X_{t}\right)-c_{i}\right] \, \left[\Pi\left({\color{blue}n_{it}^{*}+1},{\color{blue}n_{it}^{*}+1},\Gamma_{i},X_{t}\right)-c_{i}\right] \; \leq \; 0, \quad   \forall i = 1,\ldots,I .\label{eq:eqm_n_firm}
\end{align}
Lemma \ref{lemma:(Profit-Function)} in Section \ref{appendix:appendix_partial_eqm} provides an analytical characterization of the profit function under the special case of $\eta=1$. We show that the profits of any firm $j$ i) increase in its own idiosyncratic productivity $\gamma_{ij}$ and ii) decrease in the idiosyncratic productivity
of all the other firms $\gamma_{ik}$. This means that, as top firms become more productive, small firms make lower profits and become closer to their exit threshold (\textit{ceteris paribus}). This is key to understanding some of the general equilibrium properties of the model.

\subsection{General Equilibrium}
\paragraph{Equilibrium Definition}
Let $ Z_{t} \coloneqq \left[K_{t},A_{t}\right] $ be a vector containing the two state variables of this economy. We have the following definition of an equilibrium.
\begin{definition}[Equilibrium]
An equilibrium is a sequence of policies  $ \left\{ C_{t}\left(Z_{t}\right),  K_{t+1}\left(Z_{t}\right),
L_{t}\left(Z_{t}\right)\right\}_{t=0}^{\infty} $ for the household, firm policies $ \left\{ y_{ijt}\left(Z_{t}\right),  k_{ijt}\left(Z_{t}\right), l_{ijt}\left(Z_{t}\right)\right\}_{t=0}^{\infty} $, and a set of active firms $\left\{ n_{it}\left(Z_{t}\right)\right\}_{t=0}^{\infty}$ with $\forall i\in\left\{1,\ldots,I\right\}$ such that i) households optimize; ii) all active firms optimize; iii) the slackness free entry condition in eq. (\ref{eq:eqm_n_firm}) holds; iv) capital and labor markets clear.
\end{definition}

\subsubsection{Within-Period Equilibrium}
We now describe the general equilibrium of this economy. We start by focusing on a within-period equilibrium, in which production and labor supply decisions are described, taking the aggregate level of capital $K_{t}$ as given. Later on, we describe the equilibrium dynamics.
\paragraph{Aggregate Production Function} 
Given a $\left(I\times M\right)$
matrix $\mathbf{\Gamma}$ of idiosyncratic productivity draws and a $\left(I\times 1\right)$ vector $\mathbf{N}_{t} \coloneqq \left\{ n_{it}\right\}_{i=1}^{I}$ containing the number of active firms in every market, aggregate output can be written as 
\begin{align}
Y_{t}={A}_{t}\,\Phi\left(\mathbf{\Gamma},\mathbf{N}_{t} \right)\;L_{t}^{1-\alpha}K_{t}^{\alpha} .\label{eq:agg_prod_fct}  
\end{align}
The term $\Phi\left(\cdot\right)$ 
represents the endogenous component of aggregate TFP and is a function of the number of active firms, individual productivities, and market shares. An analytic expression for $\Phi\left(\mathbf{\Gamma},\mathbf{N}_{t}\right)$ is provided in Section \ref{appendix:appendix_general_eqm}.
\paragraph{Aggregate Factor Share} 
Let $\mathcal{V}_t\coloneqq W_{t}\,L_{t} + R_{t}\,K_{t}$ represent aggregate variable costs. We can write the aggregate factor share $\Omega\left(\cdot\right) \coloneqq \mathcal{V}_t /Y_{t} $ as a function of markups and market shares\footnote{The aggregate factor share is equal to the inverse of the aggregate markup $\mu\left(\cdot\right) \coloneqq Y_{t} /\mathcal{V}_t $.}
\begin{align}
\Omega\left(\mathbf{\Gamma},\mathbf{N}_{t}\right) = \sum\limits _{i=1}^{I}\sum\limits _{j=1}^{n_{it}}s_{it}\;s_{ijt}\;\mu_{ijt}^{-1}.
\label{eq:agg_factor_share}
\end{align}
%
\paragraph{Factor Prices and Factor Markets}
The aggregate demand schedules for labor $L_t$ and capital $K_t$ can be written as
\begin{align}
     W_{t} \; = &  \; \: \left(1-\alpha\right)\; \Theta\left(\mathbf{\Gamma},\mathbf{N}_{t}\right) \; L_{t}^{-\alpha}\,K_{t}^{\alpha}, 
     \label{eq:wage_eqm}
     \\[1ex]
     R_{t} \; = & \; \: \alpha\; \Theta\left(\mathbf{\Gamma},\mathbf{N}_{t}\right)\;L_{t}^{1-\alpha}\,K_{t}^{\alpha-1}.
\label{eq:rental_rate_eqm}  
\end{align}
It is convenient to write the factor price index as the product between the aggregate factor share and aggregate TFP. Combining equations (\ref{eq:factor_price_index}), (\ref{eq:agg_prod_fct}), (\ref{eq:wage_eqm}) and (\ref{eq:rental_rate_eqm}), we have
\begin{align}
\Theta\left(\mathbf{\Gamma},\mathbf{N}_{t}\right)\;=\;\underbrace{\Omega\left(\mathbf{\Gamma},\mathbf{N}_{t}\right)}_{\substack{\text{aggregate} \\ \\ \text{factor share}}
} \; \underbrace{ \: {A}_{t} \: \Phi\left(\mathbf{\Gamma},\mathbf{N}_{t}\right)}_{\text{aggregate TFP}}.
\label{eq:agg_cost_decomposition}
\end{align}
The factor demand equations in (\ref{eq:wage_eqm}) and (\ref{eq:rental_rate_eqm}) can be combined with the supply schedules
\begin{align}
     L_{t}^{S} = W_{t}^{1 / \nu}  \quad \text{and} \quad K_{t}^{S} = K_{t} 
\label{eq:factor_supply}  
\end{align}
to determine the factor market equilibrium. Combining equations (\ref{eq:agg_prod_fct}), (\ref{eq:wage_eqm}) and (\ref{eq:factor_supply}), we can write aggregate labor and output as a function of the aggregate capital stock $K_t$, the productivity distribution $\mathbf{\Gamma}$ and the set of active firms $\mathbf{N}_{t}$
\begin{align}
L_{t} \; = & \;  \left[\left(1-\alpha\right)\Theta\left(\mathbf{\Gamma},\mathbf{N}_{t}\right)\right]^{\frac{1}{\nu+\alpha}}K_{t}^{\frac{\alpha}{\nu+\alpha}},
\label{eq:agg_labor_capital}
\\
Y_{t} \; = & \; A_{t}\,\Phi\left(\mathbf{\Gamma},\mathbf{N}_{t}\right)\:\left[\left(1-\alpha\right)\Theta\left(\mathbf{\Gamma},\mathbf{N}_{t}\right)\right]^{\frac{1-\alpha}{\nu+\alpha}}\:K_{t}^{\alpha\frac{1+\nu}{\nu+\alpha}}.
\label{eq:agg_output_capital}
\end{align}
Both aggregate labor $L_{t}$ and output $Y_{t}$ are increasing in the factor price index. Higher factor prices result in higher wages (through (\ref{eq:wage_eqm})) and hence a larger labor supply (through (\ref{eq:factor_supply})). We conclude the characterization of the within-period equilibrium by determining the set of active firms $\mathbf{N}_{t}$.

\paragraph{Equilibrium Set of Firms} The number of active firms in each product market $i$ is  jointly determined by equations (\ref{eq:agg_cost_decomposition}),  (\ref{eq:agg_output_capital}) and the set of inequalities defined in (\ref{eq:eqm_n_firm}). Such a joint system does not admit a general analytical characterization. In subsection \ref{subsec:analytic-charact}, we will characterize it in a particular case of the model that admits a closed form solution.

\subsubsection{Equilibrium Dynamics}
Denoting by $s_{t}$ the aggregate savings rate out of gross output (from the household maximization problem), we can write the law of motion for capital as 
\begin{align}
    K_{t+1} = \left(1-\delta\right) K_{t} + s_{t} \cdot Y_{t}. \label{eq:law-of-motion}
\end{align}
Even though we cannot provide an analytical expression for $s_t$, we can show that, in a steady-state, it is proportional to the aggregate factor share. Therefore, a more competitive market structure will be characterized by a higher steady-state savings rate (and hence a greater supply of capital).
\begin{prop}[Steady-state savings rate] \label{prop:savings_rate}
The steady-state savings rate satisfies
\begin{align}
s^{*}= \dfrac{\beta \, \delta}{1-\left(1-\delta\right)\beta}\; \alpha\;\Omega\left(\mathbf{\Gamma},\mathbf{N}\right).
\end{align}
\end{prop}

\begin{proof}
See Appendix \ref{appendix:appendix_general_eqm}.
\end{proof}

We now restrict our attention to a specific case of the model that allows us to obtain a number of analytical results. We return to this more general version in Section \ref{sec:Quantitative-Evaluation} for the quantitative results.

\subsection{Analytic Characterization\label{subsec:analytic-charact}}
To make the analysis tractable, we focus on a deterministic version of the model and shut down aggregate shocks. We also assume that all markets are ex-ante symmetric and that, within a market, goods are homogeneous.

\begin{assumption}
\label{assumption:assumption1}
In this subsection, we assume that:
\begin{enumerate}[label=\alph*)]
    \item The exogenous component of aggregate productivity is constant: $A_{t} = 1$.
    \item Goods are perfect substitutes within a market: $\eta=1$.
    \item All markets have the same distribution of idiosyncratic productivities: $\Gamma_i=\Gamma,\,\forall i = 1,\ldots,I$.
    \item All markets have the same fixed cost: $c_i=c,\,\forall i= 1,\ldots,I$.
\end{enumerate}
\end{assumption}

The assumption that markets are \textit{ex-ante} identical does not mean that they will be identical \textit{ex-post}, as markets can differ in the equilibrium number of active firms. Proposition \ref{A-symmetric-equilibrium}
states the conditions under which a symmetric equilibrium exists and is unique.

\begin{figure}[ht]
\centering{}%
\begin{tabular}{ccc}
	\hspace{-1cm} \includegraphics*[scale=0.385]{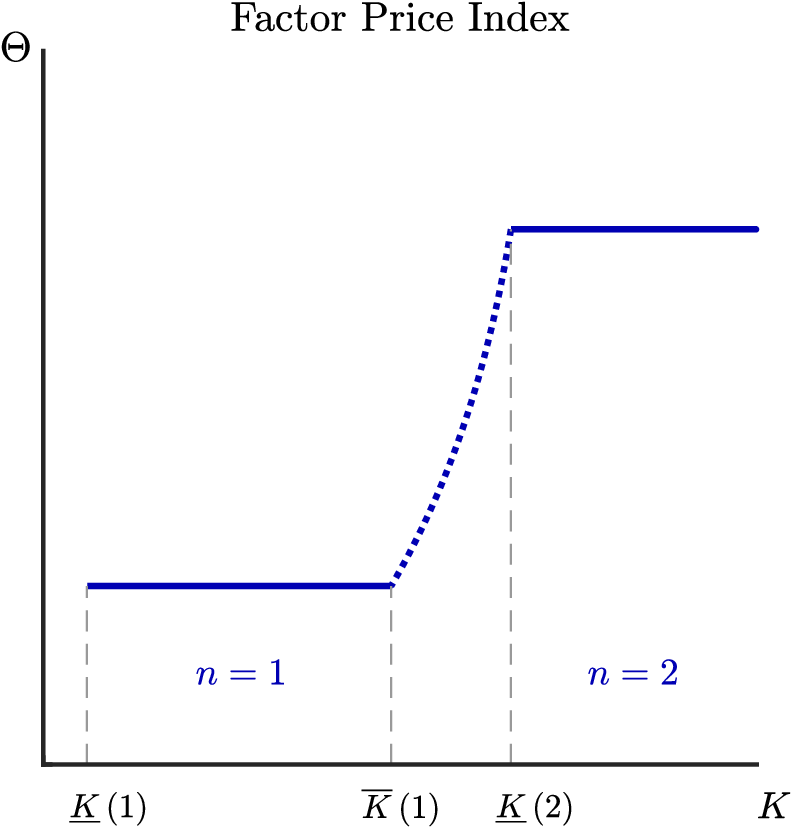} &  
	\hspace{0.25cm}
	\includegraphics*[scale=0.385]{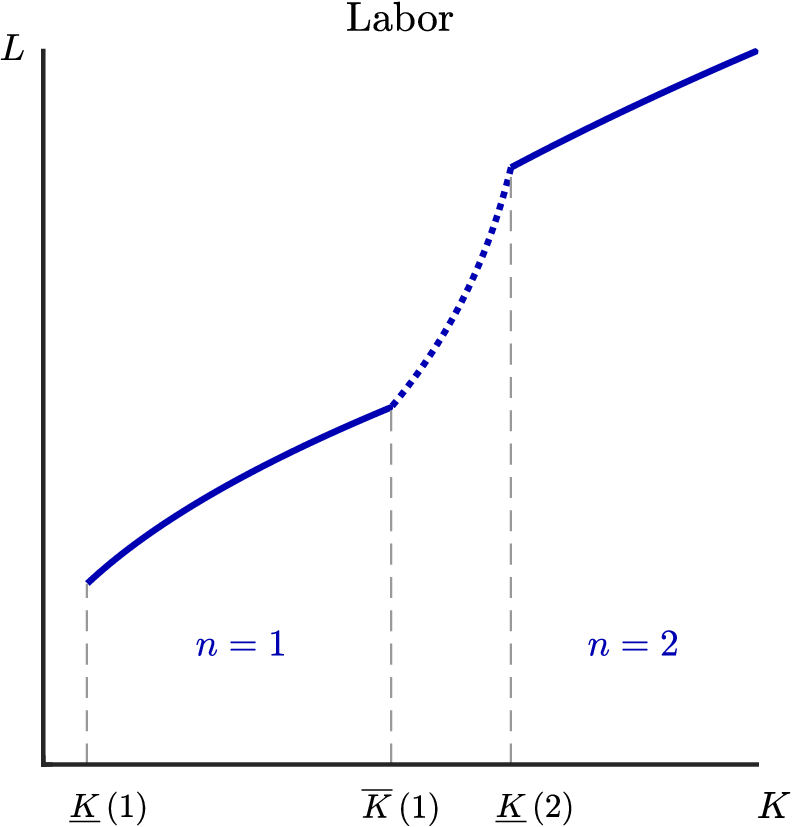}
	\hspace{0.25cm} 
	\includegraphics*[scale=0.385]{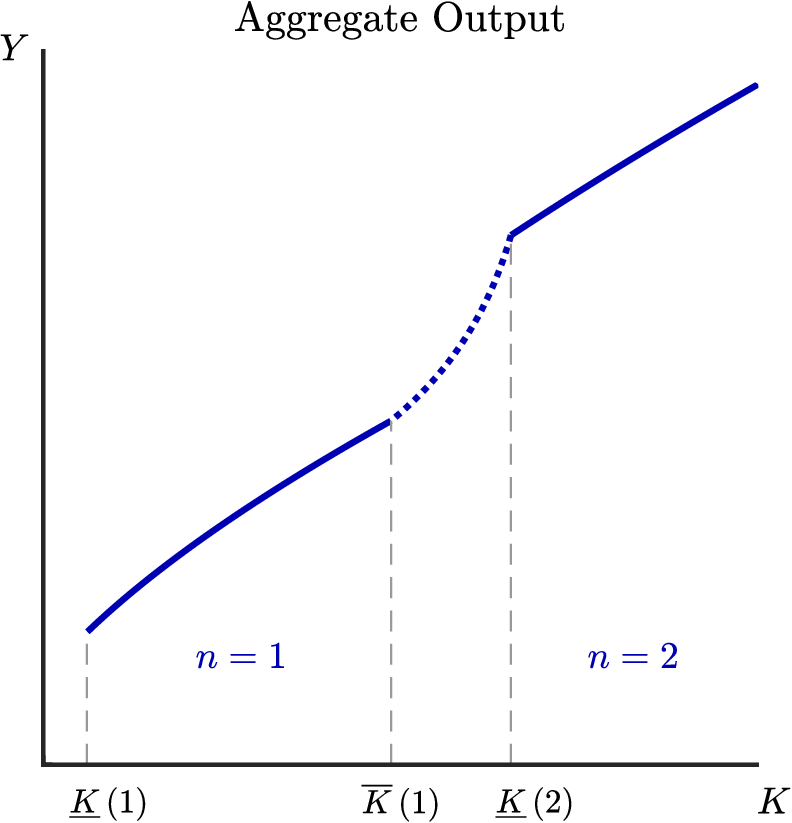} & 
\end{tabular}
\caption{Within-period equilibrium \protect \\ {Note: \small{} the figure shows how the factor price index, aggregate labor, and output move with capital. Solid segments represent regions with a symmetric equilibrium across product markets, while dotted segments represent non-symmetric regions. We use $\alpha \, = \, 1/3 \,$, $\rho \, = \, 3/4 \,$, $\eta \, = \, 1 \,$, $\nu \, = \, 2/5 \,$, $\gamma_{ij} \, = \, 1 \,$ and $ c_{i} \, = \, 0.015 \,$.}} 
\label{fig:Static-Equilibrium}
\end{figure}

\begin{prop}[Existence and uniqueness of a symmetric equilibrium]
\label{A-symmetric-equilibrium}
Under Assumption \ref{assumption:assumption1}, there exist $\underline{K}\left(\Gamma,n\right)$ and $ \overline{K}\left(\Gamma,n\right)$, with $\underline{K}\left(\Gamma,n\right)< \overline{K}\left(\Gamma,n\right)$ such that when $K_t\in[\,\underline{K}\left(\Gamma,n\right),\overline{K}\left(\Gamma,n\right)\,]$ the economy can sustain a symmetric equilibrium with $n$ firms in all markets. Furthermore, if
\[
\dfrac{\Phi\left(\Gamma,n\right)}{\Phi\left(\Gamma,n+1\right)}>\left[\dfrac{\Theta\left(\Gamma,n\right)}{\Theta\left(\Gamma,n+1\right)}\right]^{\frac{\rho}{1-\rho}-\frac{1-\alpha}{\nu+\alpha}}\, \quad \forall n,
\]
then there exists only one symmetric equilibrium for $K_{t} \in [\,\underline{K}\left(\Gamma,n\right),\overline{K}\left(\Gamma,n\right)\,], \: \forall n$. When there are no productivity differences across firms, this condition is equal to $\rho/\left(1-\rho\right)>\left(1-\alpha\right)\left(\nu+\alpha\right)$.
\end{prop}
\begin{proof}
See Appendix \ref{appendix:appendix_general_eqm}.
\end{proof}
Intuitively, $K_{t}$ must be sufficiently large so that all existing $n$ firms break even but cannot be too high, for otherwise, an additional firm could profitably enter some market. Figure \ref{fig:Static-Equilibrium} illustrates three aggregate variables as a function of $K_t$. When capital is within the bounds $\left[\underline{K}\left(\Gamma,1\right),\overline{K}\left(\Gamma,1\right)\right]$, the economy is characterized by a monopoly in every market; both labor and capital increase in the capital stock, but in a concave fashion (because of decreasing returns). When capital is above $\overline{K}\left(\Gamma,1\right)$, at least one product market can sustain a duopoly. The increase in competition translates into a higher factor price index and a higher labor supply. For this reason, output is locally convex in capital when $K_{t}\in\left[\overline{K}\left(\Gamma,1\right),\underline{K}\left(\Gamma,2\right)\right]$.\footnote{In the regions $\left[\overline{K}\left(\Gamma,n\right), \underline{K}\left(\Gamma,n+1\right)\right]$, some product markets have $n$ firms, and some others have $n+1$ firms. The number of markets with $n+1$ firms is such that the last firm exactly breaks even in these markets.}
The last part of Proposition \ref{A-symmetric-equilibrium} provides a condition for the uniqueness of symmetric equilibria in the special case in which firms are identical. If the elasticity of substitution $\sigma_I = 1/\left(1-\rho\right)$ is high if the capital elasticity $\alpha$ is large or if the inverse Frisch elasticity $\nu$ is large, it is easier for the condition for uniqueness to be satisfied. We return to the discussion on uniqueness in the calibrated version of our model.

\begin{figure}[htb]
\centering{}
\hspace{-0.2cm}
 \includegraphics*[scale=0.475]{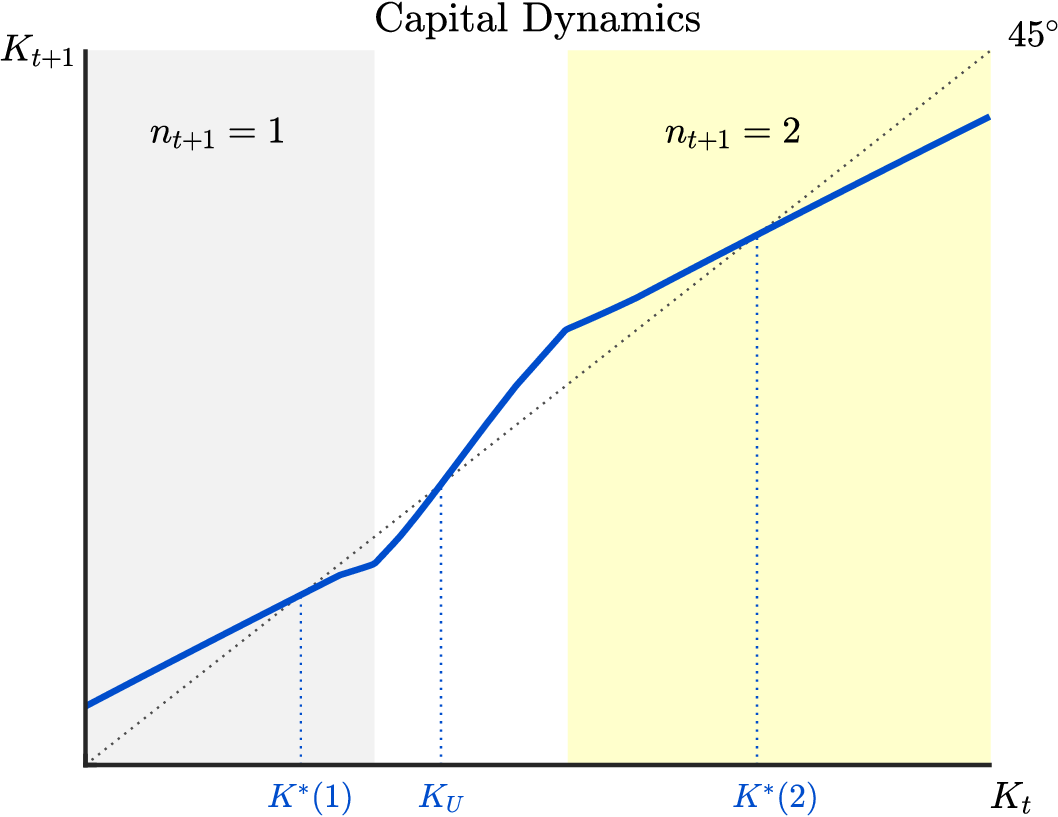}
 \hspace{0.4cm}
 \includegraphics*[scale=0.475]{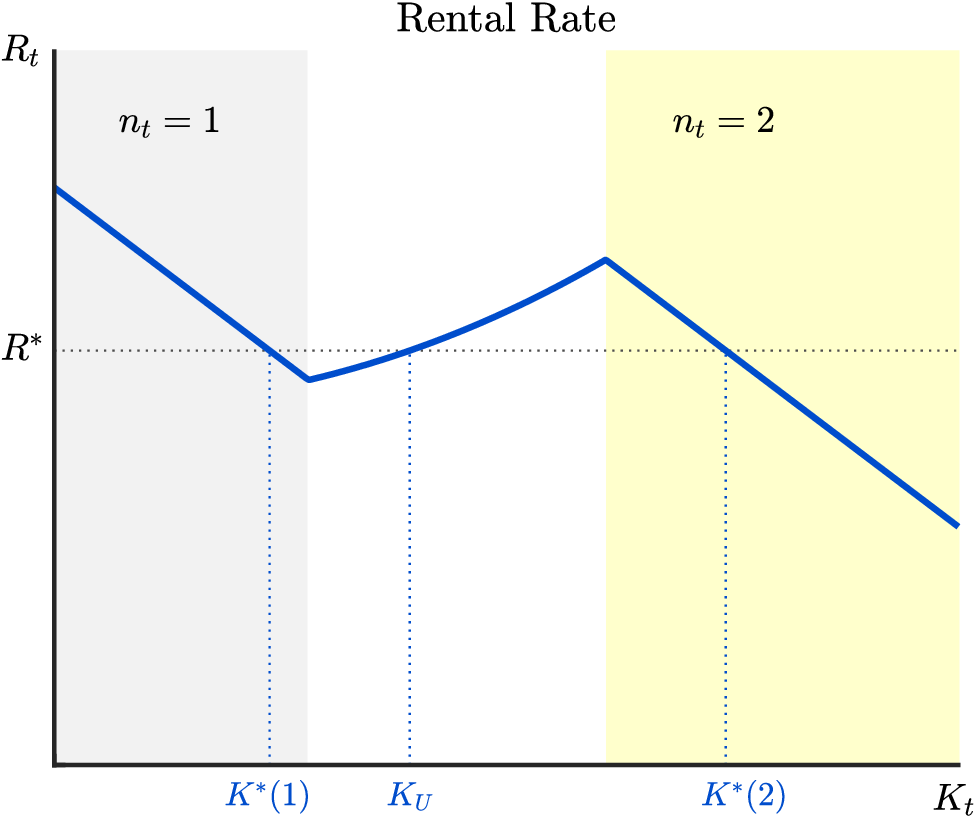}
\caption{Law of Motion and Rental Rate Map \label{fig:Law-of-Motion}
\protect \\ {Note: \small{}This example features two stable steady states and an unstable one. We use $\psi \, = \, 1 \,$,  $\rho \, = \, 3/4 \,$, $\eta \, = \, 1 \,$, $\alpha \, = \, 1/3 \,$, $\delta \, = \, 1 \,$, $\nu \, = \, 2/5 \,$ and $ c_{i} \, = \, 0.015 \,$.}} 
\end{figure}

We now focus on the equilibrium dynamics. 
The left panel of Figure \ref{fig:Law-of-Motion} represents the law of motion (\ref{eq:law-of-motion}) for a particular parameter combination. First, note that $K_{t+1}$ is not globally concave in $K_{t}$. Second, the economy features multiple steady-states: there are two stable steady-states, $K^{*}\left(1\right)$ and $K^{*}\left(2\right)$, and an unstable one, $K_{U}$.
The initial capital stock determines the steady-state to which the economy converges. If the economy starts with a capital stock $K_{0} < K_{U}$, it will monotonically converge to the low steady-state $K^{*}\left(1\right)$. Otherwise, it converges the high steady-state $K^{*}\left(2\right)$.

The shape of the law of motion (and the existence of multiple steady-states) can be explained by the interaction between competition, factor prices, and factor supply. Within the colored regions, the economy is characterized by the same market structure (same number of firms $n_{t+1}$). This ensures that the law of motion is concave within these regions, as in a standard neoclassical growth model. However, in the regions coinciding with changes in the market structure, the law of motion is convex. To understand this, consider again eq. (\ref{eq:law-of-motion}). The law of motion can be convex in capital if at least one of two conditions holds: i) $Y_{t}$ is convex in $K_{t}$ or ii) $s_{t}$ increases sufficiently fast in $K_{t}$. As already highlighted in Figure \ref{fig:Static-Equilibrium}, $Y_{t}$ can be convex in $K_{t}$ because of the positive impact of competition on labor supply. On the other hand, more intense competition can also result in a larger savings rate $s_t$ and, hence, a larger supply of capital. 
To sum up, relative to $K^{*}_{1}$, steady-state $K^{*}_{2}$ is characterized by a more competitive market structure and hence a larger supply of labor and capital by the representative household.

The right panel of Figure \ref{fig:Law-of-Motion} represents the rental rate map of this economy. Multiple steady-states occur whenever this map crosses the steady-state rental rate multiple times. A steady-state is characterized by a constant rental rate equal to $ R^{*} = \beta^{-1} - \left(1-\delta\right).$
Next, we characterize the condition for the existence of multiple steady states.

\begin{prop}[Existence of multiple steady-states]\label{prop:multiplicity} Suppose that Assumption \ref{assumption:assumption1} holds.
The economy features multiple symmetric steady-states if and only if there exists an $n\in \mathbb{N}$ such that
\begin{align}
    \Theta\left(\Gamma,n\right)^{\frac{1+\nu}{\nu+\alpha}}\:\overline{K}\left(\Gamma,n\right)^{-\nu\frac{1-\alpha}{\nu+\alpha}}<\dfrac{\beta^{-1}-\left(1-\delta\right)}{\alpha\;\left(1-\alpha\right)^{\frac{1-\alpha}{\nu+\alpha}}}<\;\Theta\left(\Gamma,n+1\right)^{\frac{1+\nu}{\nu+\alpha}}\:\underline{K}\left(\Gamma,n+1\right)^{-\nu\frac{1-\alpha}{\nu+\alpha}},
\end{align}
where $ \underline{K}\left(\Gamma,n\right)$ and $ \overline{K}\left(\Gamma,n\right)$ are defined in Appendix \ref{appendix:appendix_general_eqm}.
\end{prop}
\begin{proof}
See Appendix \ref{appendix:appendix_general_eqm}.
\end{proof}
Proposition \ref{prop:multiplicity} formalizes the idea that multiplicity obtains if there exists an increasing segment of the rental rate map and this segment crosses the steady-state interest rate $R^*=\beta^{-1} - (1-\delta)$. This condition depends on fundamental parameters such as the productivity distribution or the fixed production cost. For example, both $\underline{K}(\Gamma,n+1)$ and $\overline{K}(\Gamma,n)$ are strictly increasing in the fixed cost $c$. Therefore, as fixed costs change, the economy may enter/exit a region of steady-state multiplicity. Throughout this section, we assume that product markets are ex-ante symmetric and that the condition of Proposition \ref{prop:multiplicity} is satisfied, so that multiple steady-states exist. In Section \ref{sec:Quantitative-Evaluation}, we relax the assumption of product market symmetry and assess whether multiple steady-states arise or not under different calibrations of our model.

We can now define a key object of interest in our analysis. Suppose that the conditions of Proposition \ref{prop:multiplicity} are satisfied, so that the economy features multiple steady states. Given our focus on slumps, we are interested in understanding the likelihood of transitions between the high and low steady-states. To characterize how the likelihood of these events changes with model primitives, we define the concept of \emph{fragility}. 
Let $\mathcal{K}$ denote the ordered set of all steady-states levels of capital. Let $\mathcal{K^S}$ be the ordered collection of stable steady-states and $\mathcal{K^U}$ be the ordered collection of unstable steady-states in $\mathcal{K}$. We have the following definition.
\begin{definition}[Fragility]\label{definitionfragility}
Let $K^{*}\left(i\right)$ be the i-th element of $\mathcal{K}^S$ and $K_{U}\left(i\right)$ be the (i)-th element of $\mathcal{K}^U$. Let $\chi\left(i\right)\in\left(0,1\right)$ be defined as $\chi\left(i\right) \coloneqq K_{U}\left(i-1\right) / K^*\left(i\right)$. We say that, when $\chi\left(i\right)$ increases, $K^{*}\left(i\right)$ becomes more fragile. 
\end{definition}
To understand this definition, note that $\chi\left(i\right)$ captures the proximity of a \textit{stable} steady-state $K^{*}\left(i\right)$ to the preceding \textit{unstable} steady-state $K_{U}\left(i-1\right)$. The closer these two steady-states are to each other, the narrower the set of capital values $K_t$ that converge to $K^{*}\left(i\right)$ from below. The following thought experiment also helps us understand the concept of fragility. Suppose that the economy starts at the stable steady-state $K^{*}\left(i\right)$ and is hit by a one-time (unexpected) shock that destroys a fraction $\xi_{t}$ of the capital stock. The economy will converge back to $K^{*}\left(i\right)$ only if $\xi_{t} < 1 - \chi_{i}$. Therefore, if $\chi\left(i\right)$ is large enough (e.g., if it is close to one), even a very small negative shock can be enough to trigger a persistent downward transition.

Two further observations should be made. First, our main focus is on \textit{fragility}, which is different from the \textit{existence} of multiplicity. Although the first requires the second, these are different concepts. Second, the notion of \textit{fragility} is related to, but distinct from, the idea of the \textit{stability} of a steady-state. We think of \textit{fragility} as the possibility of downward transitions only. This is the size of the left partition of the basin of attraction of the steady-state, which is given by $\left[K_{U}\left(i-1\right), \, K^{*}\left(i\right)\right]$. The notion of stability, instead, relates to the size of the whole basin of attraction and also accounts for the possibility of upward transitions.

\subsection{Comparative Statics}

\begin{figure}[ht!]
\hspace{-0.0cm}
\begin{minipage}[b]{.5\linewidth}
\centering\includegraphics*[scale=0.5]{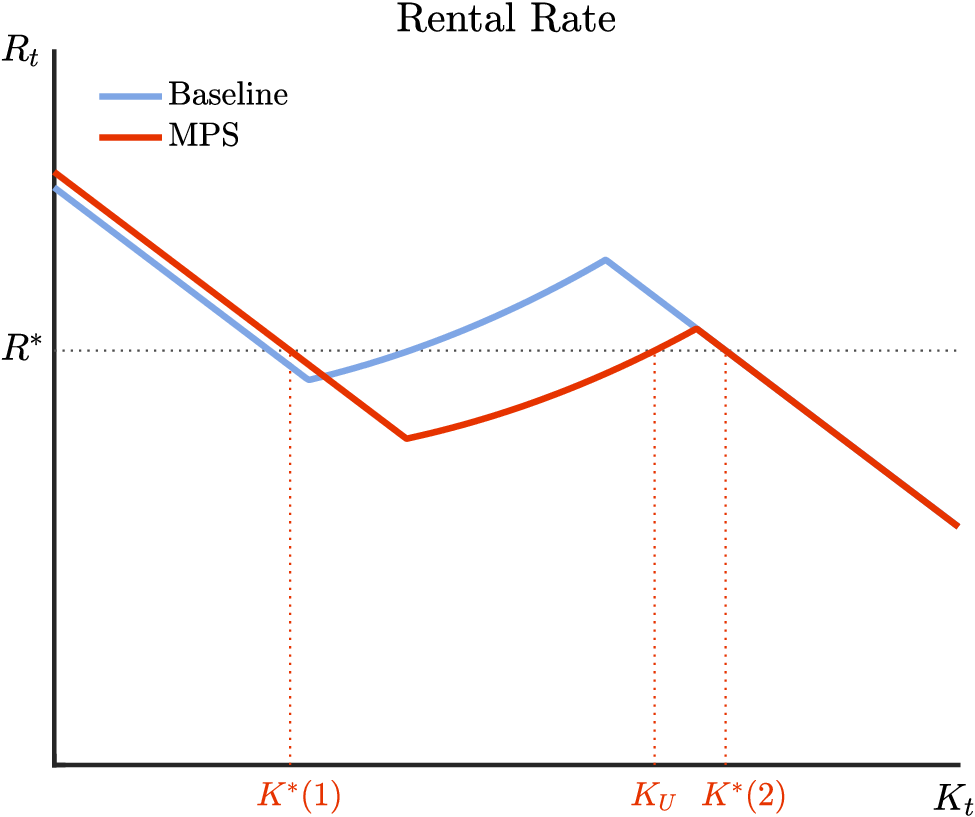}\subcaption{Effect of an MPS on capital demand} \label{fig:RR_MPS}
\end{minipage}
\vspace{0.75cm}
\begin{minipage}[b]{.5\linewidth}
\centering\includegraphics*[scale=0.5]{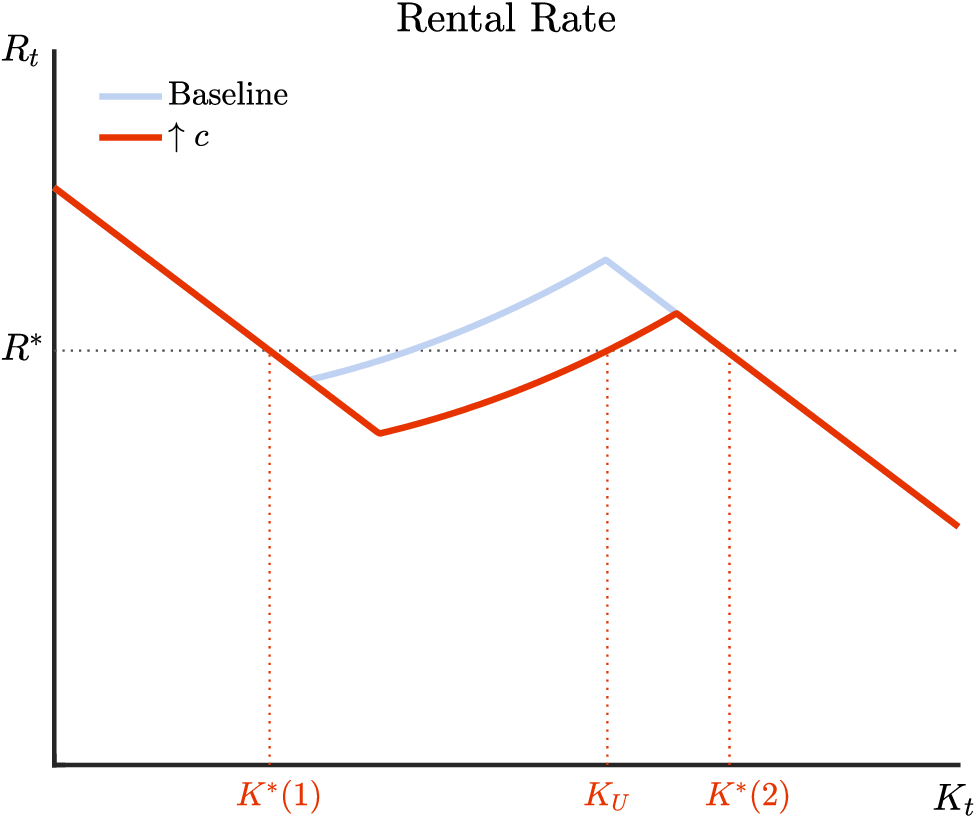}\subcaption{Effect of $\uparrow c$ on capital demand} \label{fig:LOM_MPS}
\end{minipage}
\begin{minipage}[b]{.5\linewidth}
\centering\includegraphics*[scale=0.5]{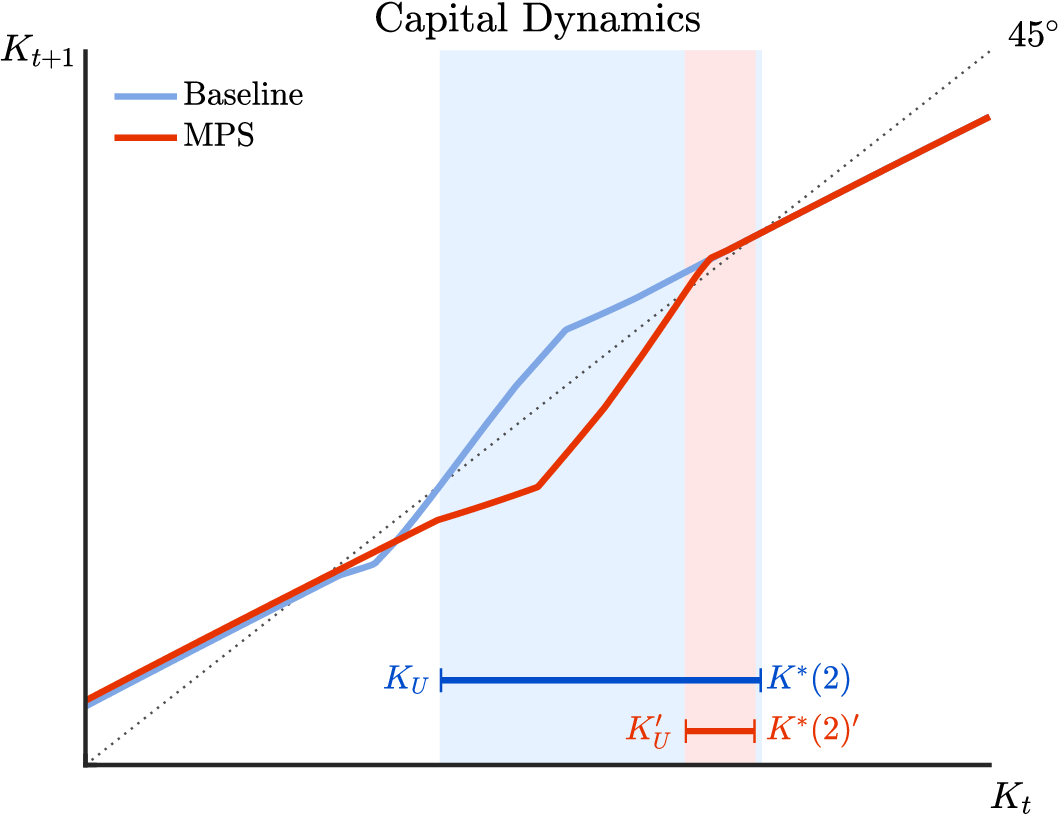}\subcaption{Effect of an MPS on the law of motion} \label{fig:RR_cf}
\end{minipage}
\begin{minipage}[b]{.5\linewidth}
\centering\includegraphics*[scale=0.5]{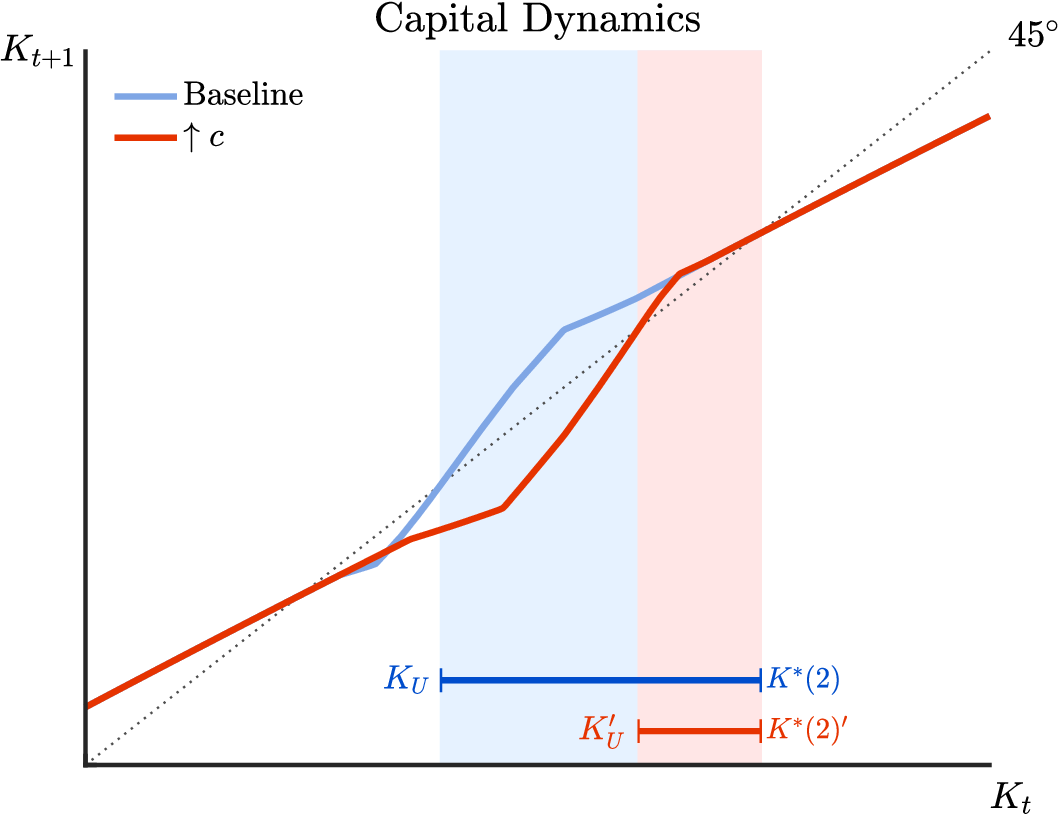}\subcaption{Effect of $\uparrow c$ on the law of motion} \label{fig:LOM_cf}
\end{minipage}
\caption{Comparative Statics \protect \\ {\small{} Panels (a) and (c) represent the effect of an MPS on capital demand and the law of motion of capital (in an economy with two steady-states). Panels (b) and (d) represent the effect of an increase in fixed costs. We use $\rho \, = \, 3/4 \,$, $\eta \, = \, 1 \,$,  $\alpha \, = \, 1/3 \,$ and $\nu \, = \, 2/5 \,$.}} \label{fig:Comparative-Statics}
\end{figure}

We next provide some comparative statics results (under the conditions of Assumption \ref{assumption:assumption1}). In particular, we study how fragility changes as i) active firms become more heterogeneous in terms of productivity and ii) fixed costs increase.

\paragraph*{Firm Heterogeneity}

We will study the impact of a mean-preserving spread (MPS) on the distribution of productivity. For tractability purposes, when analyzing an MPS, we will focus on an equilibrium with $n=2$ firms in every market. We start by showing that the factor price index $\Theta\left(\cdot\right)$ decreases in response to large productivity heterogeneity.

\begin{prop}[Firm heterogeneity, factor prices and factor shares]
\label{prop:MPS_factor_prices} 
Suppose that the conditions of Assumption \ref{assumption:assumption1} hold and that there is an equilibrium with $n = 2$ firms in all markets. Let $\Gamma=\left\{\gamma_1,\gamma_2\right\}$ be the vector of productivities and $\widetilde\Gamma$ be a mean-preserving spread on $\Gamma$. Then
\begin{enumerate}[label=\alph*)]
    \item $\Phi(\widetilde\Gamma,2)>\Phi(\Gamma,2).$
    \item $\Omega(\widetilde\Gamma,2)<\Omega(\Gamma,2).$
    \item $\Theta(\widetilde\Gamma,2)<\Theta(\Gamma,2).$
\end{enumerate}
\end{prop}
\begin{proof}
See Appendix \ref{appendix:appendix_general_eqm}.
\end{proof}

This result states that, in a steady-state with $n=2$ firms in all markets, an MPS to productivities decreases the factor price index and the aggregate factor share. As formalized by the first part of Proposition \ref{prop:MPS} below, this implies that the stable steady-state level of capital decreases.

\begin{prop} [Firm heterogeneity and fragility] \label{prop:MPS} Suppose that Assumption \ref{assumption:assumption1} is satisfied. Let ${K}^{*}\left(\Gamma, 2\right)$ be a stable steady-state with $n=2$ firms in all markets, and ${K}_{U}\left(\Gamma,1\right)$ be the preceding unstable steady-state. Let $\widetilde{\Gamma}$ be a mean-preserving spread on $\Gamma = \left\{\gamma_1,\gamma_2\right\}$. Then
\begin{enumerate}[label=\alph*)]
    \item ${K}^{*}(\widetilde{\Gamma}, 2)<{K}^{*}(\Gamma, 2)$,\label{mps_a}
    \item if $\rho>1-\dfrac{\nu\left(1-\alpha\right)}{1+\nu\alpha}$ then ${K}_{U}(\widetilde{\Gamma},1) > {K}_{U}(\Gamma,1)$. \label{mps_b}
\end{enumerate}
If condition \ref{mps_b} is satisfied, the steady-state $K^*(\widetilde\Gamma,2)$ becomes more fragile after the mean-preserving spread. Formally, $\chi^{*}(\tilde\Gamma,2)>\chi^{*}(\Gamma,2).$
\end{prop}
\begin{proof}
See Appendix \ref{appendix:appendix_general_eqm}.
\end{proof}

Before interpreting Proposition \ref{prop:MPS}, let us start by discussing the distributional consequences of an MPS. Fixing the number of firms, an MPS of idiosyncratic productivities, by increasing product market concentration, results in a lower factor share (Proposition \ref{prop:MPS_factor_prices}). This pushes the factor price index down, for a given level of aggregate TFP, as equation (\ref{eq:agg_cost_decomposition}) highlights. This is a \textit{market power effect} associated with higher firm heterogeneity, which results in lower factor prices. However, aggregate TFP will likely increase after an MPS: large, high-productivity firms become even more productive and increase their market shares. This pushes the factor price index up (equation (\ref{eq:agg_cost_decomposition})). This is an \textit{allocative efficiency effect}, which results in higher factor prices. The impact of an MPS on factor prices depends on the relative strength of these two forces.

Taken together, the two statements of Proposition \ref{prop:MPS} provide a sufficient condition under which fragility increases after an MPS.
Fragility increases after an MPS when $\rho$ is high enough. A high value of $\rho$ means that the degree of cross product market differentiation and average markups are relatively low. Small productivity differences are magnified, and entry becomes more difficult for a second player (which brings down aggregate factor shares and factor prices).

By construction, we increase the productivity of the top firms and reduce the productivity of the last one, keeping their average unchanged. However, fragility can also increase if we increase the productivity of the top firms, leaving the productivity of the last firm unchanged (see \cite{FQ_working_paper}).

\paragraph*{Rising Fixed Costs} The decline in product market competition since the 1980s may also be explained, through the lens of our model, by rising fixed costs. Higher fixed costs will trigger the exit of firms that are exactly breaking even, thus resulting in higher markups. The impact of higher fixed costs on the capital demand schedule can be understood in two steps. First, in a steady-state with $n$ firms where all firms make strictly positive profits, the capital level of such a steady-state is unaffected by a marginal increase in $c$. On the other hand, the larger fixed costs result in a rightward shift of the unstable steady-state. Recall that the unstable steady-state belongs to a region where some firms are exactly breaking even. These firms survive with a larger $c$ only if the capital stock (and hence their level of profits) is higher. Proposition \ref{prop:basin_attraction_cf} formalizes these results.

\begin{prop}[Fixed costs and fragility] \label{prop:basin_attraction_cf} Suppose that the conditions of Assumption \ref{assumption:assumption1} are satisfied. Let ${K}^{*}(c, n)$ be a stable steady-state with $n$ firms in every market and ${K}_{U}(c, n-1)$ be the preceding unstable steady-state. Then
\begin{enumerate}[label=\alph*)]
    \item $\partial K^*(c,n) \,/ \partial c \: \leq \: 0 \, \forall n$, \label{cf_a}
    \item $\partial K_U(c,n-1) / \partial c \: > \: 0  \, \forall n$. \label{cf_b}
\end{enumerate}
It follows that any stable steady-state becomes more fragile as $c$ increases, i.e. $\partial \chi(c,n) \,/ \partial c \: > \: 0 \, \forall n$.

\end{prop}
\begin{proof} See Appendix \ref{appendix:appendix_general_eqm}.
\end{proof}

Panels (b) and (d) of Figure \ref{fig:Comparative-Statics} illustrate the effects of an increase in fixed costs. The two declining segments of the rental rate map represent a situation of full monopoly and full duopoly; since all firms make strictly positive profits in these regions, the equilibrium is unchanged as $c$ increases marginally. In the increasing segment, where some firms are exactly breaking even, some are driven out of the market. The rental rate decreases, and the unstable steady-state increases. As a result, the larger steady-state becomes more fragile.

These results also shed light on the conditions of Proposition \ref{prop:multiplicity} (e.g., how a change in fixed costs can affect the existence of multiple steady-states). This is made clear in Figure \ref{fig:Comparative-Statics}: a sufficiently large increase in fixed costs can shift the rental rate map downward such that it crosses $R^{*}$ only once. In such a case, only one steady-state exists, where all product markets are monopolies: $K^{*}\left(1\right)$. The level of capital $K^{*}\left(2\right)$ is no longer a steady-state. The opposite would happen if fixed costs were to experience a sufficiently large decrease. In such a case, only $K^{*}\left(2\right)$ would be a steady-state.

\paragraph{Discussion}
We conclude by summarizing two key insights of our theory, which can be relevant to understanding the US growth experience after 2008. The first is that complementarity between competition and factor supply can generate multiple competition regimes or steady-states. A transition from a high to a low competition regime can, in many aspects, describe the 2008 recession and the subsequent great deviation. The second insight is that changes in technology that result in larger market power (e.g., larger productivity differences across firms or larger fixed costs) make high competition regimes more difficult to sustain, and transitions to low competition traps more likely to occur. Our model, therefore, suggests that the US economy, experiencing a long-run increase in markups and concentration since the 1980s, became increasingly vulnerable to transitions like the one observed after 2008.

In the next section, we use a calibrated version of our model and study its quantitative predictions. We then ask whether it can replicate the behavior of the US economy in the aftermath of the 2008 crisis and finally study the welfare gains from different policy interventions.

\section{Quantitative Results \label{sec:Quantitative-Evaluation}}
The goal of this section is to 
provide a quantification of the forces described in our model, and evaluate policy counterfactuals. To calibrate some of the model parameters, we use moments from public firms (COMPUSTAT). Since public firms represent just part of the entire economy, we assume that there are two main sectors, an \textit{uncompetitive} sector (which is calibrated to match firms in COMPUSTAT), and a \textit{competitive} sector (where firms are identical). We denote by $f_{u}$ the fraction of the total number of markets $I$ that belong to the \textit{uncompetitive} sector.

\paragraph{Uncompetitive Sector} Firms in the \textit{uncompetitive} sector draw their idiosyncratic productivities from a log normal distribution with standard deviation $\lambda$, $\log \gamma_{ij} \sim \textrm{N}\left( 0, \lambda \right).$
Each product market $i$ is characterized by $M$ such draws. Since $M$ is a finite number, product markets have different ex-post distributions of idiosyncratic productivities $\left\{ \gamma_{ij} \right\}_{j=1}^{M}$.

We assume that fixed costs can take one of two values $c_i = \left\{0,c\right\}$.\footnote{We choose this parsimonious distribution for two reasons. First, by having some markets with zero fixed costs, we greatly reduce the time required to solve the model. Second, we impose a unique value $c>0$ in the other markets for simplicity. Product markets could differ in many dimensions: fixed costs, elasticities of substitution, and number of potential players. We abstract from these sources of heterogeneity to keep the model and the calibration as simple and transparent as possible.} We denote by $x_c$ the fraction of \textit{uncompetitive} markets subject to positive fixed costs.
In product markets with a zero fixed cost, the extensive margin is muted as all potential $M$ players are always active. However, these product markets do not necessarily operate close to perfect competition, as there can be large productivity and markup differences across firms.

\paragraph{Competitive Sector} 
Firms in the \textit{competitive} sector are identical, with a common idiosyncratic productivity $\gamma_{ij}=\bar{\gamma}$, and pay no fixed costs. Therefore, in these markets, all firms have a market share of $1/M$ and charge a markup $1/\left(\eta-\left(\eta-\rho\right)/M\right)$. $\bar{\gamma}$ is set to the average productivity draw made by firms on the \textit{uncompetitive} sector (both active and inactive). 

\subsection{Calibration}
The model is calibrated to match firm moments at three different points in time: 1975, 1990, and 2007. Our calibration relies on the assumption that all parameters are time-invariant, with the exception of the productivity dispersion parameter $\lambda$ and the fixed cost $c$. This allows us to compare features of the model in economies with different levels of firm heterogeneity and fixed costs.
We calibrate the model at a quarterly frequency, under the assumption that the economy always starts at the highest steady-state (when multiple steady-states exist).

We design our calibration to not depend on assumptions about a particular level of aggregation, i.e., how to map a market $i$ in our model to a market $i$ in the data. While for some tradeable-good firms, the relevant competitive market might be a global 10-digit industry, for local service providers, it might be as narrow as a neighborhood. Our calibration choices imply that we need not take a stance on this mapping and that we can think of some product markets in our model as a combination of \textit{industry}$\times$\textit{location} in the data \citep[see][for a more complete discussion]{eeckhoutcomment, product_market_concentration}.\footnote{As \cite{product_market_concentration} report, some industries can be too broad for some products (e.g., NAICS 325620 contains products such as after-shave, mouthwash or sunscreen). In other cases, highly substitutable products belong to different industries (e.g., metal cans or glass bottles).}

\paragraph{Externally Calibrated Parameters}
Some parameters are standard and taken from the literature. For the preference parameters, we set $\beta = 0.99$ and $\psi=1$ (log utility). We set $\nu = 0.302$, which implies a Frisch elasticity of 3.31. This corresponds to the average macro elasticity of hours reported by \cite{CGMW}. Finally, we set the total number of product markets to $I=10,000$ and the maximum number of firms in a market to $M=20$.\footnote{Contrarily to \cite{de2021quantifying} we fix this parameter across the different calibrations. We have also considered $M=50$ and $M=100$ and the results were identical.} 

\paragraph{Productivity and fixed costs} 
We calibrate the standard deviation of the exogenous productivity distribution $\left(\lambda\right)$ to match dispersion in firm size. 
Specifically, we compute the standard deviation of log revenues for all firms in the \textit{uncompetitive} sector and use as a target the corresponding moment in COMPUSTAT.
Figure \ref{fig:intro_figure_2} shows the evolution of our dispersion measure in COMPUSTAT.
We calibrate the fixed cost parameter to match the average fixed to total costs ratio in COMPUSTAT. We define fixed costs as `Selling, General and Administrative Expenses'. In total costs, we include, in addition to fixed costs, the item `Cost of the Goods Sold'. We obtain a target level of 21.9\% for 1975, 31.7\% for 1990, and 36.9\% for 2007 (see Figure \ref{fig:intro_figure_2}). 

\paragraph{Elasticities of substitution} 
For the two elasticities of substitution, we assume that they are time-invariant and calibrate them to match the sales-weighted average markup of public firms in 1975, 1990, and 2007 \citep[as reported by ][]{LEU}. Conditional on the other time-varying parameters, the elasticities are informative about the level of markups charged by firms in the economy. Note that this gives us 3 moments for 2 parameters to be calibrated.

\paragraph{Sector shares} 
To calibrate $f_u$, we target the employment share of COMPUSTAT. As of 2007, public firms represented 37.8\% of aggregate nonfarm employment. We thus calibrate $f_u$ so that the \textit{uncompetitive} sector represents 37.8\% of total employment in the 2007 economy.

To calibrate $x_c$ (the share of markets with $c_{i} = c$), we target the employment share in highly concentrated industries. In the calibrated model, markets with $c_{i} = c$ will have few firms (typically less than 4). In an ideal setting, we would have information on the employment share of product markets with 4 or fewer firms. 
However, we only have information of employment shares at the industry level (6-digit NAICS). Since this will understate concentration in finer markets, we focus on industries where the top 4 firms represent more than 90\% of sales the top 8. Using data from the US Census, we find that 7\% of aggregate employment is allocated to such industries in 2007.

\paragraph{Exogenous TFP} 
We also need to calibrate the two parameters governing the dynamics of aggregate productivity: the autocorrelation parameter $\phi_A$ and the standard deviation of the innovations $\sigma_{\varepsilon}$. We do so by targeting the first-order autocorrelation and the standard deviation of output.\footnote{We compute these moments for the entire postwar period 1947\textendash{}2019. Consistent with our interpretation that the US economy moved to a different regime after 2008, we remove a linear trend computed for the period 1947\textendash{}2007.} We calibrate these parameters using our 2007 model and keep them unchanged in the 1975 and 1990 models.

\begin{table}[ht!]
\setlength{\tabcolsep}{0.1cm} \begin{center} \resizebox{0.8\textwidth}{!} {  \begin{tabular}{lccccl}
\thickhline
\\[-1ex]
{\large Description} & {\large Parameter} & \multicolumn{3}{c}{{\large Value}} & {\large Source/Target} \\[1ex] \thickhline
\\[-1ex]
[A] External Parameters \\ \\[-1ex] \hline
\\[-1ex]
Capital elasticity & $\alpha$ & \multicolumn{3}{c}{0.3} & Standard value \\
\\[-1ex]
Depreciation rate & $\delta$ & \multicolumn{3}{c}{$1 - 0.9^{1/4}$} & Standard value \\
\\[-1ex]
Discount factor & $\beta$ & \multicolumn{3}{c}{$0.96^{1/4}$} & Standard value \\
\\[-1ex]
Inverse of Frisch elasticity& $\nu$ & \multicolumn{3}{c}{0.352} & \cite{CGMW}\\
\\[-1ex]
Coefficient of risk aversion & $\psi$ & \multicolumn{3}{c}{1} & log utility \\
\\[-1ex]
Maximum number of firms per market & $M$ & \multicolumn{3}{c}{20} & \\
\\[-1ex]
Number of product markets & $I$ & \multicolumn{3}{c}{10,000} & \\
\\[-1ex]
\hline
\\[-1ex]
[B.1] Calibrated Parameters: Fixed \\ \\[-1ex]
\hline
\\[-1ex]
Between product markets ES & $\sigma_{I}$ & \multicolumn{3}{c}{1.46} & Sales-weighted average markup\\
\\[-1ex]
Within product market ES & $\sigma_{G}$ & \multicolumn{3}{c}{11.50} & Sales-weighted average markup\\
\\[-1ex]
Share of \textit{uncompetitive} sector & $f_u$ & \multicolumn{3}{c}{0.42} &  Emp share COMPUSTAT \\
\\[-1ex]
Share of \textit{uncompetitive} markets with $c_{i}>0$ & $x_c$ & \multicolumn{3}{c}{0.28} &  Emp share concentrated industries \\
\\[-1ex]
Persistence of $A_t$  & $\phi_{A}$ & \multicolumn{3}{c}{0.950} & Autocorrelation of log $Y_t$ \\
\\[-1ex]
Standard deviation of $\varepsilon_t$ & $\sigma_{\varepsilon}$ & \multicolumn{3}{c}{0.003} & Standard deviation of log $Y_t$ \\ 
\\[-1ex] \hline
\\[-1ex]
[B.2] Calibrated Parameters: Variable & & 1975 & 1990 & 2007 & \\ \\[-1ex] \hline
\\[-1ex]
Standard deviation of $\gamma_{ij}$ & $\lambda$ & 0.182 & 0.213 & 0.232 & Std log revenues \\
\\[-1ex]
Fixed cost ($\times 10^{-3}$) & $c$ & 0.351 & 0.691 & 0.751 & Average ratio fixed/total costs \\
\\[-1ex]
\thickhline
\end{tabular}  
} \end{center}
\caption{Parameter Values\label{tab:parameter_values}}
\end{table}

\begin{table}[ht]
\setlength{\tabcolsep}{0.2cm}  
\begin{center}
\resizebox{12.5cm}{3cm} {
\begin{tabular}{lcccccccc} 		\thickhline 
\\[-2ex]
& \multicolumn{2}{c}{1975} & & \multicolumn{2}{c}{1990} & & \multicolumn{2}{c}{2007} \\
\\[-1ex] 
& Data & Model & & Data & Model & & Data & Model \\ \hline  
\\[-1ex] 
Sales-weighted average markup & 1.28 & 1.28 & & 1.37 & 1.38 & & 1.46 & 1.46 \\
\\[-1ex]
Std log revenues & 1.59 & 1.59 & & 1.91 & 1.88 & & 2.04 & 2.04 \\
\\[-1ex]
Fixed to total cost ratio (average) & 0.219 & 0.220 & & 0.317 & 0.341 & & 0.369 & 0.370 \\
\\[-1ex] \hline 
\\[-1ex]
Emp share COMPUSTAT firms &  &  & &  & & & 0.378 & 0.383 \\
\\[-1ex]
Emp share \textit{concentrated} product markets &  &  & &  & & & 0.07 & 0.07 \\
\\[-1ex]
Autocorrelation log GDP  &  &  & &  &  & & 0.978* & 0.975 \\
\\[-1ex]
Standard deviation log GDP &  &  & &  &  & & 0.061* & 0.063 \\
\\[-1ex]
\thickhline
\\[-1ex]
\multicolumn{6}{l}{*computed over 1947:Q1-2019:Q4} \\
\end{tabular} } \end{center}
\caption{Targeted moments and model counterparts\label{tab:targeted_moments}}
\end{table}

\begin{table}[ht]
\setlength{\tabcolsep}{0.2cm}  
\begin{center} \resizebox{0.6\textwidth}{!} {
\begin{tabular}{lcccccccc} 		\thickhline 
\\[-2ex]
& \multicolumn{2}{c}{1975} & &
\multicolumn{2}{c}{1990} & &
\multicolumn{2}{c}{2007} \\
\\[-1ex] 
& Data & Model & & Data & Model & & Data & Model
\\ \hline
\\[-1ex]
[A] Total Economy
\\[-1ex] \\ \hline
\\[-1ex]
Cost-weighted markup (COMPUSTAT) & 1.28 & 1.26 & & 1.28 & 1.32 & & 1.33 & 1.34 \\
\\[-1ex]
HHI: 75th percentile &&&&&&& 0.335 & 0.150\\[2ex]
HHI: 90th percentile &&&&&&& 0.484 & 0.500\\[2ex]
\hline
\\[-1ex]
[B] Markets with positive fixed costs
\\[-1ex] \\ \hline
\\[-1ex]
Number of firms per product market & & 2.50 & & & 1.74 & & & 1.51 \\[2ex]
\thickhline
\end{tabular} }\end{center}
\caption{Non-targeted moments
\protect \\ {Note: \small{} The cost-weighted markup is from \cite{LEU}. The same model moment is computed for the set of COMPUSTAT firms. Data on the distribution of HHI refers to local US product markets and is from \cite{product_market_concentration}. The same model moment is computed using all product markets.}
\label{tab:nontargeted_moments}}
\end{table}

Table \ref{tab:parameter_values} reports our parameter values, while Table \ref{tab:targeted_moments} reports our targeted moments, with their model counterparts. We obtain a value of 1.46 for the cross-product market elasticity of substitution, and a value of 11.50 for the within-product market elasticity. These are in line with the estimates from similar studies using US data, such as \cite{EMX}. The model successfully matches the sales-weighted markups and dispersion in revenues in all three years. We slightly overestimate the fixed cost ratio in 1990.

Part [A] of Table \ref{tab:nontargeted_moments} reports the evolution of some non-targeted moments. Even if we target sales-weighted average markups, 
cost-weighted average markups are close to their data counterparts, as reported by \cite{LEU}. In particular, in 2007, our COMPUSTAT firms had an average cost-weighted markup of 1.34, very close to the same data moment of 1.33.
We also report the 75th and 90th percentiles of the distribution of HHIs (for all product markets) obtained under the 2007 calibration. These numbers are compared to the same moments reported by \cite{product_market_concentration}, who estimate concentration metrics for narrowly defined consumption-based product markets. The model closely matches the 90th percentile of the empirical distribution of HHIs. An HHI of 0.5 is the one that would be obtained under a symmetric duopoly. This implies that both in our model and in the data, 10\% of the markets have concentration metrics equal to or higher than that of a symmetric duopoly. Therefore, even if we do not impose a particular level of aggregation in our calibration, the model can replicate the concentration distribution for consumption-based product markets.

Product markets facing positive fixed costs play an important role in our mechanism. Part [B] of Table \ref{tab:nontargeted_moments} describes these product markets in the three calibrated economies. Product markets with positive fixed costs consist mostly of monopolies and duopolies \textemdash{} the number of firms per market is 2.50 in the 1975 economy, 1.74 in 1990, and 1.51 in 2007. This level of concentration is consistent with the high levels of HHIs observed in the data.

\subsection{Quantitative Results}

\begin{figure}[htbp]
\hspace{-1cm}
\begin{minipage}[b]{.33\linewidth}
\centering\includegraphics*[scale=0.35]{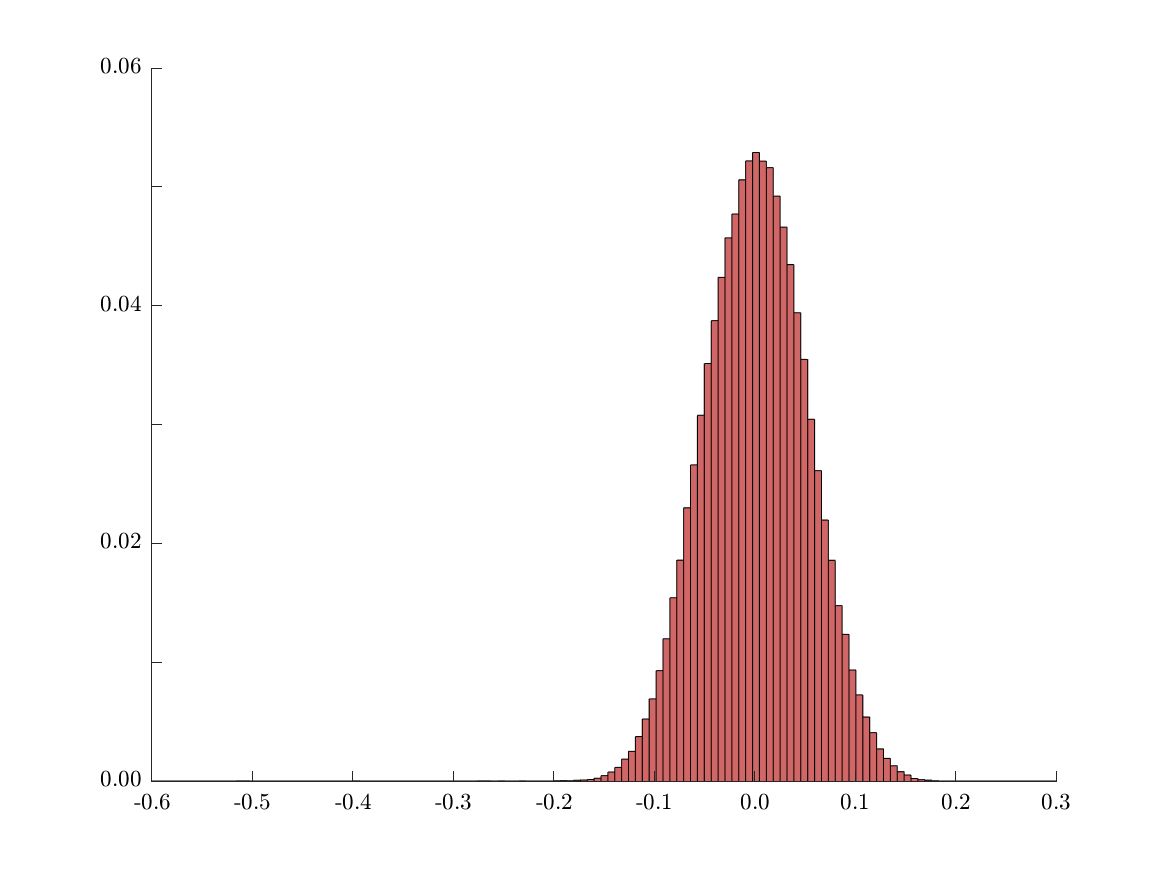}\subcaption{1975} \label{fig:y_dist_1975}
\end{minipage}%
\begin{minipage}[b]{.33\linewidth}
\centering\includegraphics*[scale=0.35]{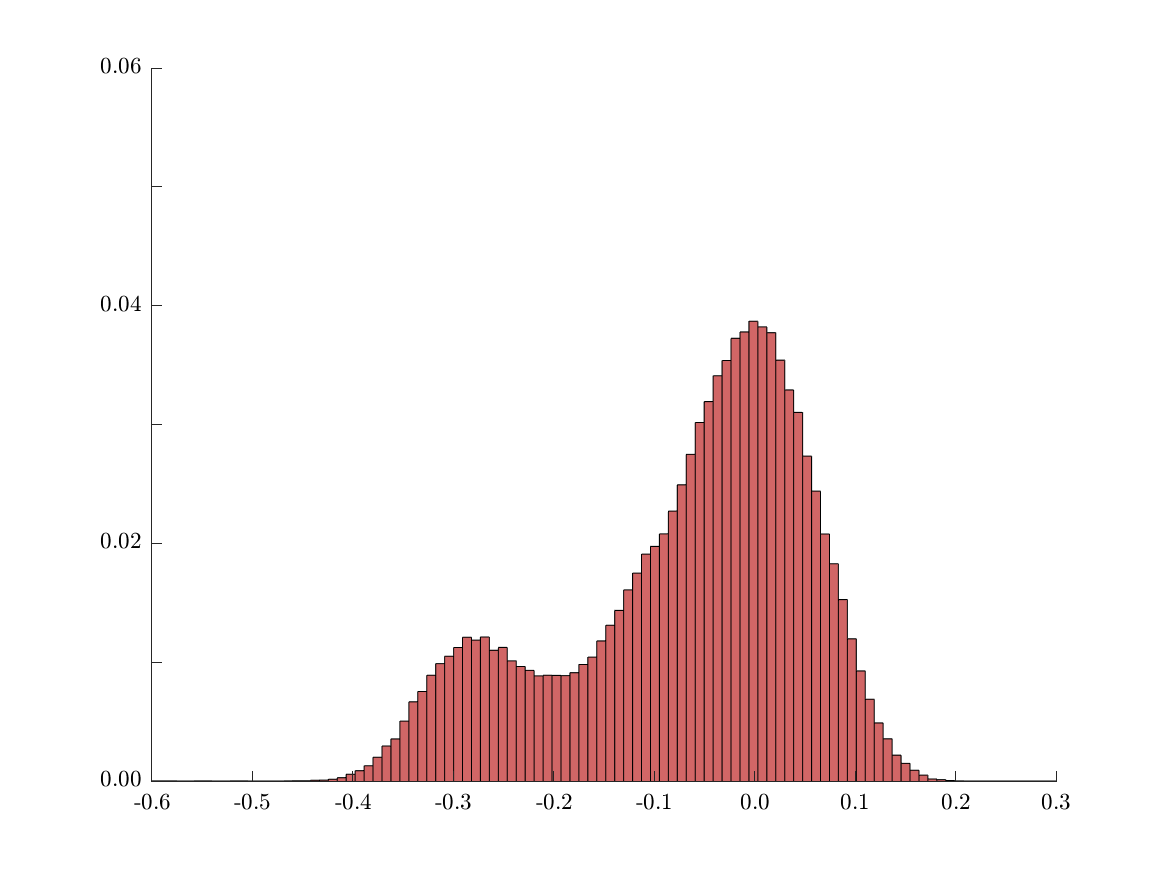}\subcaption{1990} \label{fig:y_dist_1990}
\end{minipage}
\begin{minipage}[b]{.33\linewidth}
\centering\includegraphics*[scale=0.35]{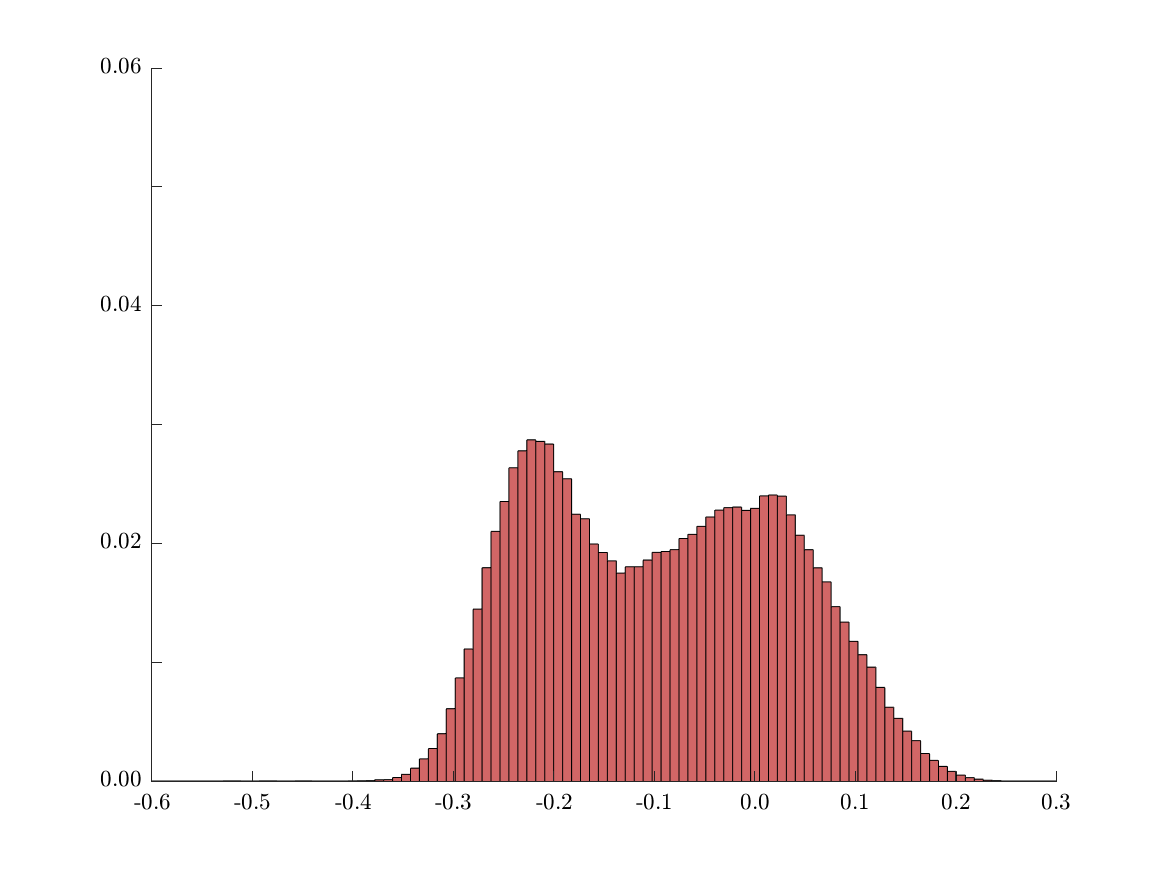}\subcaption{2007} \label{fig:y_dist_2007}
\end{minipage}
\caption{Ergodic distribution of output \protect \\ {Note: \small{}\small{} This figure shows the distribution of log output for the 1975, 1990, and the 2007 economies. We simulate each economy for 10,000,000 periods and plot output in deviation from the high steady-state.}} \label{fig:businesscycle}
\end{figure}

Having calibrated the model, we solve the full dynamic problem, approximating the policy function of the household by iterating on
the Euler equation. We describe the algorithm in the Online Appendix \ref{sec:Quantitative-Model}. 
We start by exploring the dynamic properties of the 1975, 1990 and 2007 economies. Figure \ref{fig:businesscycle} shows the ergodic distribution of log output; the distributions are centered around the highest mode so that the horizontal axis represents output in percentage deviation from the highest steady-state. We highlight three important observations. First, while the ergodic distribution of the 1975 economy is unimodal, the other two economies feature bimodal distributions, implying that these two economies are characterized by multiple steady-states.\footnote{Online Appendix \ref{sec:Quantitative-Model} provides the location of these non-stochastic steady-states and their stability properties.} Through the lens of our model, multiple competition regimes are possible in economies characterized by levels of markups and fixed costs observed in 1990 and 2007, but not in 1975, when markups and fixed costs were lower.\footnote{As highlighted in the discussion of Proposition \ref{prop:multiplicity} and later of Figure \ref{fig:Comparative-Statics}, changes in fixed costs can affect the condition for the existence of multiple steady-states.}
Second, relative to the 1990 economy, the 2007 model features a larger probability mass on the left, suggesting that the economy spends, on average, more time on the lowest regime, characterized by lower competition and output. Third, in the 2007 distribution, the two steady-states are also closer to each other \textemdash{} a transition from the high to the low regime implies a 21\% reduction in output in 2007, as opposed to approximately 30\% in 1990.
While this means that transitions are less pronounced in 2007, it also implies that they are substantially more likely in 2007 than in 1990, as discussed below. The 1975 economy, instead, behaves similarly to a standard RBC model. This economy can suffer temporary recessions, but these will not have long-lasting consequences.

\begin{figure}[ht]
\hspace{-0.0cm}
\begin{minipage}[b]{.5\linewidth}
\centering\includegraphics*[scale=0.75]{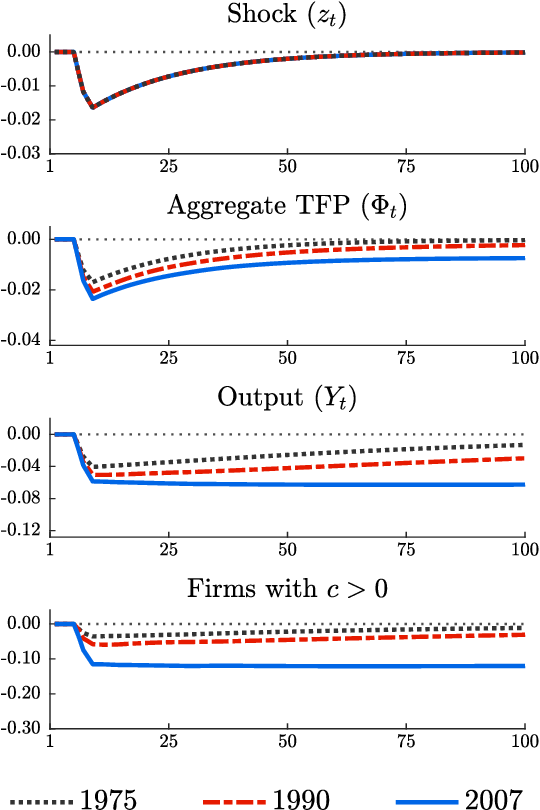}\subcaption{Small Negative Shock} \label{fig:IRF_small_shock}
\end{minipage}%
\hspace{-0.75cm}
\begin{minipage}[b]{.5\linewidth}
\centering\includegraphics*[scale=0.75]{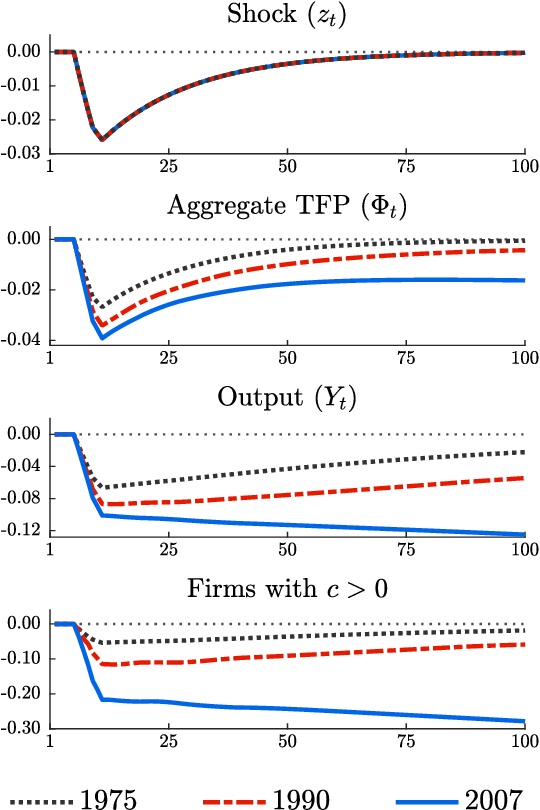}\subcaption{Large Negative Shock} \label{fig:IRF_big_shock}
\end{minipage}
\caption{Impulse Response Functions\protect \\ {Note: \small{} The graphs show the IRFs to an exogenous TFP shock. In panel (a), we feed a shock $\varepsilon_t = - \sigma_{\varepsilon}$ that lasts for four quarters. In panel (b), we feed $\varepsilon_t = - 2 \, \sigma_{\varepsilon}$ during six quarters.}} \label{fig:IRF_figure}
\end{figure}

\paragraph{Impulse Response Functions: Small Negative Shock}
We start by characterizing the reaction of the three economies to a small negative shock. We consider a shock to the innovation of the exogenous TFP process that is equal to $\varepsilon_t = - \sigma_{\varepsilon}$  and lasts for four quarters. Figure \ref{fig:IRF_small_shock} shows the impulse responses. The simulation of the transition dynamics covers 100 quarters.  This shock generates different responses for the three economies. The 2007 economy exhibits both greater amplification and persistence.  First, the 1975 economy experiences a 4.0\% reduction in aggregate output after 5 quarters, against a 5.1\% in 1990 and a 5.9\% reduction in the 2007 economy. Second, after 100 quarters, the 1975 economy is 1.3\% below the steady-state, while the 1990 economy is 3.0\% below trend, and the 2007 economy has a much more prolonged downturn, being still 6.3\% below pre-crisis output. Eventually, they converge back to the initial level.

The mechanism underlying such increased amplification and persistence can be understood by looking at the bottom panel, which plots the transition dynamics of the number of firms in markets with positive fixed cots. In 2007, there is a much more significant reduction in the number of firms due to the mechanisms outlined above: increased productivity dispersion and larger fixed costs make small, unproductive firms more sensitive to aggregate shocks. 

\paragraph{Impulse Response Functions: Large Negative Shock}

The shock introduced above was small enough to make all three economies converge to their initial steady-states, albeit in different time horizons. We now study the effect of a larger shock. We repeat the same exercise for the three economies, but now introduce a negative shock $\varepsilon_t = - 2 \sigma_{\varepsilon}$, lasting six quarters. The dynamics are shown in Figure \ref{fig:IRF_big_shock}. As before, there is greater amplification and persistence in the 2007 economy. However, this economy now experiences a permanent drop in aggregate output, i.e., it transitions to a lower steady-state. In the example we consider, after 100 quarters, output is 12.5\% below its initial value, and the gap is still widening at the end of the sample. The 1975 and 1990 economies, on the other hand, converge back to the pre-shock levels.

\begin{table}[htbp]
\setlength{\tabcolsep}{0.15cm}  
\begin{center}
\resizebox{0.7\textwidth}{!} {
\begin{tabular}{lccccccccc} 		\thickhline 
\\[-1.5ex]
&&  \multicolumn{2}{c}{1975 Model} & & \multicolumn{2}{c}{1990 Model} & & \multicolumn{2}{c}{2007 Model} \\ [0.5ex]
\hline  
\\ [-1ex]
& & T = 40 & T = 100 & & T = 40 & T = 100 & & T = 40 & T = 100
\\[-2ex]
\\ \hline
\\[-1ex] 
$\textrm{Pr} \big[$ 10\% recession $\big]$ & & 0.017 & 0.067 & & 0.124 & 0.264 & & 0.196 & 0.357 \\
\\[-1.75ex]
$\textrm{Pr} \big[$ 15\% recession $\big]$ & & 0.000 & 0.015 & & 0.024 & 0.108 & & 0.066 & 0.205 \\
\\[-1.75ex]
$\textrm{Pr} \big[$ 20\% recession $\big]$ & & 0.000 & 0.000 & & 0.003 & 0.047 & & 0.005 & 0.076 \\
\\[-1.75ex]
\thickhline
\\[-1ex] 
$\textrm{Pr} \big[$ 10\% expansion $\big]$ & & 0.019 & 0.068 & & 0.250 & 0.438 & & 0.184 & 0.355 \\
\\[-1.75ex]
$\textrm{Pr} \big[$ 15\% expansion $\big]$ & & 0.000 & 0.000 & & 0.108 & 0.287 & & 0.061 & 0.197 \\
\\[-1.75ex]
$\textrm{Pr} \big[$ 20\% expansion $\big]$ & & 0.000 & 0.000 & & 0.028 & 0.163 & & 0.010 & 0.091 \\
\\[-1.75ex]
\thickhline
\end{tabular}} \end{center}
\caption{Probability of persistent recessions and expansions \protect \\ {\small{} This table shows the probabilities of persistent recessions (expansions) in the different economies. To analyze recessions (expansions), each economy starts in the highest (lowest) steady-state and is simulated for $T=40$ and $T=100$ quarters. Each simulation is repeated 100,000 times. The table reports the fraction of simulations in which output is $ \kappa \%$ below (above) the initial value for at least 4 consecutive quarters.} \label{tab:ss_down_transition}}
\end{table}

\paragraph{Transition Probabilities}
To study the likelihood of steady-state transitions, we compute the probabilities of large and persistent recessions and expansions.
We start by focusing on recessions. We initialize each economy at the largest steady-state and simulate it 100,000 times, for windows of 40 and 100 quarters. We compute the fraction of simulations in which output experiences a 10\%, 15\% or 20\% drop relative to the high steady-state (for at least 4 consecutive quarters). The results are shown in Table \ref{tab:ss_down_transition}. When running the 2007 economy for 40 quarters, output drops by at least 10\% in 19.6\% of the simulations, whereas the same figure for the 1990 economy is 12.4\%. In the 1975 economy, this type of deep recession is extremely rare as it only occurs in 1.7\% of our simulations. Relative to 1990, the 2007 economy is 1.56 times more likely to experience a 10\% fall in output over a 10-year period. Over 100 quarters, the 2007 economy appears about 1.35 times more likely to experience a 10\% recession (35.7\% probability in 2007, against 26.4\% probability in 1990). The same probability for the 1975 economy is 6.7\%. 
This implies that, in expectation, the 2007 economy experiences a recession larger than 10\% every 70 years, the 1990 economy does so every 95 years, and the 1975 economy every 380 years. When focusing on larger recessions, the differences are even starker. 

We also investigate the likelihood of upward transitions. To this end, we initialize each economy at the lowest steady-state and repeat the same simulation exercise. We compute the fraction of simulations in which output experiences 10\%, 15\% or 20\% expansion relative to the low steady-state (for 4 consecutive quarters).
Relative to the 1990 economy, the 2007 model displays a substantially lower likelihood of an upward transition. This suggests that slumps are more likely to occur in the 2007 economy for two reasons: (i) downward transitions are more likely, and (ii) upward transitions are less likely.
Upward (as well as downward) transitions are absent in the 1975 economy because it features only one steady state.

\paragraph{Minimum shocks triggering a downward transition}

We also investigate the
minimum size of the shocks switching the economy from the high to the low steady-state.
We do this exercise for the 1990 and 2007 economies as this statistic is not defined in the 1975 economy since we find no multiplicity. We initialize these two economies at the largest (non-stochastic) steady-state and subject them to a sequence of negative TFP shocks. To compare with our Great Recession experiment, we use shocks that last 6 quarters and are identical across them. Specifically, we consider 
$\varepsilon_{t} \, = \, v \, \sigma_{\varepsilon}$
for $t=1,\ldots,6$ (and set $\varepsilon_{t}=0$ for $t\geq 7$). Table \ref{tab:min_size_shock} reports the minimum value of $v$ that can generate a downward transition to the lowest steady state.

\begin{table}[H]
    \centering
    \setlength{\tabcolsep}{0.25cm}  
    \begin{tabular}{c c c c }
    \thickhline 
    \\[-2ex]
     & 1975 & 1990 & 2007\tabularnewline
    \\[-1ex] 
    \hline  \\[-2ex]
    minimum shock & -  & $6.84 \sigma_\epsilon$& $1.62 \sigma_\epsilon$\\[1ex]
    \thickhline 
    \end{tabular}
    \caption{Minimum shock size for a downward transition}
    \label{tab:min_size_shock}
\end{table}

The values indicate that the 1990 economy requires a shock of approximately 6.8 standard deviations to experience a downward transition, while the 2007 economy only requires a 1.6 standard deviation shock and, therefore, is significantly more likely to experience prolonged recessions.

\paragraph{Discussion} As a final remark on the ergodic properties of these economies, recall that the exogenous aggregate shocks process in unchanged across the three calibrations. The large observed differences in the ergodic behavior are fully driven by the endogenous response of the economies. The three calibrations only differ in parameters related to the firm heterogeneity and market structure. 

These results suggest that rising firm differences and fixed costs are a source of non-linearity in the economy's response to aggregate shocks. This may seem inconsistent with the idea of a \textit{great moderation} \textemdash{} namely, the fact that the volatility of aggregate output declined between 1980 and 2007. Note, however, that aggregate volatility in our economy is the product of two forces \textemdash{} exogenous volatility (TFP shocks) and endogenous amplification and persistence. If exogenous volatility declined over time, it is possible that aggregate volatility also declined in spite of larger amplification. There are reasons to think that exogenous aggregate volatility may have decreased over time \textendash{} for example, because of demographic shifts \citep{JS_AER} or a rising share of low-volatility industries \citep{GC}.

\section{The 2008 Recession and Its Aftermath}\label{sec:aggregatefacts}

\begin{figure}[ht]
\begin{minipage}[b]{.5\linewidth}
\centering{}\includegraphics*[scale=0.525]{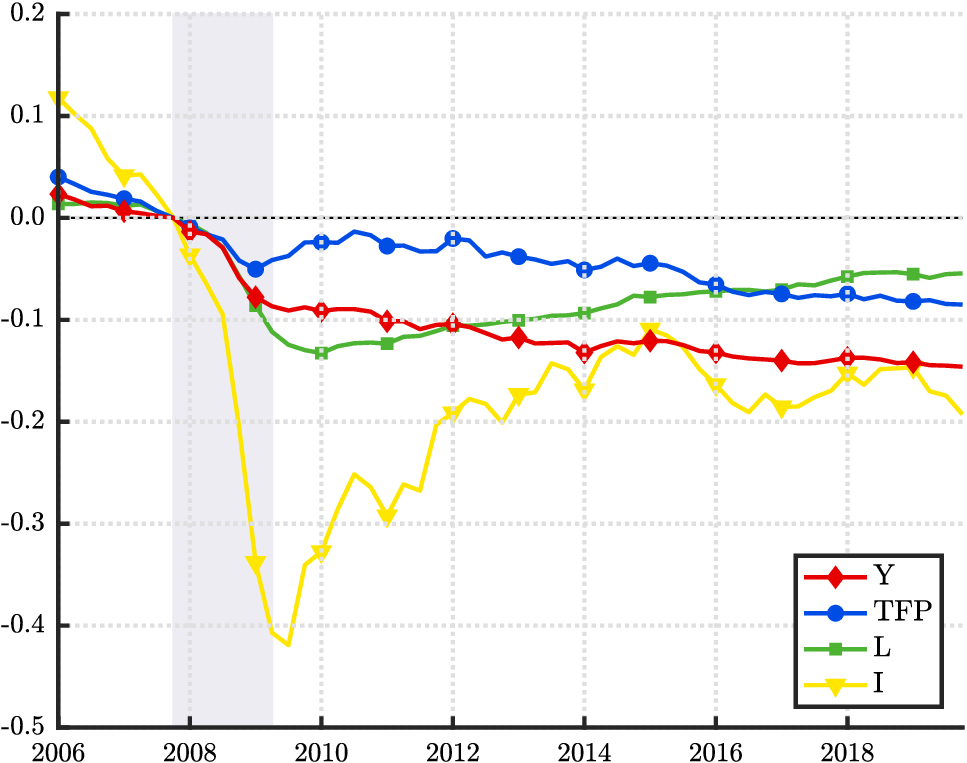}\subcaption{Data \label{fig:GR_data}}
\end{minipage}%
\begin{minipage}[b]{.5\linewidth}
\centering{}\includegraphics*[scale=0.525]{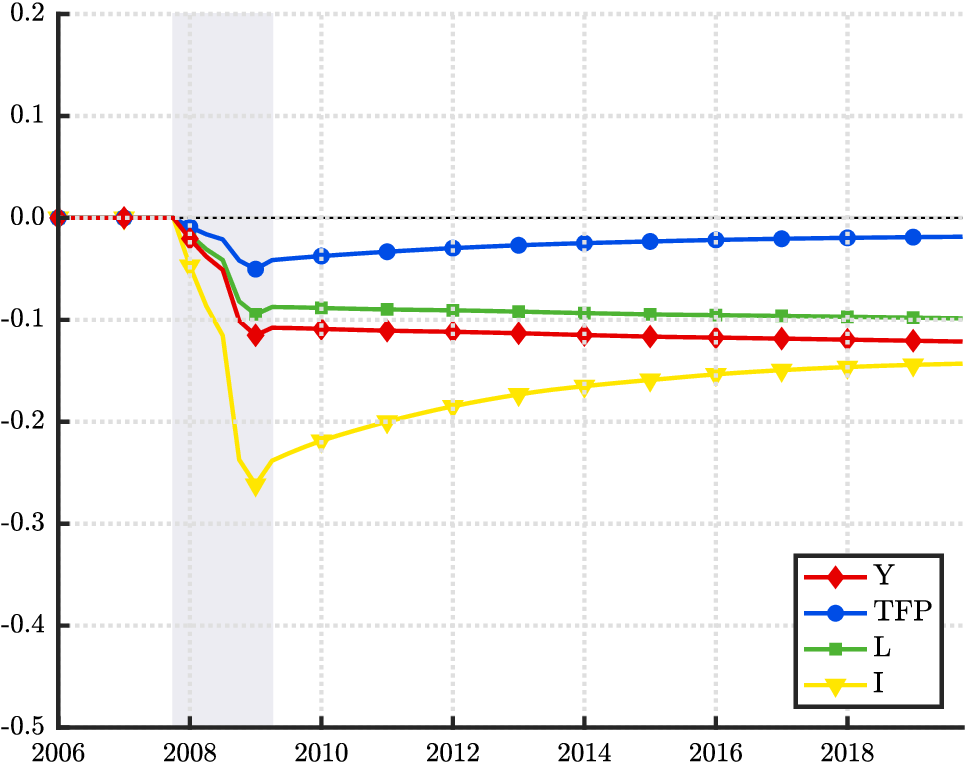}\subcaption{2007 Model \label{fig:GR_model}}
\end{minipage}
\caption{The Great Recession and its aftermath \protect \\ {Note: \small{} The figure shows the evolution of key macro aggregates in the aftermath of the 2008 recession in the data (Panel a) and the model (Panel b). The model is subjected to a sequence of six quarter shocks $\left\{\varepsilon_{t}\right\}$ to match the dynamics of aggregate TFP in the data between 2008Q1:2009Q2. See Appendix \ref{sec:Data-Appendix} for data definitions.}} \label{fig:GR_data_model}
\end{figure}

In this section, we take a closer look at the 2008 recession and its aftermath. The left panel of Figure \ref{fig:GR_data_model} shows the behavior of four aggregate variables from 2006 to 2019 \textemdash{} real GDP, real gross private investment, and total hours (all in per capita terms), as well as aggregate TFP. All variables are in logs, detrended (with a linear trend computed over 1985\textendash{}2007), and centered on 2007Q4. The four variables decline on impact and do not rebound to their pre-recession trends. For example, in the first quarter of 2019, real GDP per capita is 14.2\% below trend (Table \ref{tab:great-recession}). Aggregate TFP experienced an 8.2\% negative deviation from the trend. Investment declines by more than 40\% on impact then stabilizes at approximately 15-20\% below the pre-crisis trend.

\begin{table}[ht]
\setlength{\tabcolsep}{0.5cm} \begin{center} \resizebox{17cm}{2cm} {
\begin{tabular}{c c c c | c c c c} \thickhline
& & & & & & \\[-1.5ex]
& \multicolumn{3}{c|}{\Large Data} & \multicolumn{4}{c}{\Large 2007 Model} \\[1ex]
& {\large 2009Q4} & {\large 2015Q1} & {\large 2019Q1} & {\large 2009Q4} & {\large 2015Q1} & {\large 2019Q1} & {\large 2040Q1} \\[1.5ex]
\thickhline
& & & & & & \\[-1.25ex]
{\large Output} & -0.088 & -0.126 & -0.142 & -0.108 & -0.116 & -0.121 & -0.142 \\[1ex]
{\large TFP} & -0.045 & -0.026 & -0.082 & -0.039 & -0.023 & -0.019 & -0.018 \\[1ex]
{\large Hours} & -0.130 & -0.078 & -0.055 & -0.088 & -0.095 & -0.098 & -0.115 \\[1ex]
{\large Investment} & -0.340 & -0.109 & -0.147 & -0.225 & -0.159 & -0.144 & -0.153 \\[1ex]

\thickhline 
\end{tabular}}
\end{center}

\caption{The Great Recession and its aftermath: data versus model \label{tab:great-recession}}
\end{table}

We then ask whether our model can replicate the behavior of these four variables. To this end, we feed the model a sequence of shocks $\varepsilon_{t}$ that lasts for six quarters (2008Q1:2009Q2); these shocks are calibrated so that aggregate measured TFP in our model (i.e., $A_{t} \: \Phi_{t}$) matches the dynamics of its data counterpart over the same period. The economy starts at the high steady-state
(with $z_{t}=0$). We set the innovations to productivity to zero after 2009Q2 and let the economy recover. The right panel of Figure \ref{fig:GR_data_model} shows the implied responses of output, aggregate TFP, employment, and investment generated by our model. As the figure shows, this series of shocks triggers a transition to the low steady-state. Our model provides a very good description of the evolution of the four variables. Output declines by 12.1\% after 10 years, whereas hours drop by 9.8\% (Table \ref{tab:great-recession}). Both numbers are close to their data counterparts.\footnote{In the data, hours worked seem to recover faster than output, which seems inconsistent with a jobless recovery. However, i) we are showing variables in deviation from the trend, and ii) hours worked were characterized by slower growth (and hence a flatter trend) before the crisis. For example, total hours worked were stagnant between 2000-2007.} Additionally,  our model predicts a 22.5\% decline in impact for investment and a 14.4\% drop by 2019 (14.7\% in the data). Finally, we observe a long-run decline of aggregate TFP, representing approximately 1/4 of the drop in the data (1.9\% in the model, 8.2\% in the data). We discuss the mechanisms underlying this result in the next section. The crisis experiment in our model can also replicate the still widening gap between output and its trend throughout our sample. Importantly, the model shows a persistent deviation from the pre-crisis trend 30 years after the recession (last column of Table \ref{tab:great-recession}). Output is 14.2\% lower while hours and investment are 11.5\% and 15.3\% below trend. The economy, therefore, features a transition to a lower steady-state.

\begin{figure}[ht!]
\begin{minipage}[b]{.5\linewidth}
\centering{}\includegraphics*[scale=0.525]{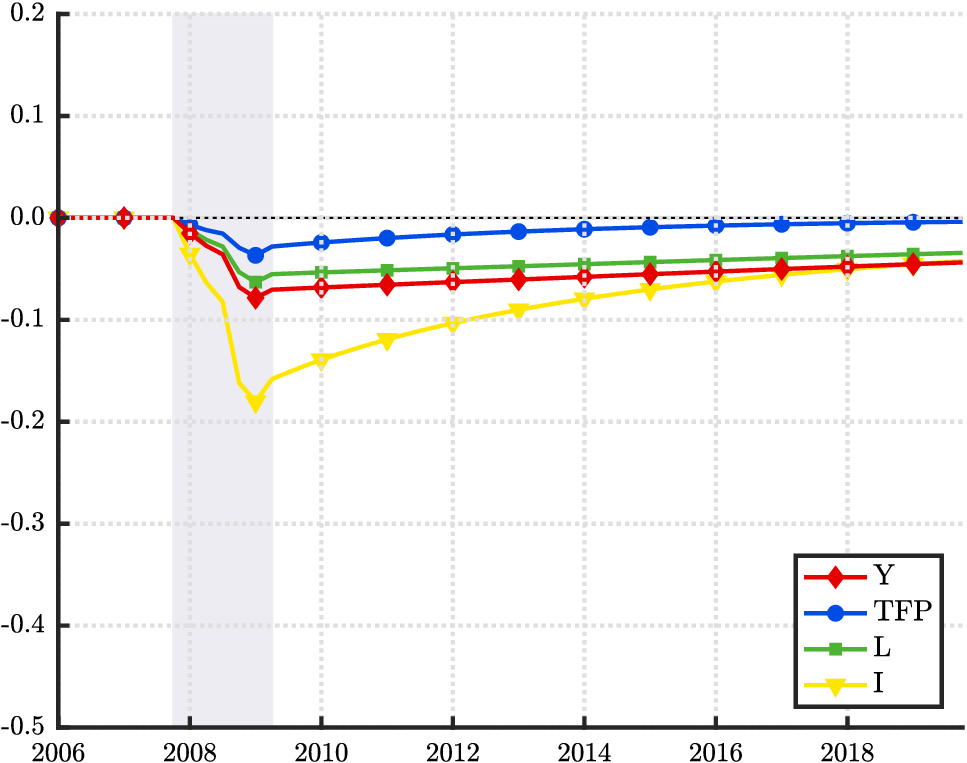}\subcaption{1975 Model \label{fig:GR_1975}}
\end{minipage}%
\begin{minipage}[b]{.5\linewidth}
\centering{}\includegraphics*[scale=0.525]{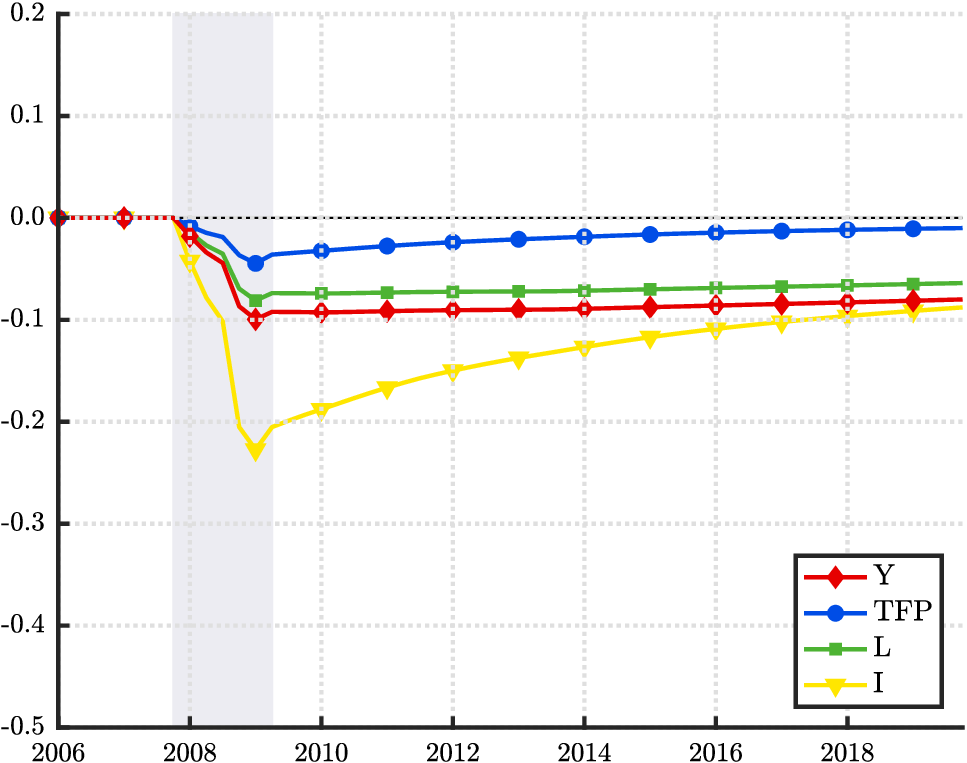}\subcaption{1990 Model \label{fig:GR_1990}}
\end{minipage}
\caption{The Great Recession in the 1975 and 1990 models \protect \\ {Note: \small{}This figure shows the response of the 1975 and 1990 economies to the sequence of shocks used in Figure \ref{fig:GR_model}.} \label{fig:GR_model_75_90}}
\end{figure}

Next, as our first counterfactual, we ask whether the same sequence of aggregate TFP shocks used in the 2007 economy can also trigger a transition to the low steady-state in the 1975 and 1990 economies. We study this experiment to ask whether the deviation our model predicts for the 2007 economy is driven by an unusually large shock or by the inherent fragility of the economy itself. Figure  \ref{fig:GR_model_75_90} and Table \ref{tab:great-recession-counterfactual-models} show the transitional dynamics. Both economies exhibit substantially less amplification, and, importantly, they also revert to their pre-crisis steady-states. 
\begin{table}[ht!]
\setlength{\tabcolsep}{0.4cm} \begin{center} \resizebox{16cm}{2.15cm} {
\begin{tabular}{c c c c | c c c} \thickhline
& & & & & & \\[-1.5ex]
& \multicolumn{3}{c|}{\Large 1975 Model} & \multicolumn{3}{c}{\Large 1990 Model} \\[1ex]
& {\large 2009Q4} & {\large 2019Q1} & {\large 2040Q1} & {\large 2009Q4} & {\large 2019Q1} & {\large 2040Q1} \\[1.5ex]
\thickhline
& & & & & & \\[-1.25ex]
{\large Output} & -0.069 & -0.045 & -0.015 & -0.092 & -0.081 & -0.047 \\[1ex]
{\large TFP} & -0.025 & -0.004 & -0.000 & -0.033 & -0.011 & -0.003 \\[1ex]
{\large Hours} & -0.054 & -0.036 & -0.011 & -0.074 & -0.065 & -0.037 \\[1ex]
{\large Investment} & -0.145 & -0.045 & -0.010 & -0.193 & -0.091 & -0.042 \\[1ex]

\thickhline 
\end{tabular}}
\end{center}

\caption{The Great Recession and its aftermath in the 1975 and 1990 models \label{tab:great-recession-counterfactual-models}}
\end{table}
In economies with the 1975 and 1990 features, a negative aggregate shock of the magnitude required in our model to generate the 2008-2009 recession would not be large enough to induce a persistent deviation from the trend. These economies would have experienced a faster reversal of trend. For example, as of 2019, the output would be 4.5\% and 8.1\% below trend (in the 1975 and 1990 economies, respectively), against 12.1\% of the 2007 model and 14.2\% as observed in the data. When we evaluate the longer-run behavior in 2040, we find that the 1975 economy would experience a 1.5\% drop, while for the 1990 economy, this figure is 4.7\%. We conclude that the structural differences between the 1975, 1990, and the 2007 economies (namely larger productivity differences and higher fixed costs) are key to understanding the 2008 crisis and the subsequent \textit{great deviation}.
\subsection{Discussion and Extensions}
\paragraph{The 1990 Recession}
Through the lens of our model, the 2008 crisis made the US economy transition to a new steady state. This fact has not been observed after any other postwar recession. This raises a natural question: what was special about the 2008 crisis? Was the shock hitting the economy in 2008 larger than in previous recessions? Or was the economy more fragile in 2008 and, therefore, more prone to experience a transition even for moderate shocks? Earlier in this section, we showed that the same shocks underlying the 2008 recession in our model do not trigger transitions in the 1975 and 1990 economies. We show that this holds for other recessions. We repeat the experiment of Section \ref{sec:aggregatefacts} using the 1990 crisis. We feed the 1990 economy a sequence of shocks that replicates the dynamics of aggregate TFP during the 1990\textendash{}1991 recession (1990Q3:1991Q1). We then take this same sequence of exogenous shocks and apply them to the 2007 economy.
The results of this experiment are shown in Online Appendix \ref{sec:90_recession}. When looking at the response of the 1990 economy, we observe a temporary decline in all variables, followed by a gradual recovery to the previous steady state.
This contrasts with the response of the 2007 economy, which eventually experiences a transition to the lower steady-state. These results suggest that, rather than the consequence of an unusually large shock, the post-2008 deviation can be linked to an underlying market structure that made the US economy more fragile to negative shocks.

\paragraph{Alternative sources of fluctuations} In the discussion so far, we have considered business cycle fluctuations driven by changes in the exogenous component of aggregate TFP.  Here we entertain an alternative source of shocks. We study an economy in which the exogenous component of TFP is constant, but the fraction of industries subject to a fixed cost fluctuates. We assume that $x_c^\prime=\frac{\bar{x}_c}{1-\rho_x}+\rho_x x_c+ \epsilon_x$ with $\epsilon_x\sim N(0,\sigma_x^2)$ and $\bar x_c=0.28$ as we calibrated in the previous section. Agents form expectations on $x_c^\prime$ when making their intertemporal choices. We calibrate $\rho_x$ and $\sigma_x$ to match the autocorrelation and standard deviation of log GDP. We consider this source of fluctuations as a stand-in for the fraction of firms financially constrained at the onset of the Great Recession. We replicate our crisis experiment, reverse engineering a series of $x_c$ shocks to match the empirical behavior of measured aggregate TFP in the first 6 quarters of the recession. We then allow $x_c$ to converge back to its unconditional mean according to the AR(1) process. The results of the crisis experiment are reported in   Figure \ref{fig:GR_model_fshock} and Table \ref{tab:great-recession-counterfactual-fshock}.

We find that this source of fluctuations closely replicates the empirical behavior of the US economy in the medium run. On impact, the economy suffers a recession of approximately 15\%, almost twice as large as the 9\% output drop in the data. However, the economy then displays a qualitatively similar behavior by further deviating from the pre-crisis trend. In the long run, the economy is -18\% below trend. The on-impact effect on hours and investment is very similar to that in the data at -14.4\% and -35\%, as reported in Table \ref{tab:great-recession-counterfactual-fshock}.

\paragraph{What drives persistence and amplification?} Our model has three main channels contributing to the endogenous persistence and amplification of business cycles: i) endogenous market structure, ii) oligopolistic competition, and iii) elastic factor supply.\footnote{In \cite{FQ_working_paper}, we provide a detailed discussion of these channels.} To understand the contribution of each element, we solve three alternative versions of the model in which we eliminate these elements one at a time. We detail our calibration strategy for each of these models in Online Appendix \ref{sec:different-channels}.

First, we consider an economy like the one described above but where the market structure is exogenous. Namely, we assume that the number of firms is fixed. Thus, both the love-for-variety and the market power effects are suppressed, and endogenous TFP is constant over the business cycle. The only source of endogenous amplification is given by the response of elastic factor supply.
Second, we consider an economy where each firm is monopolistically competitive on its own variety, as in \cite{dixit_stiglitz}. By construction, markups are constant and exogenous shocks can be amplified via the love-for-variety effect and by the elastic factor supply.
Finally, we study an economy with an inelastic labor supply. In this model, business cycle amplification is driven by endogenous changes in the market structure due to cyclical entry and exit as well as love-for-variety effects on endogenous TFP.

We report more details about each model in Online Appendix \ref{sec:different-channels}. In summary, we find that the three economies feature an unimodal ergodic distribution of output. This is a sign of steady-state uniqueness, which suggests that these models cannot replicate the large, persistent, and increasing deviation from trend that characterizes the Great Recession. Quantitatively, we find that they also exhibit different levels of amplification and persistence. We report these numbers in Table \ref{tab:submodels}.

\begin{table}[ht]
\begin{center}\resizebox{0.85\textwidth}{!} {
\begin{tabular}{ c c c c c }
\thickhline  \\[-1ex]
& Baseline & Fixed Number of Firms & Monopolistic Competition & Fixed Labor Supply \\[2ex]
\thickhline  \\[-1ex]
{\large $\rho\left(Y\right)$ } & 0.975 & 0.968 & 0.974 & 0.950 \\[2ex]{\large $\sigma\left(Y\right)$} & 0.063 & 0.035 & 0.049 & 0.030 \\[2ex]
\thickhline 
\end{tabular}}
\caption{Volatility and persistence in different models \protect \\ {\small $\rho\left(Y\right)$ and $\sigma\left(Y\right)$ represent the first order autocorrelation and the standard deviation of log output (for a simulation window of 100,000 periods).}}
\label{tab:submodels}
\end{center}
\end{table}
Comparing the volatility and persistence across models, we conclude that the elastic factor supply drives approximately half of the amplification we observe. In an economy with a fixed number of firms, this is the only active source of amplification, and it accounts for 55\% of the volatility in our baseline economy. Next, we note that when we shut down the endogenous market structure channel by assuming monopolistic competition, we recover approximately 78\% of the baseline economy volatility. This suggests that the endogenous markup channel accounts for about one-fourth of the amplification in our economy. Finally, the love-for-variety channel also explains approximately one-fourth of the amplification in our economy. 
\paragraph{Other extensions and discussion}
In our model, firms pay a fixed cost $c > 0$ in units of
the final good. This assumption implies that the cost of entry is independent of the state of the
economy. If fixed costs were to change
with factor prices, entry could be cheaper in a low-competition regime,
which could, in principle, eliminate steady-state multiplicity. In \cite{FQ_working_paper}, we show that steady-state multiplicity is preserved when fixed costs depend on factor prices.

Finally, in our model, firms are not subject to sunk entry costs as in \cite{hopenhayn} and \cite{clementi_palazzo}. The existence of sunk costs can make entry and exit less sensitive to the business cycle and can, in principle, have an impact on fragility. While this alternative setting is certainly richer and more realistic, developing a model with oligopolistic competition, endogenous entry, and sunk costs is outside the scope of this paper.\footnote{In the current model, firms make their entry and exit decisions based on the current realization of the vector of productivities. In the presence of sunk entry costs, firms must also consider all future realizations of that vector. If there are $N=20$ productivity draws per market, and each draw can take one of 10 possible values, there are $20^{10}$ partial equilibria to be solved.}

Next we provide empirical evidence consistent with our proposed mechanism both at the aggregate and at the industry level.

\subsection{Empirical Evidence}

According to our model, a transition between steady states is driven by a change in the competitive regime of the economy. We now provide evidence consistent with our model's mechanism. In particular, we show that after 2008, the US economy experienced i) a persistent decline in the number of active firms, ii) a persistent decline in the aggregate labor share, and iii) an acceleration in the aggregate profit share and markup. Furthermore, the model has stark predictions about how product markets starting with different levels of concentration should feature different responses to the negative shock. The extensive margin is more elastic in product markets with higher concentration, to begin with; as a consequence, we should observe that these product markets experience a larger drop in output. We provide cross-industry evidence corroborating this prediction.

\subsubsection{Aggregate Level Evidence}
We begin by reviewing the evidence at the aggregate level on the number of firms in the economy, the behavior of markups, labor, and profit shares, and the dynamics of aggregate TFP.

\begin{figure}[ht]
\begin{centering}
\includegraphics*[scale=0.485]{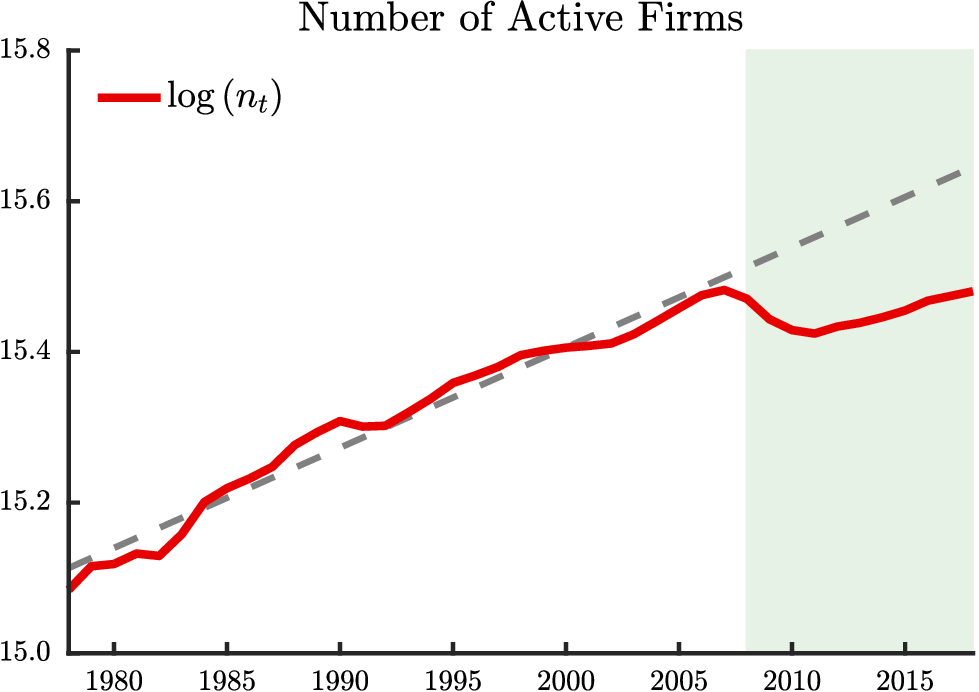} 
\par\end{centering}
\caption{\textbf{Number of Firms: 1978-2018}\protect \\ {Note: \small{} The red line shows the number of firms with at least one employee (in logs). The dashed grey line shows a linear trend computed for the period 1978-2007. Data is from the US Business Dynamics Statistics. \label{fig:nfirms_total}}}
\end{figure}

\paragraph{Number of Firms}
Firms in our model are single-product and operate in only one market. This contrasts with the definition of firms in the data, which are often multi-product and operate in several markets (e.g. geographic segmentation). 
In fact, \cite{Broda_Weinstein_10} show that net product creation is strongly procyclical (more products are introduced in expansions, and fewer products are destroyed) and more volatile than plant entry and exit. 
With this caveat in mind, we ask whether the evolution of the number of firms in the data qualitatively matches the prediction of our model. Figure \ref{fig:nfirms_total} shows the evolution of the number of active firms (with at least one employee). As the figure shows, the number of active firms experiences a persistent deviation from trend after 2007. As of 2016, the number of active firms was 0.151 log points below trend. Such a persistent decline can also be observed within most sectors of activity (Section \ref{sec:numberfirms}). 
With the caveat that a firm in our model does not necessarily represent a firm in the data, we report the evolution of the number of firms in markets with positive fixed costs in our crisis experiment. As shown in Table \ref{tab:great-recession-labor-share}, these markets experience a persistent decline in the number of firms of 0.134 log points.

\begin{figure}[ht!]
\centering
\includegraphics*[scale=1]{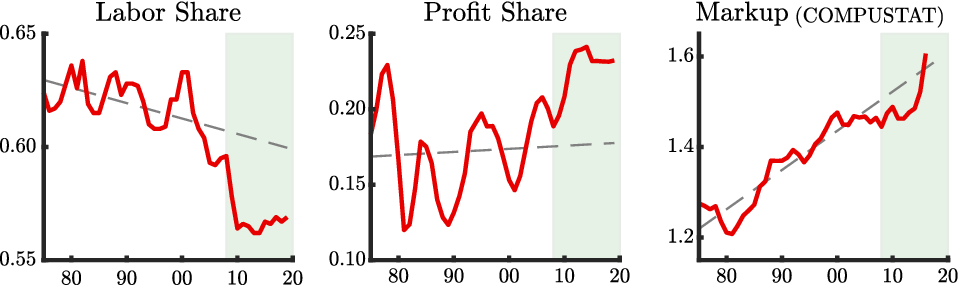}\caption{\textbf{Aggregate Markup and Labor and Profit Share: 1975-2019}\protect \\ {Note: \small{}This figure shows (i) labor share of the US business sector (from the BLS), (ii) the profit share of the US business sector (defined in Appendix \ref{sec:Data-Appendix}) and (iii) aggregate markups for COMPUSTAT from \cite{LEU}. The dashed grey lines represent linear trends computed for the 
period 1975\textendash{}2007.\label{fig:US-LS-Markup-Trend}}}
\end{figure}

Alternative models could replicate this figure only if they featured an active extensive margin, multiple steady states, and predicted a transition during the 2007-08 recession. We argue that these features and the changes in the market structure, which we discuss next, corroborate our proposed mechanism.
\paragraph{Aggregate Markups, Labor and Profit Shares}
We compute our model's labor and profit shares and compare them to the data. Figure \ref{fig:US-LS-Markup-Trend} shows the evolution of the labor share, the profit share (both computed for the US business sector), and the aggregate markup series for publicly listed firms from \cite{LEU}. The grey dashed lines represent linear trends computed for the period 1975\textendash{}2007.

Table \ref{tab:great-recession-labor-share} compares the evolution between 2007 and 2016 observed in the data and obtained in our model. Our model predicts a 0.4pp decline in the aggregate labor share, approximately 14\% of the observed decline between 2007 and 2016. If we account for a pre-crisis trend, we explain 17\% of the deviation in 2016. For the profit share, we can explain 30\% of its 3pp increase from the trend. Markups increase by 3.7 points in our model, which represents 26\% of the observed increase (14.2 points) and 58\% of the deviation from the pre-crisis trend (6.4 points).\footnote{ Other studies have also documented a sharp rise in markups in the post-crisis years. For example, \cite{rising_markups} use US product-level data and document a 25\% increase in markups between 2006 and 2019.}

\begin{table}[ht!]
\setlength{\tabcolsep}{0.2cm} \begin{center}
\resizebox{0.7\textwidth}{!}{
\begin{tabular}{ccc c c} \thickhline
& & & \\
& \multicolumn{2}{c}{ [A] Data}  & & [B] 2007 Model \\[1ex]
& $\Delta_{2007-2016}$ & $\Delta_{2007-2016}- \Delta_{\textrm{trend}}$ & \hspace{0.5cm} & $\Delta_{2007-2016}$ \\[1ex]
\hline
& & & \\[-1.25ex]
Number of firms  & -0.014 & -0.151 & & -0.134 \\[0.25ex]
& & & \\[-1.25ex]
Aggregate labor share  & -0.029 & -0.023 & & -0.004 \\[1ex]
Aggregate profit share  & 0.032 & 0.030 & & 0.009\\[2ex]
COMPUSTAT markup & 14.2 & 6.4 & & 3.7 \\[1ex]
\thickhline 
\end{tabular}
}
\end{center}
\vspace{-0.5cm}
\caption{Change in the number of firms and in income shares
\protect \\ {\small Note: Part [A] shows data moments for the series described in Figures \ref{fig:nfirms_total} and \ref{fig:US-LS-Markup-Trend}. Part [B] shows model moments, computed for the experiment described in Figure \ref{fig:GR_data_model}. We report the evolution of the number of firms subject to fixed costs between 2007 and 2016 (in logs). The aggregate labor and profit shares are computed using all markets in the model; the profit share is net of fixed costs. The COMPUSTAT markup refers to the sales-weighted markup for the set of COMPUSTAT firms (i.e., uncompetitive sector).
} \label{tab:great-recession-labor-share}}
\end{table}

\paragraph{Aggregate Productivity}\label{aggrtfp}

As shown in Figure \ref{fig:GR_data_model}, the post-2008 growth experience has also been characterized by a negative deviation of aggregate TFP from the trend. Our model predicts a persistent fall in aggregate TFP (though of a smaller magnitude than in the data). This happens in spite of a ``cleansing effect'' of recessions (\cite{Caballero_Hammour}) \textendash{} as small unproductive firms exit, average firm-level TFP increases (see Figure \ref{fig:TFP_Model_transition}).
There are two reasons explaining the decline in aggregate TFP. First, there is a \textit{love-for-variety} effect, associated with the reduction in the number of firms. This can best be seen in the limit case in which all product markets have $n$ firms with identical productivity $\tau$. In such a case, aggregate TFP is equal to $\Phi=I^{\frac{1-\rho}{\rho}}n^{\frac{1-\eta}{\eta}}\tau$. Second, in a low competition trap, there is higher misallocation. This happens because markets with a larger contraction are the ones with positive fixed costs, i.e. whose output is already low.\footnote{Figure \ref{fig:output_dispersion_transition} (Online Appendix \ref{sec:Quantitative-Model}) shows that the standard deviation of  outputs, $\text{std}_i \left( \log y_{it} \right)$, increases.}


These macro trends suggest that consistent with our model, market power accelerated after 2008. 
We next review the cross-sectional implications of our model and test them in the data.

\subsubsection{Industry Level Evidence}\label{sec:empiricalevidence}
According to our model, product markets featuring a larger concentration in 2007 should have experienced a larger contraction in 2008. This prediction follows from equation (\ref{eq:firm_FOC}), which establishes a positive link between productivity, market shares, and markups (for a given number of active firms). Therefore, if we take two markets with the same number of firms, the one featuring a more uneven distribution of productivities will have a larger dispersion in market shares and hence a larger concentration. In these markets, firms at the bottom of the distribution will be smaller and charge lower markups, and will hence be more likely to exit upon a negative shock. 

We build a dataset combining the 2002 and 2007 US Census data on industry concentration to the Statistics of US Businesses (SUSB) and the Bureau of Labor Statistics (BLS) to obtain outcomes such as employment, total wage bill, and the number of firms at the industry level (6-digits NAICS). The final dataset includes 791 6-digit industries. In 2016, the median industry had 1,316 firms, 36,910 workers, and a total payroll of \$1,880 million.

To assess whether industries with a larger concentration before the crisis experienced a larger post-crisis decline, we estimate the following regression 
\[
\dfrac{\Delta y_{i,07-16}}{y_{i,07}}=\beta_{0}+\beta_{1}\;\textrm{concent}_{i,07}+\beta_{2}\; \log \textrm{firms}_{i,07} +\beta_3 \dfrac{\Delta y_{i,03-07}}{y_{i,03}} + a_s \mathbbm{1}\{i\in s\}+u_{i}.
\]
$y_i$ is an outcome for industry $i$ (for example, total employment, total wage bill, or total number of firms), and $\textit{concent}_i$ is the share of the 4 largest firms (scaled by the share of the largest 50); we also control for the number of firms before the crisis ($\textrm{firms}_{i,07}$). Importantly, to control for possible differences in the pre-crisis dynamics, e.g., different growth opportunities in different industries, we include pre-trends in the regression. The outcomes always take the form of the annualized growth rate between 2007 and 2016 in a specific industry. We will also include sector fixed effects $\left(a_s\right)$ in all regressions. The unit of observation is a 6-digit industry.

We start by studying the correlation between the change in employment between 2007 and 2016 and the concentration in 2007. The results are compactly presented in Part [A] of Table \ref{empirics_main}, and robustness is shown in Appendix \ref{sec:reg-tables}. We find that more concentrated industries experienced lower employment growth in the aftermath of the Great Recession. Quantitatively, a 1pp higher pre-crisis concentration is associated with a 2pp lower employment growth rate between 2007 and 2016. This pattern holds irrespective of the inclusion of the number of firms in 2007. To address the concern that industries with larger concentration in 2007 could have already exhibited lower growth before the crisis, we include cumulative employment growth between 2003 and 2007 as a control; the results do not change. Finally, the results are also robust to the inclusion of macro-sector fixed effects. While these results concern the evolution of employment growth, a similar pattern is found if we use the total wage bill instead. We also study the correlation between concentration and net entry after the crisis. Our findings suggest that a 1pp increase in the concentration measure is associated with a 2-3pp decrease in the post-crisis net entry.
These results suggest that industries with larger concentration in 2007 experienced a larger contraction in activity after the crisis. 

\begin{table}[ht]
\setlength{\tabcolsep}{0.2cm}  \begin{center}
\resizebox{1\textwidth}{!} {
\begin{tabular}{l cccc c ccc} 				\thickhline
			\\[-1ex]
			& \multicolumn{4}{c}{{\large [A] Data}} & & \multicolumn{3}{c}{{\large [B] 2007 Model}} \\
            \\[-1ex]
			 & (1) & (2) & (3) & (4) & \hspace{0.65cm} & (5) & (6) & (7) \\
			 \\[-1.5ex]
			VARIABLES & $\Delta \log \textrm{emp}_{07-16}$ & $\Delta \log \textrm{payroll}_{07-16}$ & $\Delta \log \textrm{firms}_{07-16}$ & $\Delta \textrm{labor share}_{07-16}$ & & $\Delta \log \textrm{emp}$ & $\Delta \log \textrm{firms}$ & $\Delta \textrm{labor share}$ \\
			\\[-1.5ex]
            \thickhline
			&  &  &  \\
			$ \textrm{concent}_{07} $ & -0.0177*** & -0.0189*** &-0.0406*** & -0.0314* & & -0.0537*** &  -0.0369*** &     -0.0112*** \\
			& (0.00682) & (0.00697) & (0.00635) & (0.0167) & & (0.00344) &   (0.00237)&  (0.000720) \\
			&  &  &  & \\
			$ \log \textrm{firms}_{07} $ & 0.00193*** &0.00164** & 0.00119* &-0.00120 & & 0.0414*** &      0.0284*** &     0.00864***\\
			& (0.000706) & (0.000725)&(0.000661) & (0.00240) & & (0.00307)  &   (0.00211) &  (0.000641)\\
			&  &  &  & \\
			$\Delta \log \textrm{y}_{03-07}$ &  0.0984***  &0.0823*** & 0.0881*** & 0.169*\\
			&   (0.0241) & (0.0219)&(0.0270)&(0.0867) \\
			&  &  &  &  \\
			Observations & 769 &773 & 791&98 & & 9875 & 9875 & 9875 \\
			R-squared  & 0.050 & 0.043& 0.078&0.075 & & 0.115 & 0.115 & 0.115 \\
			\\[-2ex] \thickhline
			\\[-2ex]
			\multicolumn{9}{c}{ Standard errors in parentheses} \\
			\multicolumn{9}{c}{ *** p$<$0.01, ** p$<$0.05, * p$<$0.1} \\
\end{tabular} }
\caption{Cross-industry regressions\protect \\ {Note: \small{} Part [A] shows the results of regressing the growth rate of sectoral employment, payroll, number of firms, and labor share between 2007 and 2016 on the measure of concentration in 2007 and controls. Robustness on these regressions is displayed in Appendix \ref{sec:reg-tables}. Part [B] shows the results of regressing the growth rate of employment, number of firms, and labor share between 2007 and an aggregate state featuring a 10\% lower capital stock. This regression is estimated on model-generated data.}}
\label{empirics_main}
\end{center}
\end{table}

We conclude this section by providing evidence on the evolution of the labor share across industries. While the US Census of Firms provides data on total employment and the total number of firms for all 6-digit industries, it does not contain data on the labor share. We rely on data from the BLS `Labor Productivity and Cost' program (see Appendix \ref{sec:Data-Appendix} for details). This database, however, only provides data on the labor share for a restricted group of industries. We find a negative relationship between the post-crisis change in the labor share and the pre-crisis level of concentration. Industries with larger concentration in 2007 experienced a larger drop in labor share between 2008 and 2016.

In summary, these results suggest that the structure of US product markets in 2007 is important to understand the consequences of the 2008 crisis. The results presented are purely cross-sectional \textemdash{} industries with a larger concentration in 2007, displayed a larger post-crisis contraction. Nonetheless, they support one of the main insights of the model \textemdash{} rising concentration can have made the US economy more vulnerable to aggregate shocks.
Taken together, we view the empirical evidence reported here as corroborating our proposed mechanism. The aggregate economy shows a long-lasting deviation and a widening gap from the pre-crisis trend and a change in the distribution of incomes from factor suppliers to firms, suggesting an increase in market power. At the industry level we observe that industries respond differentially depending on the pre-existing market structure. 

We conclude by comparing these cross-industry results to the equivalent regressions in our model. To do so, we eliminate 10\% of the steady-state capital stock in the 2007 economy and compute the difference in employment, number of firms, and labor share. The results are identical for larger drops. We then estimate the same regressions as on the data, with the only difference being the exclusion of the pre-crisis trend since, in our model, there is no trend. We report the results in Part [B] of Table \ref{empirics_main}.
We find that markets with higher concentration, measured by the market share of the largest 4 firms, experience a 5.4\% larger drop in employment, have 3.7\% higher firm exit, and a 1.1\% larger decline in the labor share. These results are both qualitatively and quantitatively very similar to our findings on the actual cross-industry data.
\section{Policy Experiment}\label{policy}

We conclude by studying the role of policy in our model. 
The presence of imperfect competition distorts our economy in two separate ways: i) each steady state is inefficient as markups are both too high and heterogeneous, thereby inducing cross-sectional misallocation; ii) transitions across steady states are costly for risk-averse agents. The policies we consider affect both inefficiencies.

We study two simple policy interventions. First, we study a revenue subsidy $\tau_R$ such that for every dollar sold, firms obtain $1+\tau_R$ in revenues. This subsidy distorts firms' optimal size, increasing the output level. The policy applies to all firms in the economy. 
Second, we consider an entry subsidy $\tau_f$ such that firms with positive fixed costs $c$ only have to pay $(1-\tau_f)c$ upon entry. By design, the entry subsidy affects only product markets with $c>0$, which are approximately 11.8\% of the markets in our economy.
In both cases, the government raises the necessary tax revenues by taxing the household lump sum to balance the budget every period.

The two policies affect the degree of market power in our economy differently. The revenue subsidy increases the optimal size of each firm, even when the number of firms is unchanged. It also indirectly induces entry as the profits of new firms increase. The latter effect generates additional welfare gains through a love-for-variety. Differently, the entry subsidy directly distorts the entry margin by making entry cheaper. As more firms enter, the degree of competition increases inducing firms to expand output. Both policies trade off higher output and love-for-variety gains with the entry of lower-productivity firms and more resources used to cover fixed costs.

\begin{figure}[ht!]
\centering{}
\hspace{-0.2cm}
 \includegraphics*[scale=0.5]{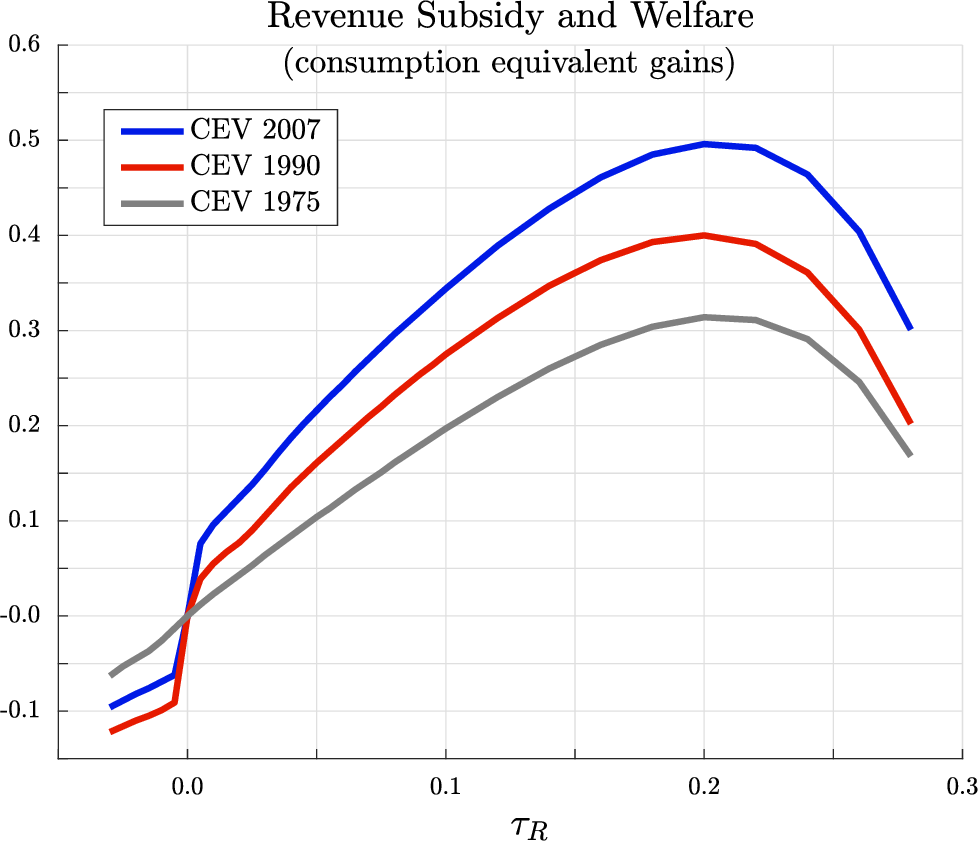}
 \hspace{0.4cm}
 \includegraphics*[scale=0.5]{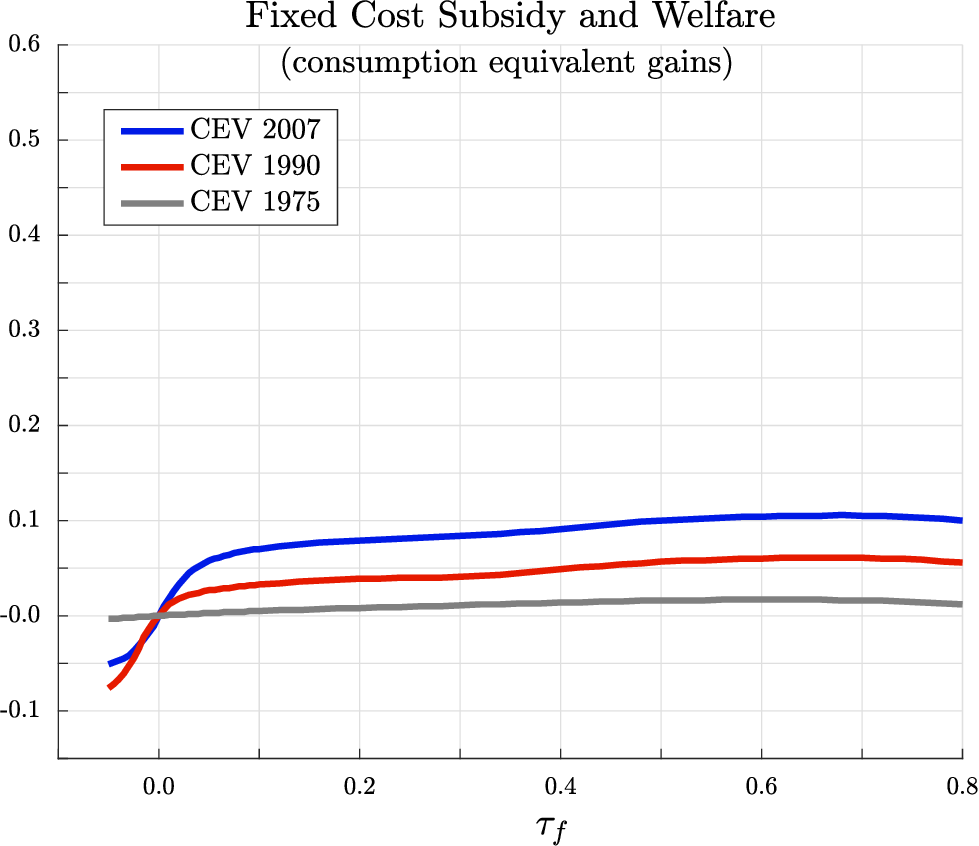}
\caption{Welfare: consumption equivalent gain
\protect \\ {Note:  \small{} the figure shows the welfare impact (in consumption equivalent gains) of (i) an \textit{ad valorem} subsidy equal to a fraction $\tau_R$ of revenues and (ii) a lump subsidy equal to a fraction $\tau_f$ of fixed costs. For each subsidy level, we simulate the economy 100,000 times and calculate average welfare.}}
\label{fig:welfare_tau}
\end{figure}

Figure \ref{fig:welfare_tau} shows the consumption equivalent welfare effects of the two policy designs for each of our three calibrated economies. Introducing the revenue subsidy generates large welfare gains in all economies, with the highest gains between 30\% and 50\% for a 20\% subsidy. The effect of the policy is highly nonlinear for the 1990 and 2007 economies, which feature multiple steady states. For example, in the 2007 economy, a 5\% subsidy can induce a 20\% welfare gain. The reason is that even
a relatively small subsidy can significantly shift the probability mass from the low to the high
steady state. This intuition explains why the effect is partially and fully muted in the 1990 and
1975 economies. As the likelihood of moving to a low competition regime reduces, the welfare gain
from the policy declines. In the 1975 economy, where the probability of quasi-permanent recessions
is zero, the welfare effects are solely driven by trading off lower markups and more
varieties with more resources absorbed by fixed costs. Interestingly, the three economies also
differ in terms of the welfare costs associated with a revenue tax (i.e. $\tau_R < 0$). In particular, such a tax can have two complementary effects: i) it can force the economy into the low competition regime, and ii) it can worsen the welfare in the low steady state as fewer firms can survive. This rationale explains the large costs for the 1990 economy. Recall that quasi-permanent recessions are unlikely in this calibration but feature large output losses relative to the high steady state. These welfare costs are smaller for the 2007 economy as downward transitions entail smaller output losses and are more likely even before the tax. Our large welfare gains are in line with the findings of \cite{EMX}, who find that the welfare cost of markups can be as high as 50\%. Our model features an additional cost of market power, which is associated with persistent recessions or slumps. Production or entry subsidies can act as a steady-state selection device, and bring about dynamic welfare gains when multiple steady-states exist.

The entry subsidy has qualitatively similar but quantitatively different effects. First, since it applies to approximately 1/8 of the economy, its welfare effects are much more muted. Second, since it cannot directly distort the size of firms, a large subsidy (around 70\%) is optimal. Importantly, it behaves similarly to the revenue subsidy around the status quo: a 5\% subsidy is sufficient to eliminate the welfare costs associated with multiplicity. Higher subsidies are targeted to improve the allocation within the high steady state.

\section{Conclusion \label{sec:Conclusion}}
The US economy appears to have experienced a fundamental change over the past decades, with several studies and data sources indicating a reallocation of activity towards large, high markup firms. This observation has raised concerns in academic and policy circles about increasing market power, and it has been proposed as an explanation for recent macroeconomic \textit{puzzles}, such as low aggregate investment, low wage growth, or declining labor shares. Our model suggests that, besides their impact on factor shares and factor prices, rising firm differences and greater market power can also have an impact on business cycles and provide an amplification and persistence mechanism to aggregate fluctuations. In particular, larger firm heterogeneity may have rendered the US economy more vulnerable to aggregate shocks and more likely to experience quasi-permanent slumps. Through the lens of our theory, such increased fragility may have been difficult to identify, as it manifests itself only in reaction to large shocks.

Our work also provides further motivation for policies that curb market power. As we have shown, the endogenous response of the market structure to aggregate shocks acts as an accelerant. On top of the standard static effects, any policy that reduces market power can have dynamic benefits in terms of the persistence and amplitude of aggregate fluctuations. These effects are particularly large if the economy is at risk of quasi-permanent slumps.

To keep the analysis simple, we have abstracted from a number of important features. One such example is that, in our model, firms do not pay sunk entry costs, and their idiosyncratic productivities are time-invariant. 
Developing a model of firm dynamics with oligopolistic competition, sunk entry costs, and endogenous entry is a promising avenue for future research.  Further, our model features one-sided market power. Recent models of oligopoly \citep[see][]{azar2021general} lend themselves to the study of the interaction between two-sided market power and the likelihood of quasi-permanent slumps. Lastly, recent work studies the interplay between competition and monetary policy \citep[see][]{mongey2017market,wang2020dynamic,fabianiJMP}. The question of how monetary policy, by changing the market structure, shapes the dynamic properties of an economy is an avenue for future research. 

\newpage
\appendix

\renewcommand{\thesection}{A.\arabic{section}}
\renewcommand{\thelemma}{A.\arabic{lemma}}
\renewcommand{\theprop}{A.\arabic{prop}}
\renewcommand{\theremark}{A.\arabic{remark}}
\renewcommand{\thecor}{A.\arabic{cor}}

\setcounter{figure}{0}
\setcounter{table}{0}
\setcounter{lemma}{0}
\setcounter{prop}{0}
\setcounter{cor}{0}

\renewcommand{\thesection}{A.\arabic{section}}
\renewcommand{\thefigure}{A.\arabic{figure}}
\renewcommand{\thetable}{A.\arabic{table}}
\renewcommand{\theprop}{A.\arabic{prop}}
\renewcommand{\thecor}{A.\arabic{cor}}
\renewcommand{\theremark}{A.\arabic{remark}}
\renewcommand{\thelemma}{A.\arabic{lemma}}

\begin{center}
{\Huge {Appendix A} }
\end{center}

\section{Data Appendix \label{sec:Data-Appendix}}

\begin{figure}[ht!]
\centering{}\includegraphics*[scale=0.7]{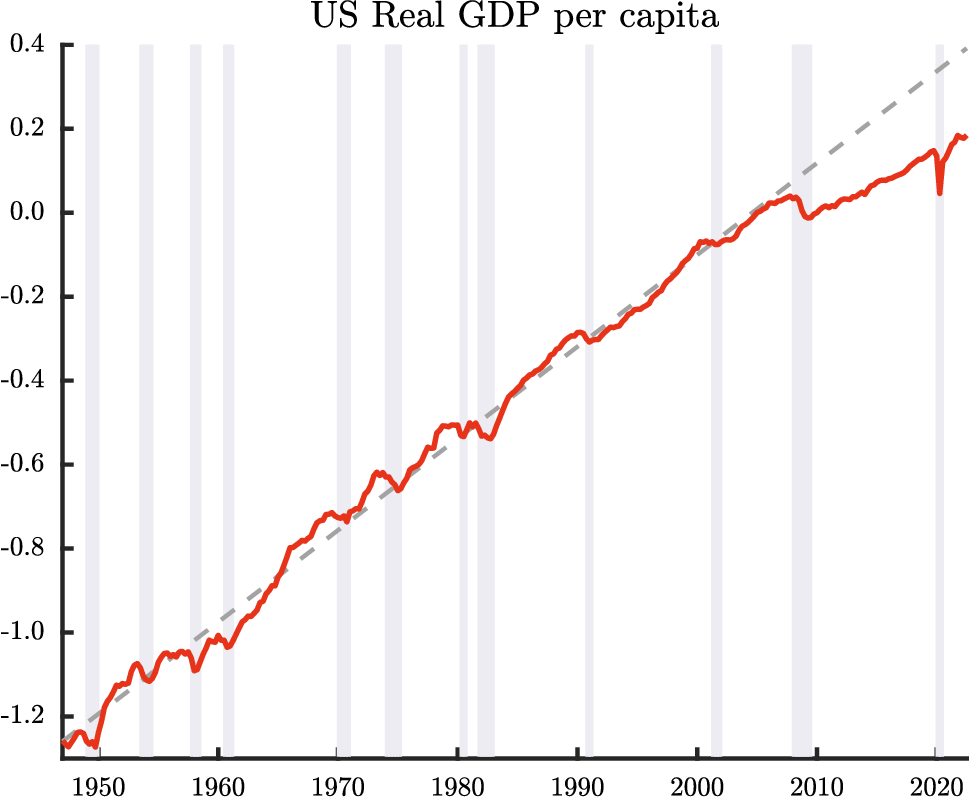}
\caption{The \textit{Great Deviation} \protect \\ {Note: \small{} This figure shows real GDP per capita (from BEA). The series is in logs, undetrended and centered around 2007. The linear trend is computed for the 1947-2007 period. Shaded areas indicate NBER recessions.}} 
\label{fig:great_deviation_figure}
\end{figure}

\paragraph{Data Definition}

Table \ref{tab:data_sources} provides information on all the data sources used in Section \ref{sec:aggregatefacts}.

\begin{table}[ht!]
\setlength{\tabcolsep}{0.25cm} \begin{center} \begin{tabular}{cc} \thickhline
\\[-2ex]
{\large Variable} & {\large Source} \\ [0.5ex] \thickhline
\\[-1ex]
Real GDP & BEA \textendash{} NIPA Table 1.1.3 (line 1)  \\[0.75ex]
Real Personal Consumption Expenditures & BEA \textendash{} NIPA Table 1.1.3 (line 2) \\[0.75ex]
Real Gross Private Domestic Investment & BEA \textendash{} NIPA Table 1.1.3 (line 7) \\[0.75ex]
Total Hours & BLS \textendash{} Nonfarm Business sector: Hours of all persons \\[0.75ex]
Aggregate TFP & \cite{F}: Raw Business Sector TFP \\[0.75ex]
Population & BEA \textendash{} NIPA Table 2.1 (line 40) \\ [1.5ex]
\thickhline
\end{tabular}  
\end{center}
\caption{Data sources}
\label{tab:data_sources}
\end{table}

\justifying

\paragraph{Aggregate Profit Share }

The aggregate profit share is computed as
\[
\text{profit share}_{t} \; = \; 1 \; - \; \text{labor share}_{t} \; - \; \underbrace{\dfrac{ R_{t}\cdot\textrm{K}_{t} \: - \: \textrm{DEP}_{t}}{\textrm{VA}_{t}}}_{\text{capital share}}
\]
where $\text{labor share}_{t}$ is the labor share of the US business sector (from BLS),
$\textrm{VA}_{t}$ is the total value added of the US business sector (NIPA Table 1.3.5, line 2). $\textrm{K}_{t}$ is the value of private fixed assets (including intangibles) of the US business sector (NIPA Table 6.1, line 1 - line 9 - line 10) and $\textrm{DEP}_{t}$ is depreciation (NIPA Table 6.4, line 1 - line 9 - line 10). Finally, $\textrm{R}_{t}$ is the required rate of return. We follow \cite{ERW} and compute it as the difference between Moody's Seasoned BAA Corporate Bond Yield and a 5-year moving average of past CPI inflation (from BLS, used as a proxy for expected inflation).

\paragraph{Industry-level Labor Share }

We obtain data on the labor share at the industry level from the BLS `Labor Productivity and Costs' (LPC) database. We calculate the labor share as the ratio of `Labor compensation' to `Value of Production'. Note that this ratio gives the share of labor compensation in total revenues, and not in value added.\footnote{This ratio coincides with the `Labor cost share' provided by the BLS. This variable is, however, available just for a restricted number of industries.}

\newpage

\section{Partial Equilibrium: Additional Results}\label{appendix:appendix_partial_eqm}

When $\eta=1$ and $n$ firms produce in a given market, we have a system of $n$ first order conditions
\begin{align*}
p\left[1-\left(1-\rho\right)s_{j}\right] \;= \; \dfrac{\Theta}{\gamma_{j}}.
\end{align*}
Each firm has a market share
\begin{align*}
s_{j,n}\;\coloneqq\;\dfrac{y_j}{\sum\limits_{k=1}^{n} y_{k}}\;=\;\dfrac{1}{1-\rho}\left[1-\dfrac{n-\left(1-\rho\right)}{\sum\limits_{k=1}^{n}\dfrac{1}{\gamma_{k}}}\dfrac{1}{\gamma_{j}}\right].
\end{align*}
A firm with productivity $\gamma_{j}$ makes production profits
\begin{align*}
\Pi\left(j,n,\Gamma,X\right) 
 & \; = \;  \underbrace{\dfrac{1}{1-\rho}\left[1-\dfrac{n-\left(1-\rho\right)}{\sum\limits_{k=1}^{n}\dfrac{1}{\gamma_{k}}}\dfrac{1}{\gamma_{j}}\right]^{2}\left[\dfrac{n-\left(1-\rho\right)}{\sum\limits_{k=1}^{n}\dfrac{1}{\gamma_{k}}}\right]^{\frac{\rho}{1-\rho}}}_{\coloneqq \varLambda\left(\gamma_{j},n,\Gamma\right)}\Theta^{-\frac{\rho}{1-\rho}}Y.
\end{align*}
\begin{lemma}
\label{lemma:(Profit-Function)} When $\eta=1$, the profit function \textup{$\Pi\left(j,n,\Gamma,X\right) $}
satisfies
\begin{align*} 
1) \quad & \dfrac{\partial\Pi\left(j,n,\Gamma,X\right)}{\partial Y}\vphantom{\vphantom{\dfrac{\dfrac{1}{1}}{\frac{_{1}}{1}}}}>0  &
2) \quad & \dfrac{\partial\Pi\left(j,n,\Gamma,X\right)}{\partial n}\vphantom{\vphantom{\dfrac{\dfrac{1}{1}}{\frac{_{1}}{1}}}}<0  \quad,\;n>j \\
3) \quad & \dfrac{\partial\Pi\left(j,n,\Gamma,X\right)}{\partial\gamma_{j}}\vphantom{\dfrac{\dfrac{1}{1}}{\frac{_{1}}{1}}}>0 &
4) \quad & \dfrac{\partial\Pi\left(j,n,\Gamma,X\right)}{\partial\gamma_{k}}\vphantom{\vphantom{\dfrac{\dfrac{1}{1}}{\frac{_{1}}{1}}}}<0  \quad,\;\forall k\neq j.
\end{align*}
\end{lemma}
\begin{proof} [Proof of Lemma \ref{lemma:(Profit-Function)}]
We start by showing that $\Pi\left(\cdot\right)$ increases in $\gamma_{j}$
\begin{align*}
 & \dfrac{\partial\Pi\left(j,n,\Gamma,X\right)}{\partial\gamma_{j}}\vphantom{\dfrac{\dfrac{1}{1}}{\frac{_{1}}{1}}}>0 \\
\Leftrightarrow & \; \: 2\left[1-\dfrac{n-\left(1-\rho\right)}{\sum\limits_{k=1}^{n}\dfrac{1}{\gamma_{k}}}\dfrac{1}{\gamma_{j}}\right]^{-1}\left(\sum\limits_{k\neq j}^{n}\dfrac{1}{\gamma_{k}}\right)+\dfrac{\rho}{1-\rho}\left[\dfrac{n-\left(1-\rho\right)}{\sum\limits_{k=1}^{n}\dfrac{1}{\gamma_{k}}}\right]^{-1}>0.
\end{align*}
To prove points 2) and 4) it suffices to show that $\varLambda\left(\cdot\right)$
is decreasing in $[{n-\left(1-\rho\right)}]/\left[{\sum\limits_{k=1}^{n}\dfrac{1}{\gamma_{k}}}\right]$
\begin{align*}
 & 2\left[1-\dfrac{n-\left(1-\rho\right)}{\sum\limits_{k=1}^{n}\dfrac{1}{\gamma_{k}}}\dfrac{1}{\gamma_{j}}\right]^{-1}\left(-\dfrac{1}{\gamma_{j}}\right)+\dfrac{\rho}{1-\rho}\left[\dfrac{n-\left(1-\rho\right)}{\sum\limits_{k=1}^{n}\dfrac{1}{\gamma_{k}}}\right]^{-1}<0\\
 \\
\Leftrightarrow & \; \: \gamma_{j}\sum\limits_{k=1}^{n}\dfrac{1}{\gamma_{k}}<\dfrac{2-\rho}{\rho}\left[n-\left(1-\rho\right)\right].
\end{align*}
The last condition is implied by $s_{n+1,n+1}<1$.
\end{proof}

\section{General Equilibrium: Additional Results and Proofs}\label{appendix:appendix_general_eqm}
\subsubsection*{Aggregate TFP and Factor Price Index}
Aggregate TFP is given by
\begin{equation}
\Phi\left(\mathbf{\Gamma},\mathbf{N}_{t}\right)=\left[\sum_{i=1}^{I}\left(\sum_{j=1}^{n_{it}}\omega_{ijt}^{\eta}\right)^{\frac{\rho}{\eta}}\right]^{\frac{1}{\rho}}\left(\sum_{i=1}^{I}\sum_{j=1}^{n_{it}}\dfrac{\omega_{ijt}}{\tau_{ijt}}\right)^{-1},\label{eq:agg_tfp}
\end{equation}
where
\vspace{-0.15cm}
\begin{align*}
\omega_{ijt}\coloneqq\left[\sum_{k=1}^{n_{it}}\left(\dfrac{\mu_{ikt}}{\tau_{ikt}}\right)^{\frac{\eta}{1-\eta}}\right]^{\frac{\eta-\rho}{\eta}\frac{1}{1-\rho}}\left(\dfrac{\tau_{ijt}}{\mu_{ijt}}\right)^{\frac{1}{1-\eta}}.
\end{align*}
Aggregating over all firms' first order condition (\ref{eq:firm_FOC}), we can express $\Theta\left(\mathbf{\Gamma},\mathbf{N}_{t}\right)$ as a function of the number of active firms $(n_{it})$, markups $(\mu_{ijt})$ and individual TFP $(\gamma_{ij})$
\begin{align}
\Theta\left(\mathbf{\Gamma},\mathbf{N}_{t}\right)=A_{t} \, \left\{ \sum_{i=1}^{I}\left[\sum_{j=1}^{n_{it}}\left(\dfrac{\gamma_{ijt}}{\mu_{ijt}}\right)^{\frac{\eta}{1-\eta}}\right]^{\frac{1-\eta}{\eta}\frac{\rho}{1-\rho}}\right\} ^{\frac{1-\rho}{\rho}}.\label{eq:agg_cost}
\end{align}

\subsubsection*{Proof of Proposition \ref{prop:savings_rate}}

\begin{proof}
In a steady-state, there is a constant rental rate $R^{*} = \beta^{-1} - \left(1-\delta\right)$.
Combining this with equations (\ref{eq:rental_rate_eqm}), (\ref{eq:agg_cost_decomposition}) and $\delta  K = s Y$ we obtain
\begin{align*}\nonumber
 & \beta^{-1}-\left(1-\delta\right)=\alpha\;\Omega\left(\mathbf{\Gamma},\mathbf{N}\right)\;\dfrac{Y^{*}}{K^{*}}\\[2ex] \nonumber
\Leftrightarrow & \; s^{*} \:= \; \dfrac{\delta\;\alpha}{\beta^{-1}-\left(1-\delta\right)}\Omega\left(\mathbf{\Gamma},\mathbf{N}\right).
\end{align*}
\end{proof}

\subsubsection*{Proof of Proposition \ref{A-symmetric-equilibrium}}
\begin{proof}
When $n$ firms produce, the profits of a firm with productivity $\gamma_{j}$ are equal to
\begin{align}
\Pi\left(\gamma_{j},n,\Gamma,\Theta,Y\right) = \varLambda\left(\gamma_{j},n,\Gamma\right) \Theta^{-\frac{\rho}{1-\rho}}Y
\end{align}
where
$\varLambda\left(\gamma_{j},n,\Gamma\right)$ has been defined in Appendix \ref{appendix:appendix_partial_eqm}. A symmetric equilibrium with $n$ firms in all markets is possible provided that
\vspace{-0.15cm}
\begin{align*}
\varLambda\left(\gamma_{n},n,\Gamma\right) \Theta^{-\frac{\rho}{1-\rho}}Y & \geq c, \\[2ex]
    \varLambda\left(\gamma_{n+1},n+1,\Gamma\right) \Theta^{-\frac{\rho}{1-\rho}}Y & \leq c.
\end{align*}
Using equation (\ref{eq:agg_output_capital}), we can write the above inequalities as 
\begin{align}
\underline{K}\left(\Gamma,n\right)\leq K_{t}\leq\overline{K}\left(\Gamma,n\right),
\end{align}
where
\begin{align}
\underline{K}\left(\Gamma,n\right) \, \coloneqq \, \left\{ \dfrac{c}{\varLambda\left(\gamma_{n},n,\Gamma\right)}\left(1-\alpha\right)^{-\frac{1-\alpha}{\nu+\alpha}}\left[\Phi\left(\Gamma,n\right)\right]^{-1}\left[\Theta\left(\Gamma,n\right)\right]^{\frac{\rho}{1-\rho}-\frac{1-\alpha}{\nu+\alpha}}\right\} ^{\frac{\nu+\alpha}{\alpha\left(1+\nu\right)}}, \\[2ex]
\overline{K}\left(\Gamma,n\right) \, \coloneqq \, \left\{ \dfrac{c}{\varLambda\left(\gamma_{n+1},n+1,\Gamma\right)}\left(1-\alpha\right)^{-\frac{1-\alpha}{\nu+\alpha}}\left[\Phi\left(\Gamma,n\right)\right]^{-1}\left[\Theta\left(\Gamma,n\right)\right]^{\frac{\rho}{1-\rho}-\frac{1-\alpha}{\nu+\alpha}}\right\} ^{\frac{\nu+\alpha}{\alpha\left(1+\nu\right)}}.
\end{align}
Equilibrium uniqueness arises if and only if
\[
\begin{array}{cl}
&\underline{K}\left(\Gamma,n+1\right)> \overline{K}\left(\Gamma,n\right) \; \, \forall \, n \\[2ex]
\Leftrightarrow & \dfrac{\Phi\left(\Gamma,n\right)}{\Phi\left(\Gamma,n+1\right)}>\left[\dfrac{\Theta\left(\Gamma,n\right)}{\Theta\left(\Gamma,n+1\right)}\right]^{\frac{\rho}{1-\rho}-\frac{1-\alpha}{\nu+\alpha}} \; \, \forall \, n.
\end{array}
\]
When there are no productivity differences across firms, this condition is equivalent to
\[
\begin{array}{cl}
& \left[\dfrac{\Theta\left(\Gamma,n\right)}{\Theta\left(\Gamma,n+1\right)}\right]^{\frac{\rho}{1-\rho}-\frac{1-\alpha}{\nu+\alpha}}<1\\[3ex]
\Leftrightarrow & \dfrac{\rho}{1-\rho}>\dfrac{1-\alpha}{\nu+\alpha}\, ,
\end{array}
\]
since $\Theta\left(\Gamma,n+1\right)>\Theta\left(\Gamma,n\right)$.
\end{proof}

\subsubsection*{Proof of Proposition \ref{prop:multiplicity}}
\begin{proof}
We have
\[
R_{t}\;=\;\alpha\;\left(1-\alpha\right)^{\frac{1-\alpha}{\nu+\alpha}}\;\Theta\left(\Gamma,n\right)^{\frac{1+\nu}{\nu+\alpha}}\:K_{t}^{-\nu\frac{1-\alpha}{\nu+\alpha}}.
\]
Let $\underline{R}\left(\Gamma,n\right)$ and $\overline{R}\left(\Gamma,n\right)$ be the rental rates at $\underline{K}\left(\Gamma,n\right)$ and $\overline{K}\left(\Gamma,n\right)$ respectively. Multiple steady-states exist if there exists an $n \in \mathbb{N}$ such that

\[
\begin{array}{rl}
 & \overline{R}\left(\Gamma,n\right)<\underbrace{\beta^{-1}-\left(1-\delta\right)}_{R^{*}}<\underline{R}\left(\Gamma,n+1\right)\\[1ex]
\Leftrightarrow & \;\Theta\left(\Gamma,n\right)^{\frac{1+\nu}{\nu+\alpha}}\:\overline{K}\left(\Gamma,n\right)^{-\nu\frac{1-\alpha}{\nu+\alpha}}<\dfrac{\beta^{-1}-\left(1-\delta\right)}{\alpha\;\left(1-\alpha\right)^{\frac{1-\alpha}{\nu+\alpha}}}<\;\Theta\left(\Gamma,n+1\right)^{\frac{1+\nu}{\nu+\alpha}}\:\underline{K}\left(\Gamma,n+1\right)^{-\nu\frac{1-\alpha}{\nu+\alpha}}.
\end{array}
\]
\end{proof}

\subsubsection*{Proof of Proposition \ref{prop:MPS_factor_prices}}
\begin{proof}
Under $\eta=1$ and when all markets are identical and have $n$ firms, we have
\begin{align*}
\begin{array}{rl}
\Theta\;= & \;\dfrac{n-\left(1-\rho\right)}{\sum\limits _{j=1}^{n}\dfrac{1}{\gamma_{j}}},\\
\Phi\;= & \;\dfrac{\Theta}{1-\dfrac{1}{1-\rho}\sum\limits _{j=1}^{n}\left(1-\dfrac{\Theta}{\gamma_{j}}\right)^{2}},\\
\Omega\;= & \;\dfrac{\Theta}{\Phi}.
\end{array}
\end{align*}
Suppose that we have $n=2$ and $\gamma_{1} \geq \gamma_{2}$. To consider an MPS on the distribution of productivities, pick some  $x>0$ and let $\widetilde{\gamma}_{1} \, = \, \gamma_{1} + x$ and $\widetilde{\gamma}_{2} \, = \, \gamma_{2} - x$.
We have that
\[
\dfrac{1-\rho}{\widetilde{\Phi}} \, = \, \dfrac{1}{\overline{\gamma}+x}+\dfrac{1}{\overline{\gamma}-x}-\dfrac{1+\rho}{2\overline{\gamma}}\left(\dfrac{\overline{\gamma}-x}{\overline{\gamma}+x}+\dfrac{\overline{\gamma}+x}{\overline{\gamma}-x}\right).
\]
To prove statement (a), we need to show that 
\[
\begin{array}{rl}
 & \dfrac{\partial\left(1/\widetilde{\Phi}\right)}{\partial x}<0\\[2ex]
\Leftrightarrow & \left(-1\right)\left(\overline{\gamma}+x\right)^{-2}+\left(-1\right)\left(\overline{\gamma}-x\right)^{-2}\left(-1\right)\\[2ex]
 & -\dfrac{1+\rho}{2\overline{\gamma}}\left[\left(-1\right)\left(\overline{\gamma}+x\right)^{-1}+\left(\overline{\gamma}-x\right)\left(-1\right)\left(\overline{\gamma}+x\right)^{-2}+\left(\overline{\gamma}-x\right)^{-1}+\left(\overline{\gamma}+x\right)\left(-1\right)\left(\overline{\gamma}-x\right)^{-2}\left(-1\right)\right]<0\\[2ex]
\Leftrightarrow & \dfrac{1}{\left(\overline{\gamma}+x\right)^{2}}-\dfrac{1}{\left(\overline{\gamma}-x\right)^{2}}+\dfrac{1+\rho}{2\overline{\gamma}}\Biggl[\underbrace{\dfrac{1}{\overline{\gamma}-x}-\dfrac{1}{\overline{\gamma}+x}\vphantom{\dfrac{\overline{\gamma}+x}{\left(\overline{\gamma}-x\right)^{2}}}}_{>0}+\underbrace{\dfrac{\overline{\gamma}+x}{\left(\overline{\gamma}-x\right)^{2}}-\dfrac{\overline{\gamma}-x}{\left(\overline{\gamma}+x\right)^{2}}}_{>0}\Biggr]>0.
\end{array}
\]
It suffices to show that the previous condition is weakly satisfied for $\rho=0$. We have
\[
\begin{array}{rl}
 & \dfrac{2\overline{\gamma}}{\left(\overline{\gamma}+x\right)^{2}}-\dfrac{2\overline{\gamma}}{\left(\overline{\gamma}-x\right)^{2}}+\dfrac{1}{\overline{\gamma}-x}-\dfrac{1}{\overline{\gamma}+x}+\dfrac{\overline{\gamma}+x}{\left(\overline{\gamma}-x\right)^{2}}-\dfrac{\overline{\gamma}-x}{\left(\overline{\gamma}+x\right)^{2}}\geq0\\[3ex]
\Leftrightarrow & \dfrac{2\overline{\gamma}-\overline{\gamma}+x}{\left(\overline{\gamma}+x\right)^{2}}+\dfrac{-2\overline{\gamma}+\overline{\gamma}+x}{\left(\overline{\gamma}-x\right)^{2}}+\dfrac{1}{\overline{\gamma}-x}-\dfrac{1}{\overline{\gamma}+x}\geq0\\[2ex]
\Leftrightarrow & \underbrace{\dfrac{1}{\overline{\gamma}+x}-\dfrac{1}{\overline{\gamma}-x}+\dfrac{1}{\overline{\gamma}-x}-\dfrac{1}{\overline{\gamma}+x}}_{0}\geq0.
\end{array}
\]
This proves statement (a).
We will now prove statement (c). We have that
\[
\widetilde{\Theta} \, = \, \left(1+\rho\right)\left(\dfrac{1}{\overline{\gamma}+x}+\dfrac{1}{\overline{\gamma}-x}\right)^{-1},
\]
and
\vspace{-0.25cm}
\[
\begin{array}{rl}
 & \dfrac{\partial\widetilde{\Theta}}{\partial x}<0\\
\Leftrightarrow & \left(1+\rho\right)\left(-1\right)\left(\dfrac{1}{\overline{\gamma}+x}+\dfrac{1}{\overline{\gamma}-x}\right)^{-2}\left[\left(-1\right)\left(\overline{\gamma}+x\right)^{-2}+\left(-1\right)\left(\overline{\gamma}-x\right)^{-2}\left(-1\right)\right]<0\\[2ex]
\Leftrightarrow & \left(\overline{\gamma}-x\right)^{-2}-\left(\overline{\gamma}+x\right)^{-2}>0.
\end{array}
\]
This condition is always satisfied, and proves (c).
Statement (b) follows from (a) and (c).
\end{proof}

\subsubsection*{Proof of Proposition \ref{prop:MPS}}

\begin{proof}
a) Let $K^{*}$ be a steady-state where all markets have $n=2$ firms and a common productivity distribution $\Gamma$. Using equations (\ref{eq:rental_rate_eqm}) and (\ref{eq:agg_labor_capital}), we can define $K^{*}$ as
\begin{align}
\underbrace{\beta^{-1} - \left(1-\delta\right)}_{=R^{*}}\:=\:\alpha\:\left(1-\alpha\right)^{\left(1-\alpha\right)/\left(\nu+\alpha\right)}\:\Theta\left(\Gamma\right)^{\left(\nu+1\right)/\left(\nu+\alpha\right)}\:\left(K^{*}\right)^{-\nu\left(1-\alpha\right)/\left(\nu+\alpha\right)}.
\end{align}
From Proposition \ref{prop:MPS_factor_prices}, $\Theta\left(\Gamma\right)$ declines after a MPS on $\Gamma$. Then $K^{*}$ must also decline.

b)
We provide a sufficient condition under which the unstable steady-state increases after an MPS. Note that the unstable steady-state increases whenever the increasing segment of the rental rate map lies strictly underneath the original one.

From the proof of Proposition \ref{prop:MPS} \ref{mps_a} that the new rental rate at $\underline{K}\left(2\right)$ is strictly lower than before. The proof involves two steps:

[A] We derive a sufficient condition under which the new rental rate at $\overline{K}\left(1\right)$ is lower;

[B] We show that, after an MPS, the increasing segment of the rental rate map cannot cross the previous one more than once. Thus, if the new segment starts and ends below the previous one, it can never go above it.

\subsubsection*{Proof of Part [A]}

The free entry condition is
\begin{align}
\Lambda\left(n\right)\Theta\left(n-1\right)^{-\frac{\rho}{1-\rho}}\:\Phi\left(n-1\right)K^{\alpha}\:L^{1-\alpha}\;=\;c.
\end{align}
Using
$
L=\left[\left(1-\alpha\right)\Theta\left(n-1\right)\right]^{\frac{1}{\nu+\alpha}}K^{\frac{\alpha}{\nu+\alpha}}
$, we can rewrite the free-entry condition as
\begin{align}\nonumber
 & \Lambda\left(n\right)\Theta\left(n-1\right)^{-\frac{\rho}{1-\rho}}\:\Phi\left(n-1\right)\;K^{\alpha}\:\left[\left(1-\alpha\right)\Theta\left(n-1\right)\right]^{\frac{1-\alpha}{\nu+\alpha}}K^{\alpha\frac{1-\alpha}{\nu+\alpha}}\;=\;c\\[2ex] \nonumber
\Leftrightarrow & \; K\;=\;\left[\dfrac{c\left(1-\alpha\right)^{-\frac{1-\alpha}{\nu+\alpha}}}{\Lambda\left(n\right)\Theta\left(n-1\right)^{\frac{1-\alpha}{\nu+\alpha}-\frac{\rho}{1-\rho}}\:\Phi\left(n-1\right)}\right]^{\frac{1}{\alpha}\frac{\nu+\alpha}{\nu+1}}.
\end{align}
The rental rate is
\begin{align}\nonumber
R \; = \; & \alpha\:\Theta\left(n-1\right)\:K^{\alpha-1}\:\left[\left(1-\alpha\right)\Theta\left(n-1\right)\right]^{\frac{1-\alpha}{\nu+\alpha}}K^{\alpha\frac{1-\alpha}{\nu+\alpha}}\\[2ex]
= \; & \alpha\:\left(1-\alpha\right)^{\frac{1-\alpha}{\nu+\alpha}}\:\Theta\left(\mathbf{\Gamma},\mathbf{N}_{t}\right)^{\frac{\nu+1}{\nu+\alpha}}\:K^{\nu\frac{\alpha-1}{\nu+\alpha}}.
\end{align}
Putting the two together
\begin{equation}
R \; = \; \alpha\:\left(1-\alpha\right)^{\frac{1-\alpha}{\nu+\alpha}}\:\Theta\left(n-1\right)^{\frac{\nu+1}{\nu+\alpha}}\:\left(\dfrac{\Lambda\left(n\right)\Theta\left(n-1\right)^{\frac{1-\alpha}{\nu+\alpha}-\frac{\rho}{1-\rho}}\:\Phi\left(n-1\right)}{c\left(1-\alpha\right)^{-\frac{1-\alpha}{\nu+\alpha}}}\right)^{\frac{\nu}{\nu+1}\frac{1-\alpha}{\alpha}},\label{eq:R_1}
\end{equation}
implying
\begin{align}
R^{\frac{\nu+1}{\nu}\frac{\alpha}{1-\alpha}}=\propto\:\Theta\left(n-1\right)^{\frac{\nu+1}{\nu}\frac{\alpha}{1-\alpha}\frac{\nu+1}{\nu+\alpha}}\:\Lambda\left(n\right)\Theta\left(n-1\right)^{\frac{1-\alpha}{\nu+\alpha}-\frac{\rho}{1-\rho}}\:\Phi\left(n-1\right)
\end{align}
where
\begin{align}
\Theta\left(n\right) \: = \: & g\left(n\right), \\[2ex]
\Phi\left(n\right) \: = \: & \dfrac{1}{\sum\limits _{j=1}^{n}\dfrac{s_{j}}{\gamma_{j}}}\\[2ex]
\Lambda\left(n\right) \: = \: & s_{n}^{2}\left[g\left(n\right)\right]^{\frac{\rho}{1-\rho}}.
\end{align}
We can thus rewrite (\ref{eq:R_1}) as
\begin{equation}
R^{\frac{\nu+1}{\nu}\frac{\alpha}{1-\alpha}}=g\left(n-1\right)^{\frac{\nu+1}{\nu}\frac{\alpha}{1-\alpha}\frac{\nu+1}{\nu+\alpha}+\frac{1-\alpha}{\nu+\alpha}}\:\dfrac{s_{n}^{2}\left[\dfrac{g\left(n\right)}{g\left(n-1\right)}\right]^{\frac{\rho}{1-\rho}}}{\sum\limits _{j=1}^{n-1}\dfrac{\hat{s}_{j}}{\gamma_{j}}},\label{eq:R_2}
\end{equation}
where $\hat{s}_{j}$ is the market share of firm $j$ in an industry
with $n-1$ firms and $s_{j}$ is the market share of that firm when
there are $n$ player in the industry. We want to show that the expression
in (\ref{eq:R_2}) goes down when we do an MPS on $n=2$ firms. Under $n=2$, we have
\begin{align}
g\left(1\right) \; = \; & \rho\,\gamma_{1},\\[2ex]
g\left(2\right) \; = \; & \dfrac{1+\rho}{\dfrac{1}{\gamma_{1}}+\dfrac{1}{\gamma_{2}}},
\end{align}
so that
\begin{align}\nonumber
R^{\frac{\alpha}{1-\alpha}}= & \left(\rho\,\gamma_{1}\right)^{\frac{\nu+1}{\nu}\frac{\alpha}{1-\alpha}\frac{\nu+1}{\nu+\alpha}+\frac{1-\alpha}{\nu+\alpha}-\frac{\rho}{1-\rho}}\:\left(\dfrac{1+\rho}{\dfrac{1}{\gamma_{1}}+\dfrac{1}{\gamma_{2}}}\right)^{\frac{\rho}{1-\rho}}\dfrac{\left[1-\dfrac{1+\rho}{\dfrac{1}{\gamma_{1}}+\dfrac{1}{\gamma_{2}}}\dfrac{1}{\gamma_{2}}\right]^{2}}{\dfrac{1}{\gamma_{1}}}\\
= & \propto\gamma_{1}^{\frac{\nu+1}{\nu}\frac{\alpha}{1-\alpha}\frac{\nu+1}{\nu+\alpha}+\frac{1-\alpha}{\nu+\alpha}-\frac{\rho}{1-\rho}-1}\:\left[\gamma_{1}-\gamma_{1}\dfrac{1+\rho}{\dfrac{2x}{\gamma_{1}}}\right]^{2}\left[\dfrac{1+\rho}{\dfrac{1}{\gamma_{1}}+\dfrac{1}{2x-\gamma_{1}}}\right]^{\frac{\rho}{1-\rho}}.
\end{align}
The last term is decreasing on an MPS, since it is simply $g\left(2\right)$.
The first term is decreasing on an MPS provided that
\begin{align}
1+\dfrac{\rho}{1-\rho}>\dfrac{\nu+1}{\nu}\dfrac{\alpha}{1-\alpha}\dfrac{\nu+1}{\nu+\alpha}+\dfrac{1-\alpha}{\nu+\alpha}
\Leftrightarrow  \alpha<\dfrac{\nu+\rho-1}{\nu\left(2-\rho\right)},
\end{align}
since this MPS must result in higher $\gamma_{1}$. We just need to evaluate
the term in the middle. Note that we can rewrite it as
\begin{align}
\gamma_{1}-\gamma_{1}^{2}\dfrac{1+\rho}{2x},
\end{align}
where $2x\coloneqq\gamma_{1}+\gamma_{2}$ is fixed by construction. The derivative
of the expression above is
\begin{align}
\dfrac{\partial}{\partial\gamma_{1}} \; = \; & 1-2\gamma_{1}\dfrac{1+\rho}{2x}\\
\; = \; & 1-\underbrace{\dfrac{\gamma_{1}}{x}}_{>1}\left(1+\rho\right)<0.
\end{align}
Therefore, for $n=2$, the interest rate is always declining on an
MPS provided that
\begin{align}
\rho>1-\nu\dfrac{1-\alpha}{1+\nu\alpha}.
\end{align}

\subsubsection*{Proof of Part [B]}
In a market with $n$ firms, the free entry condition is
\begin{align}
\Lambda_{n}\Theta^{-\frac{\rho}{1-\rho}}\:\Phi\:K^{\alpha}\:L^{1-\alpha} \; = \; c.
\end{align}
Aggregate TFP can be written as
\hspace{-5cm}
\begin{align*}
\begin{array}{rl}
\Phi \; = \; & \dfrac{\left\{ \left(1-m\right)\left[g\left(n-1\right)\right]^{\frac{\rho}{1-\rho}}+m\left[g\left(n\right)\right]^{\frac{\rho}{1-\rho}}\right\} ^{\frac{1}{\rho}}}{\left(1-m\right)\left[g\left(n-1\right)\right]^{\frac{1}{1-\rho}}h\left(n-1\right)+m\left[g\left(n\right)\right]^{\frac{1}{1-\rho}}h\left(n\right)} \\[3ex]
\; = \; & \dfrac{\Theta^{\frac{1}{1-\rho}}}{\left(1-m\right)\left[g\left(n-1\right)\right]^{\frac{1}{1-\rho}}h\left(n-1\right)+m\left[g\left(n\right)\right]^{\frac{1}{1-\rho}}h\left(n\right)},
\end{array}
\end{align*}
where
\begin{align}
g\left(n\right) \; = \; \dfrac{n-\left(1-\rho\right)}{\sum\limits _{j=1}^{n}\dfrac{1}{\gamma_{j}}} \quad \text{and} \quad
h\left(n\right) \; = \; \sum\limits _{j=1}^{n}\dfrac{s_{j}}{\gamma_{j}}.
\end{align}
Now suppose that we do the MPS and have $\tilde{\Theta}=\Theta$ at
the same $K$.\footnote{We also have $\tilde{L}=L$, since $L$ is a function of $\Theta$
and $K$.} From the free entry condition and the expression for $\Phi$, this
is possible if
\begin{align}\nonumber
 & \dfrac{\Lambda_{n}}{\tilde{\Lambda}_{n}}=\dfrac{\left(1-m\right)\left[g\left(n-1\right)\right]^{\frac{1}{1-\rho}}h\left(n-1\right)+m\left[g\left(n\right)\right]^{\frac{1}{1-\rho}}h\left(n\right)}{\left(1-\tilde{m}\right)\left[\tilde{g}\left(n-1\right)\right]^{\frac{1}{1-\rho}}\tilde{h}\left(n-1\right)+\tilde{m}\left[\tilde{g}\left(n\right)\right]^{\frac{1}{1-\rho}}\tilde{h}\left(n\right)}\\
\Leftrightarrow & \dfrac{\Lambda_{n}}{\tilde{\Lambda}_{n}}=\dfrac{\left[g\left(n-1\right)\right]^{\frac{1}{1-\rho}}h\left(n-1\right)+m\left\{ \left[g\left(n\right)\right]^{\frac{1}{1-\rho}}h\left(n\right)-\left[g\left(n-1\right)\right]^{\frac{1}{1-\rho}}h\left(n-1\right)\right\} }{\left[\tilde{g}\left(n-1\right)\right]^{\frac{1}{1-\rho}}\tilde{h}\left(n-1\right)+\tilde{m}\left\{ \left[\tilde{g}\left(n\right)\right]^{\frac{1}{1-\rho}}\tilde{h}\left(n\right)-\left[\tilde{g}\left(n-1\right)\right]^{\frac{1}{1-\rho}}\tilde{h}\left(n-1\right)\right\} }.
\end{align}
Rearranging this equation, we can write
\begin{equation}
\tilde{m}=a_{1}+b_{1}\cdot m , \label{eq:eq_1}
\end{equation}
where $a_{1}$ and $b_{1}$ are some numbers (independent of $K$).
Furthermore, from $\tilde{\Theta}=\Theta$ we have
\begin{align}\nonumber
 & \left(1-m\right)\left[g\left(n-1\right)\right]^{\frac{\rho}{1-\rho}}+m\left[g\left(n\right)\right]^{\frac{\rho}{1-\rho}}=\left(1-\tilde{m}\right)\left[\tilde{g}\left(n-1\right)\right]^{\frac{\rho}{1-\rho}}+\tilde{m}\left[\tilde{g}\left(n\right)\right]^{\frac{\rho}{1-\rho}}\\[3ex]
\Leftrightarrow & \left[g\left(n-1\right)\right]^{\frac{\rho}{1-\rho}}+m\left\{ \left[g\left(n\right)\right]^{\frac{\rho}{1-\rho}}-\left[g\left(n-1\right)\right]^{\frac{\rho}{1-\rho}}\right\} =\left[\tilde{g}\left(n-1\right)\right]^{\frac{\rho}{1-\rho}}+\tilde{m}\left\{ \left[\tilde{g}\left(n\right)\right]^{\frac{\rho}{1-\rho}}-\left[\tilde{g}\left(n-1\right)\right]^{\frac{\rho}{1-\rho}}\right\}.
\end{align}
Rearranging this equation, we can write
\begin{equation}
\tilde{m}=a_{2}+b_{2}\cdot m. \label{eq:eq_2}
\end{equation}
Combining (\ref{eq:eq_1}) and (\ref{eq:eq_2}), there is at most
one pair $\left(m,\tilde{m}\right)$ such that $\tilde{\Theta}=\Theta$.
This establishes that $\tilde{\Theta}$ cannot cross $\Theta$ twice.
\end{proof}

\subsubsection*{Proof of Proposition \ref{prop:basin_attraction_cf}}

\begin{proof}

From equation (\ref{eq:rental_rate_eqm}) we can write
\begin{align}
R_{t}\:=\:\alpha\:\left(1-\alpha\right)^{\left(1-\alpha\right)/\left(\nu+\alpha\right)}\:\Theta\left(\mathbf{\Gamma},\mathbf{N}_{t}\right)^{\left(\nu+1\right)/\left(\nu+\alpha\right)}\:K_{t}^{-\nu\left(1-\alpha\right)/\left(\nu+\alpha\right)},
\end{align}
where $\Theta \left(\mathbf{\Gamma}, \mathbf{N}_{t}\right) $ is increasing in the number of active firms (as explained above). For a given steady-state $K^*$, the slackness free entry condition may or may not hold exactly. If it does hold exactly then, in response to a marginal increase in $c$, the number of firms will necessarily decrease and so will the level of capital at the steady-state. If it does not hold exactly then the level of capital will be unchanged as no firm will leave the market. The statement of part a) follows.

Second, at an unstable steady-state $K_{U}$, the rental rate is increasing in the capital stock. For this to happen, $\Theta\left(\mathbf{\Gamma}, \mathbf{N}_{t}\right)$ must be increasing in $K$ at that point. Assuming $I$ large, this only happens if some firm is exactly breaking even. Therefore, the rental rate at an unstable steady-state $K_{U}$ necessarily declines after an increase in $c$, as stated in part b).

\end{proof}

\newpage


\addcontentsline{toc}{section}{References} 
\bibliographystyle{aer}
{\footnotesize\bibliography{input/biblct}}

\newpage 

\appendix

\begin{center}  
{\huge{Online Appendix} \\[2ex] \large{\textit{Firm Heterogeneity, Market Power and Macroeconomic Fragility} }\\[2ex] Not for Publication}
\end{center}

\raggedright
\justify

\setcounter{page}{1}
\setcounter{section}{0}
\setcounter{figure}{0}
\setcounter{table}{0}
\setcounter{lemma}{0}
\newcounter{mysection}
\makeatletter
\@addtoreset{section}{mysection}
\makeatother

\renewcommand{\thesection}{B.\arabic{section}}
\renewcommand{\thetable}{B.\arabic{table}}
\renewcommand{\theprop}{B.\arabic{prop}}
\renewcommand{\thecor}{B.\arabic{cor}}
\renewcommand{\theremark}{B.\arabic{remark}}
\renewcommand{\thefigure}{B.\arabic{figure}}

\renewcommand{\thelemma}{B.\arabic{lemma}}

\section{The Quantitative Model \label{sec:Quantitative-Model}}

\subsection*{Calibration \label{subsec:Calibration} }

\paragraph{Steady-State}

We perform three different calibrations of our model \textendash{} to match the average level of markups and its dispersion in 1975, 1990 and in 2007. We need to calibrate five technology parameters: the elasticity of substitution $\sigma_{I}$ and $\sigma_{G}$ (which are time-invariant), the log-normal standard deviation $\lambda$, the fixed production cost $c$, the fraction of markets in the uncompetitive sector $f_{u}$ and the fraction of uncompetitive markets with fixed costs $x_c$.

We start by specifying a grid for these six different parameters $\left(\sigma_{I}, \sigma_{G}, \lambda, c, f_{u}, x_{c} \right)$, and then construct a vector with different values for the capital stock $K$. We then compute the aggregate equilibrium for each parameter combination and for each value $K$.\footnote{Aggregate TFP $e^{z_{t}}$ is assumed to be constant and equal to one.} We start by assuming that all firms are active, so that there are $N$ firms in each of the $I$ industries. We compute the aggregate equilibrium using equations (\ref{eq:agg_tfp}) and (\ref{eq:agg_cost}). We then compute the profits net of the fixed cost that each firm makes
\[
\left(p_{ijt} - \dfrac{\Theta_{t}}{\tau_{ijt}} \right)\:y_{ijt} - c_{i}
\]
and identify the firm with the largest negative value. We exclude this firm and recompute the aggregate equilibrium. We repeat this iterative procedure until all firms have non-negative profits (net of the fixed production cost). If equilibrium multiplicity arises, this algorithm allows us to consistently select the equilibrium that features the largest number of  firms.

For each triplet $\left(\lambda, c, f_{comp} \right)$, we then have the general equilibrium computed for all possible capital values. The steady-state(s) of our economy correspond to the value(s) of $K$ for with the rental rate $R_{t}$ is equal to $\dfrac{1}{\beta}-\left(1-\delta\right)$. 

When multiple steady-states arise (as in the 1990 and 2007 economies), we compute model moments in the highest steady-state.

\subsection*{Solution Algorithm for the Dynamic Optimization Problem}

We now explain the algorithm we use for the dynamic optimization problem of the representative household. We take the calibrated parameters $\left(\lambda, c \right)$ and form a grid for aggregate capital with $n_K = 70 $ points. This grid is centered around the highest steady-state $K^{\text{ss}}_H$, with a lower-bound $0.5 \times K^{\text{ss}}_H$ and upper bound $1.5 \times K^{\text{ss}}_H$. We also form a grid for aggregate TFP, $A$. We use Tauchen's algorithm with $n_A = 11$ points, autocorrelation parameter $\phi_A$ and standard deviation for the innovations $\sigma_{\varepsilon}$ (the last two parameters are calibrated, as explained in the main text). We compute the aggregate equilibrium for each value of $K$ and $A$. 

We next compute a numerical approximation for the household policy function, by iterating on the Euler equation. We start by making a guess about the savings rate
\[
s\left(X_t\right) \coloneqq \dfrac{C\left(X_t\right) }{Y\left(X_t\right) }
\]
for every combination of the vector of state-variables $X_t \coloneqq \left(K_t, A_t\right)$. Given a guess $s^{\left(n\right)}\left(X_{t}\right) \; \forall X_t$ for the savings rate, we use the Euler equation to obtain a new guess $s^{\left(n+1\right)}\left(X_{t}\right)$ as follows

{\footnotesize
\[
\begin{array}{rl}
 & \dfrac{1}{\left(1-s^{\left(n+1\right)}\left(X_{t}\right)\right)Y\left(X_{t}\right)-\dfrac{W\left(X_{t}\right)^{\left(1+\nu\right)/\nu}}{1+\nu}}=\mathbb{E}_{t}\left\{ \dfrac{\beta\left[R\left(X_{t+1}\right)+\left(1-\delta\right)\right]}{\left(1-s^{\left(n\right)}\left(X_{t+1}\right)\right)Y\left(X_{t+1}\right)-\dfrac{W\left(X_{t+1}\right)^{\left(1+\nu\right)/\nu}}{1+\nu}}\right\} \\[8ex]
\Leftrightarrow & s^{\left(n+1\right)}\left(X_{t}\right)=1-\dfrac{1}{Y\left(X_{t}\right)}\left\{ \dfrac{W\left(X_{t}\right)^{\left(1+\nu\right)/\nu}}{1+\nu}+\left[\mathbb{E}_{t}\left\{ \dfrac{\beta\left[R\left(X_{t+1}\right)+\left(1-\delta\right)\right]}{\left(1-s^{\left(n\right)}\left(X_{t+1}\right)\right)Y\left(X_{t+1}\right)-\dfrac{W\left(X_{t+1}\right)^{\left(1+\nu\right)/\nu}}{1+\nu}}\right\} \right]^{-1}\right\}.
\end{array}
\]
}
We iterate on this procedure until

\[
\left|s^{\left(n+1\right)}\left(X_{t}\right)- s^{\left(n\right)}\left(X_{t}\right) \right| \; < \;  \epsilon \quad \forall X_t .
\]

\subsection*{Steady States \label{subsec:steady_state_stability}}

\paragraph{Stady state location}

\setlength{\tabcolsep}{15pt}
\renewcommand{\arraystretch}{1.5}
\begin{table}[H]
    \centering
    \begin{tabular}{|c|c|c|c|}
    \hline 
     & $K_{L}^{*}$ & $K_{U}$ & $K_{H}^{*}$\tabularnewline
    \hline 
    \hline 
    1975 & - & - & 0 \\
    \hline 
    1990 & -0.10 & -0.04 & 0.17 \\
    \hline 
    2007 & -0.06 & 0.06 & 0.15 \\
    \hline 
    \end{tabular}
    \caption{Steady State Capital Values \protect \\ {Note:
    $K_{L}^{*}$ and $K_{H}^{*}$ denote the two stable steady-states (low and high), while $K_{U}$ denotes the unstable steady-state.}}
\label{tab:steady_state_location}
\end{table}

Table \ref{tab:steady_state_location}
reports the values of the capital stock in all (non-stochastic) steady-states of our calibrated economies. Since the levels are themselves uninformative, we report them in log deviations from the steady-state capital of the 1975 economy. This allows for a better comparison across calibrations.

We normalize the log steady-state capital in 1975 to zero, so that the values in the other steady-states can be interpreted as growth rates. The 1990 economy fluctuates between two steady-states with capital levels -10\% and 17\% greater than the 1975 economy. For the 2007 economy, the same numbers are -6\% and 15\%. These numbers indicate that the level of $K_{H}^{*}$ (that is, the largest steady state of each economy) was greater in 1990 than in 2007. These results are consistent with our theory (see, for example, the discussion of the comparative statics exercises in Figure 5 of the paper). We also highlight that our theory does not feature any source of growth, which could make $K_{H}^{*}$ larger in 2007 than in 1990.

\paragraph{Log-linearization of the Euler Equation}

We report here the derivation of the dynamic system, log-linearized around the steady state values. We show that the system is saddle-path stable in the steady states we analyze. 

We start with the Euler equation
\[
\begin{array}{rl}
 & U_{c}\left(C_{t},L_{t}\right)\:=\:\beta\,\mathbb{E}_{t}\left\{ \left[R_{t+1}+\left(1-\delta\right)\right]U_{c}\left(C_{t+1},L_{t+1}\right)\right\} \\[2ex]
\Leftrightarrow & \dfrac{1}{C_{t}-\dfrac{L_{t}^{1+\nu}}{1+\nu}}\:=\:\beta\,\mathbb{E}_{t}\left\{ \dfrac{\alpha\Theta_{t+1}K_{t+1}^{\alpha-1}L_{t+1}^{1-\alpha}+\left(1-\delta\right)}{C_{t+1}-\dfrac{L_{t+1}^{1+\nu}}{1+\nu}}\right\}.
\end{array}
\]
Using
\[
L_{t}=\left[\left(1-\alpha\right)\;\Theta_{t}\,K_{t}^{\alpha}\right]^{1/\left(\alpha+\nu\right)},
\]
and $\Theta_{t}\,=\,\Theta^{*}$, $\Phi_{t}\,=\,\Phi^{*}$, the Euler
equation becomes (and assuming no uncertainty)
\[
\begin{array}{rl}
 & \dfrac{\left(1+\nu\right)C_{t+1}-L_{t+1}^{1+\nu}}{\left(1+\nu\right)C_{t}-L_{t}^{1+\nu}}\:=\:\beta\,\left[\alpha\Theta^{*}K_{t+1}^{\alpha-1}L_{t+1}^{1-\alpha}+\left(1-\delta\right)\right]\\[2ex]
\Leftrightarrow & \dfrac{\left(1+\nu\right)C_{t+1}-\left[\left(1-\alpha\right)\;\Theta^{*}\,K_{t+1}^{\alpha}\right]^{\left(1+\nu\right)/\left(\alpha+\nu\right)}}{\left(1+\nu\right)C_{t}-\left[\left(1-\alpha\right)\;\Theta^{*}\,K_{t}^{\alpha}\right]^{\left(1+\nu\right)/\left(\alpha+\nu\right)}}\:=\:\beta\,\left[\alpha\Theta^{*}K_{t+1}^{\alpha-1}\left[\left(1-\alpha\right)\;\Theta^{*}\,K_{t+1}^{\alpha}\right]^{\left(1-\alpha\right)/\left(\alpha+\nu\right)}+\left(1-\delta\right)\right]\\[2ex]
\Leftrightarrow & \beta^{-1}\left(1+\nu\right)C_{t+1}-\beta^{-1}\omega_{1}K_{t+1}^{\sigma_{1}}\:=\:\left(1+\nu\right)\omega_{2}C_{t}K_{t+1}^{-\sigma_{2}}-\omega_{1}\omega_{2}K_{t}^{\sigma_{1}}K_{t+1}^{-\sigma_{2}}+\left(1-\delta\right)\left[\left(1+\nu\right)C_{t}-\omega_{1}K_{t}^{\sigma_{1}}\right].
\end{array}
\]

Taking a first order Taylor expansion around $k_{0}=k^{*} \coloneqq \log\left(K^{*}\right)$ and $c_{0}=c^{*}\coloneqq \log\left(C^{*}\right)$
\[
\begin{array}{rl}
K_{t+1}^{\sigma_{1}}\approx & \exp\left(\sigma_{1}k^{*}\right)+\sigma_{1}\exp\left(\sigma_{1}k^{*}\right)\left(k_{t+1}-k^{*}\right)\\[1ex]
= & \exp\left(\sigma_{1}k^{*}\right)\left[1+\sigma_{1}\left(k_{t+1}-k^{*}\right)\right]
\end{array}
\]
\[
\begin{array}{rl}
K_{t+1}^{-\sigma_{2}}\approx & \exp\left(-\sigma_{2}k^{*}\right)-\sigma_{2}\exp\left(-\sigma_{2}k^{*}\right)\left(k_{t+1}-k^{*}\right)\\[1ex]
= & \exp\left(-\sigma_{2}k^{*}\right)\left[1-\sigma_{2}\left(k_{t+1}-k^{*}\right)\right]
\end{array}
\]
\[
\begin{array}{rl}
K_{t}^{\sigma_{1}}K_{t+1}^{-\sigma_{2}}\approx & \exp\left(\sigma_{1}k^{*}-\sigma_{2}k^{*}\right)+\sigma_{1}\exp\left(\sigma_{1}k^{*}-\sigma_{2}k^{*}\right)\left(k_{t}-k^{*}\right)-\sigma_{2}\exp\left(\sigma_{1}k^{*}-\sigma_{2}k^{*}\right)\left(k_{t+1}-k^{*}\right)\\[1ex]
= & \exp\left(\sigma_{1}k^{*}-\sigma_{2}k^{*}\right)\left[1+\sigma_{1}\left(k_{t}-k^{*}\right)-\sigma_{2}\left(k_{t+1}-k^{*}\right)\right]
\end{array}
\]
\[
\begin{array}{rl}
C_{t}K_{t+1}^{-\sigma_{2}}\approx & \exp\left(c^{*}-\sigma_{2}k^{*}\right)+\exp\left(c^{*}-\sigma_{2}k^{*}\right)\left(c_{t}-c^{*}\right)-\sigma_{2}\exp\left(c^{*}-\sigma_{2}k^{*}\right)\left(k_{t+1}-k^{*}\right)\\
= & \exp\left(c^{*}-\sigma_{2}k^{*}\right)\left[1+\left(c_{t}-c^{*}\right)-\sigma_{2}\left(k_{t+1}-k^{*}\right)\right]
\end{array}
\]
\[
C_{t}\approx\exp\left(c^{*}\right)\left(1+c_{t}-c^{*}\right)
\]
\[
C_{t+1}\approx\exp\left(c^{*}\right)\left(1+c_{t+1}-c^{*}\right).
\]
The log-linearized version of the Euler equation is then
\[
\begin{array}{rl}
 & \beta^{-1}\left(1+\nu\right)\exp\left(c^{*}\right)\left(1+c_{t+1}-c^{*}\right)-\beta^{-1}\omega_{1}\exp\left(\sigma_{1}k^{*}\right)\left[1+\sigma_{1}\left(k_{t+1}-k^{*}\right)\right]\\
= & \left(1+\nu\right)\omega_{2}\exp\left(c^{*}-\sigma_{2}k^{*}\right)\left[1+c_{t}-c^{*}-\sigma_{2}\left(k_{t+1}-k^{*}\right)\right]-\\
&-\omega_{1}\omega_{2}\exp\left(\sigma_{1}k^{*}-\sigma_{2}k^{*}\right)\left[1+\sigma_{1}\left(k_{t}-k^{*}\right)-\sigma_{2}\left(k_{t+1}-k^{*}\right)\right]+\\
 & +\left(1-\delta\right)\left\{ \left(1+\nu\right)\exp\left(c^{*}\right)\left(1+c_{t}-c^{*}\right)-\omega_{1}\exp\left(\sigma_{1}k^{*}\right)\left[1+\sigma_{1}\left(k_{t}-k^{*}\right)\right]\right\}.
\end{array}
\]
We can put all the constants together and define
\[
\begin{array}{rl}
 & -\beta^{-1}\left(1+\nu\right)\exp\left(c^{*}\right)\left(1-c^{*}\right)+\beta^{-1}\omega_{1}\exp\left(\sigma_{1}k^{*}\right)\left(1-\sigma_{1}k^{*}\right)\\
+ & \left(1+\nu\right)\omega_{2}\exp\left(c^{*}-\sigma_{2}k^{*}\right)\left(1-c^{*}+\sigma_{2}k^{*}\right)-\omega_{1}\omega_{2}\exp\left(\sigma_{1}k^{*}-\sigma_{2}k^{*}\right)\left(1-\sigma_{1}k^{*}+\sigma_{2}k^{*}\right)\\
+ & \left(1-\delta\right)\left[\left(1+\nu\right)\exp\left(c^{*}\right)\left(1-c^{*}\right)-\omega_{1}\exp\left(\sigma_{1}k^{*}\right)\left(1-\sigma_{1}k^{*}\right)\right]\\
\coloneqq & a_{1}.
\end{array}
\]
We can put all time $t$ variables together and define
\[
\begin{array}{rl}
 & \left(1+\nu\right)\omega_{2}\exp\left(c^{*}-\sigma_{2}k^{*}\right)c_{t}-\omega_{1}\omega_{2}\exp\left(\sigma_{1}k^{*}-\sigma_{2}k^{*}\right)\sigma_{1}k_{t}+\\
+ & \left(1-\delta\right)\left[\left(1+\nu\right)\exp\left(c^{*}\right)c_{t}-\omega_{1}\exp\left(\sigma_{1}k^{*}\right)\sigma_{1}k_{t}\right]\\
= & \left[\left(1+\nu\right)\omega_{2}\exp\left(c^{*}-\sigma_{2}k^{*}\right)+\left(1-\delta\right)\left(1+\nu\right)\exp\left(c^{*}\right)\right]c_{t}+\\
+&\left[-\omega_{1}\omega_{2}\exp\left(\sigma_{1}k^{*}-\sigma_{2}k^{*}\right)\sigma_{1}-\left(1-\delta\right)\omega_{1}\exp\left(\sigma_{1}k^{*}\right)\sigma_{1}\right]k_{t}\\
\coloneqq & b_{1c}c_{t}+b_{1k}k_{t}.
\end{array}
\]
Finally, we can also put all time $t+1$ variables together and define
\[
\begin{array}{rl}
 & \beta^{-1}\left(1+\nu\right)\exp\left(c^{*}\right)c_{t+1}-\beta^{-1}\omega_{1}\exp\left(\sigma_{1}k^{*}\right)\sigma_{1}k_{t+1}\\
 & +\left(1+\nu\right)\omega_{2}\exp\left(c^{*}-\sigma_{2}k^{*}\right)\sigma_{2}k_{t+1}+\omega_{1}\omega_{2}\exp\left(\sigma_{1}k^{*}-\sigma_{2}k^{*}\right)\sigma_{2}k_{t+1}\\
= & \beta^{-1}\left(1+\nu\right)\exp\left(c^{*}\right)c_{t+1}+\left[-\beta^{-1}\omega_{1}\exp\left(\sigma_{1}k^{*}\right)\sigma_{1}+\right.\\
+&\left.\left(1+\nu\right)\omega_{2}\exp\left(c^{*}-\sigma_{2}k^{*}\right)\sigma_{2}+\omega_{1}\omega_{2}\exp\left(\sigma_{1}k^{*}-\sigma_{2}k^{*}\right)\sigma_{2}\right]k_{t+1}\\
\coloneqq & d_{1c}c_{t+1}+d_{1k}k_{t+1}
\end{array}.
\]
Then, the log-linearized version of the Euler equation can be written
as
\[
d_{1c}c_{t+1}+d_{1k}k_{t+1}=a_{1}+b_{1c}c_{t}+b_{1k}k_{t}.
\]
\paragraph{Log-linearization of the aggregate resource constraint}
The aggregate resource constraint is
\[
K_{t+1}=\left(1-\delta\right)K_{t}+\Phi_{t}K_{t}^{\alpha}L_{t}^{1-\alpha}-N_{t}^{f}f-C_{t}.
\]
Using
\[
L_{t}=\left[\left(1-\alpha\right)\;\Theta_{t}\,K_{t}^{\alpha}\right]^{1/\left(\alpha+\nu\right)}.
\]
The aggregate resource constraint becomes
\[
\begin{array}{rl}
 & K_{t+1}=\left(1-\delta\right)K_{t}+\Phi^{*}\left[\left(1-\alpha\right)\;\Theta^{*}\right]^{\left(1-\alpha\right)/\left(\alpha+\nu\right)}K_{t}^{\alpha\left(1+\nu\right)/\left(\alpha+\nu\right)}-N^{f*}f-C_{t}\\
\Leftrightarrow & K_{t+1}=\left(1-\delta\right)K_{t}+\gamma K_{t}^{\sigma_{1}}-N^{f*}f-C_{t}
\end{array}.
\]
We have
\[
K_{t+1}\approx\exp\left(k^{*}\right)\left(1+k_{t+1}-k^{*}\right)
\]
\[
K_{t}\approx\exp\left(k^{*}\right)\left(1+k_{t}-k^{*}\right)
\]
\[
C_{t}\approx\exp\left(k^{*}\right)\left(1+c_{t}-c^{*}\right)
\]
\[
K_{t}^{\sigma_{1}}\approx\exp\left(\sigma_{1}k^{*}\right)\left[1+\sigma_{1}\left(k_{t}-k^{*}\right)\right],
\]
and so
\[
\begin{array}{rl}
 & \exp\left(k^{*}\right)\left(1+k_{t+1}-k^{*}\right)=\left(1-\delta\right)\exp\left(k^{*}\right)\left(1+k_{t}-k^{*}\right)+\\
 +&\gamma\exp\left(\sigma_{1}k^{*}\right)\left[1+\sigma_{1}\left(k_{t}-k^{*}\right)\right]-N^{f*}f-\exp\left(k^{*}\right)\left(1+c_{t}-c^{*}\right)\\
\Leftrightarrow & \exp\left(k^{*}\right)k_{t+1}=-\exp\left(k^{*}\right)\left(1-k^{*}\right)+\left(1-\delta\right)\exp\left(k^{*}\right)\left(1-k^{*}\right)+\\
+&\gamma\exp\left(\sigma_{1}k^{*}\right)\left(1-\sigma_{1}k^{*}\right)-\exp\left(k^{*}\right)\left(1-c^{*}\right)-N^{f*}f\\
 & +\left(1-\delta\right)\exp\left(k^{*}\right)k_{t}+\gamma\exp\left(\sigma_{1}k^{*}\right)\sigma_{1}k_{t}-\exp\left(k^{*}\right)c_{t}\\
\Leftrightarrow & d_{2k}k_{t+1}=a_{2}+b_{2k}k_{t}+b_{2c}k_{t}c_{t}
\end{array}.
\]

\paragraph*{Log-linear system}

In matrix form, we have
\[
\left[\begin{array}{cc}
d_{1k} & d_{1c}\\
d_{2k} & 0
\end{array}\right]\left[\begin{array}{c}
k_{t+1}\\
c_{t+1}
\end{array}\right]=\left[\begin{array}{c}
a_{1}\\
a_{2}
\end{array}\right]+\left[\begin{array}{cc}
b_{1k} & b_{1c}\\
b_{2k} & b_{2c}
\end{array}\right]\left[\begin{array}{c}
k_{t}\\
c_{t}
\end{array}\right],
\]
which is equivalent to
\[
\left[\begin{array}{c}
k_{t+1}\\
c_{t+1}
\end{array}\right]=\left[\begin{array}{cc}
d_{1k} & d_{1c}\\
d_{2k} & 0
\end{array}\right]^{-1}\left[\begin{array}{c}
a_{1}\\
a_{2}
\end{array}\right]+\left[\begin{array}{cc}
d_{1k} & d_{1c}\\
d_{2k} & 0
\end{array}\right]^{-1}\left[\begin{array}{cc}
b_{1k} & b_{1c}\\
b_{2k} & b_{2c}
\end{array}\right]\left[\begin{array}{c}
k_{t}\\
c_{t}
\end{array}\right].
\]

\paragraph*{Eigenvalues}

We study numerically the properties of the matrix $\left[\begin{array}{cc}
d_{1k} & d_{1c}\\
d_{2k} & 0
\end{array}\right]^{-1}\left[\begin{array}{cc}
b_{1k} & b_{1c}\\
b_{2k} & b_{2c}
\end{array}\right],$ around the stable steady-states of our economy. Table \ref{tab:roots_dynamic_system} reports the eigenvalues of the system. We find that for all the steady states we study in the paper for the three economies of 1975, 1990, and 2007, the system has one eigenvalue that is inside the unit circle and one that is outside. We conclude that the steady states we refer to as \textit{stable} in the paper are all saddle-path stable: starting in a neighborhood of these steady states, there is only one consumption-savings path that leads the economy to that steady state. 

\begin{table}[H]
    \centering
    \begin{tabular}{|c|c|c|c|c|c|}
    \hline 
     & $\lambda_{1L}$ & $\lambda_{2L}$ &  & $\lambda_{1H}$ & $\lambda_{2H}$\tabularnewline
    \hline 
    \hline 
    1975 & - & - &  & 0.67 & 1.07 \tabularnewline
    \hline 
    1990 & 0.67 & 1.06 &  & 0.66 & 1.06\tabularnewline
    \hline 
    2007 & 0.67 & 1.06 &  & 0.67 & 1.06\tabularnewline
    \hline 
    \end{tabular}
    \caption{Roots of the dynamic system}
    \label{tab:roots_dynamic_system}
\end{table}

\vspace{0.25cm}

\section{The 1990 Recession \label{sec:90_recession}}

\subsubsection*{The response in the 1990 economy}

\begin{figure}[H]
\begin{minipage}[b]{.5\linewidth}
\centering{}\includegraphics*[scale=0.5]{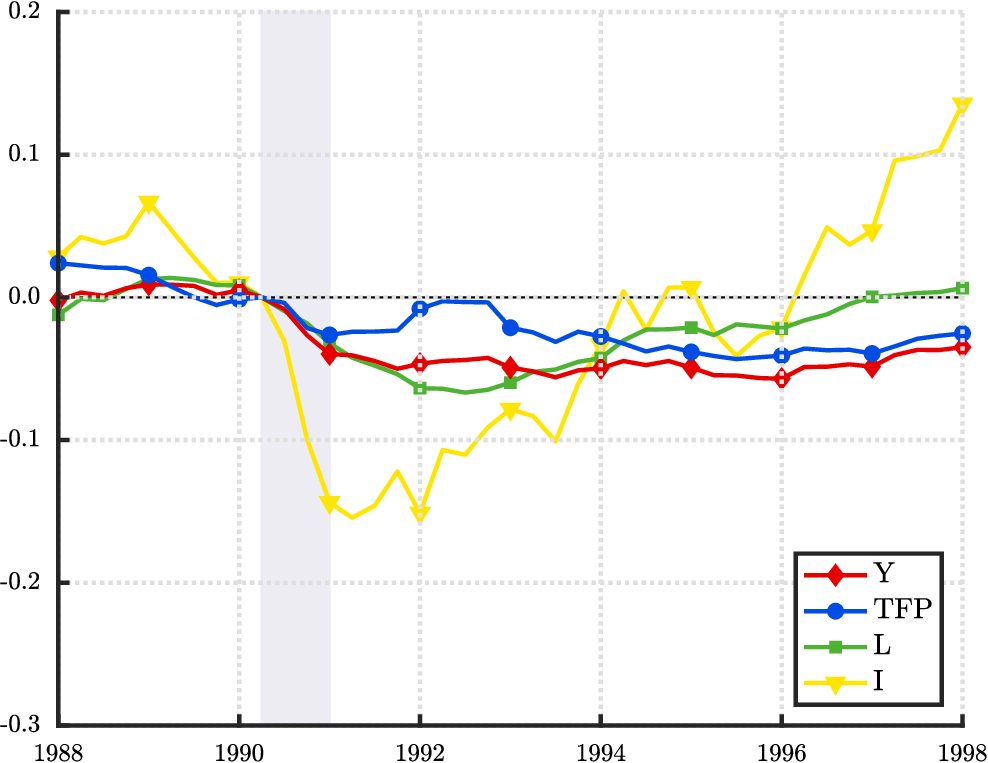}\subcaption{1990-1991 recession (data) \label{fig:data_90_recession}}
\end{minipage}
\begin{minipage}[b]{.5\linewidth}
\centering{}\includegraphics*[scale=0.5]{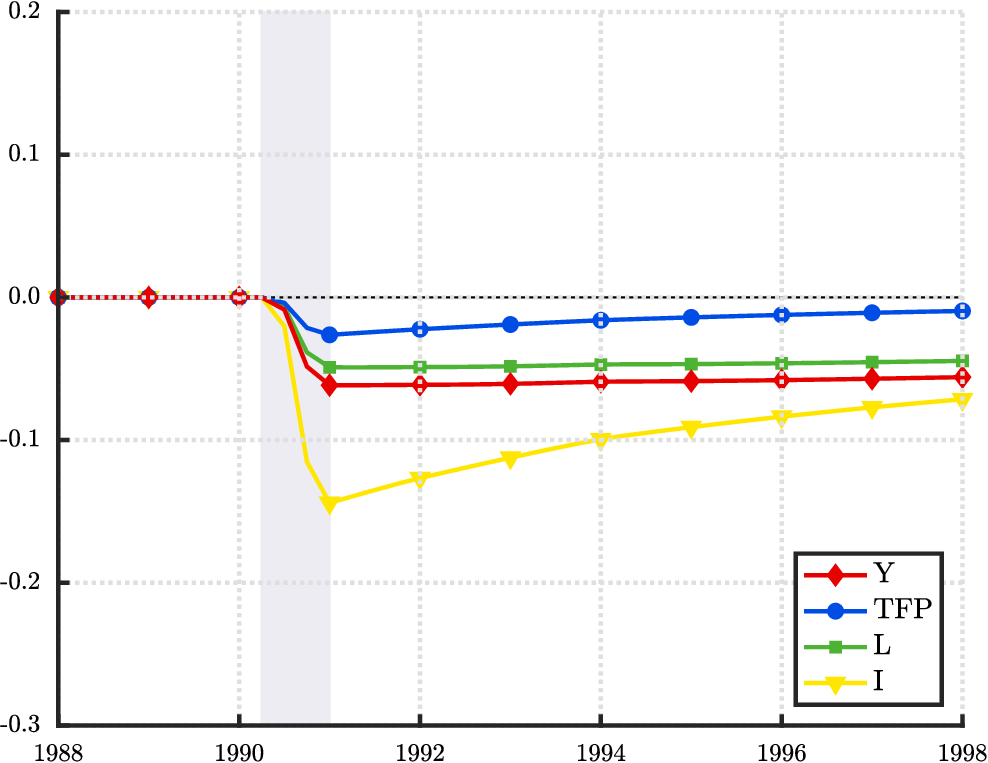}\subcaption{The 1990-1991 shock in the 1990 model \label{fig:85_model_82_crisis}}
\end{minipage}
\caption{The 1990-1991 recession \label{fig:90_crisis}}
\end{figure}

\subsubsection*{The response in the 2007 economy}

\begin{figure}[H]
\centering{}\includegraphics*[scale=0.65]{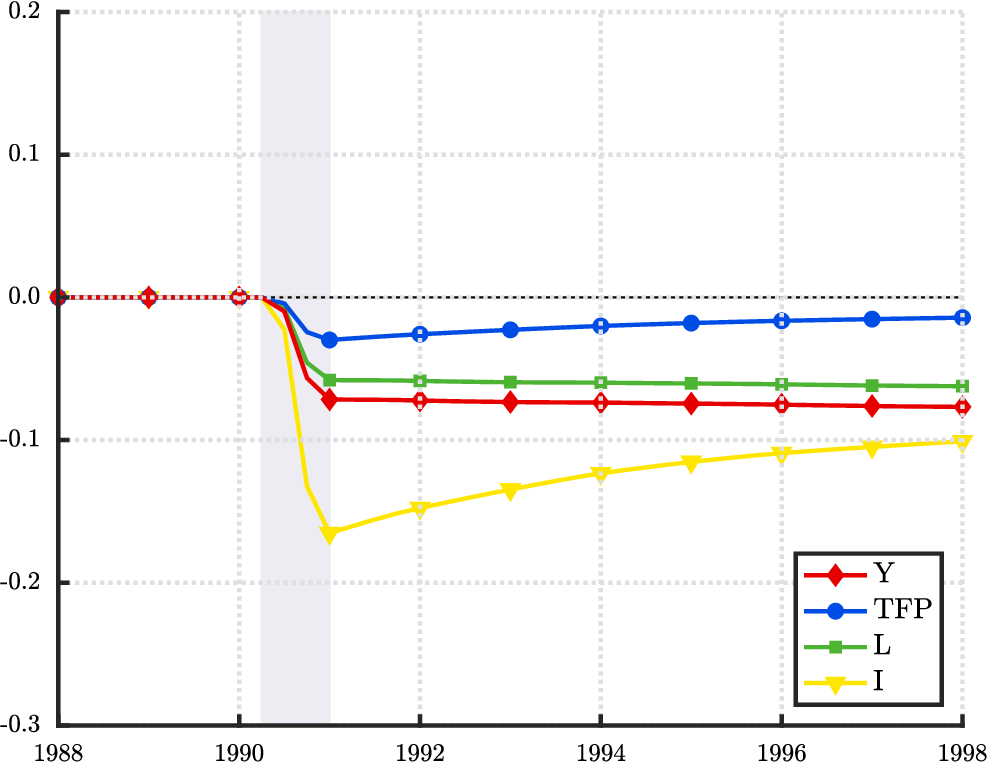}
\caption{The 1990-1991 shock in the 2007 model
\label{fig:07_model_90_crisis}}
\end{figure}

\section{Number of Firms per Sector}
\label{sec:numberfirms}
\begin{figure}[H]
\begin{centering}
\includegraphics*[scale=0.9]{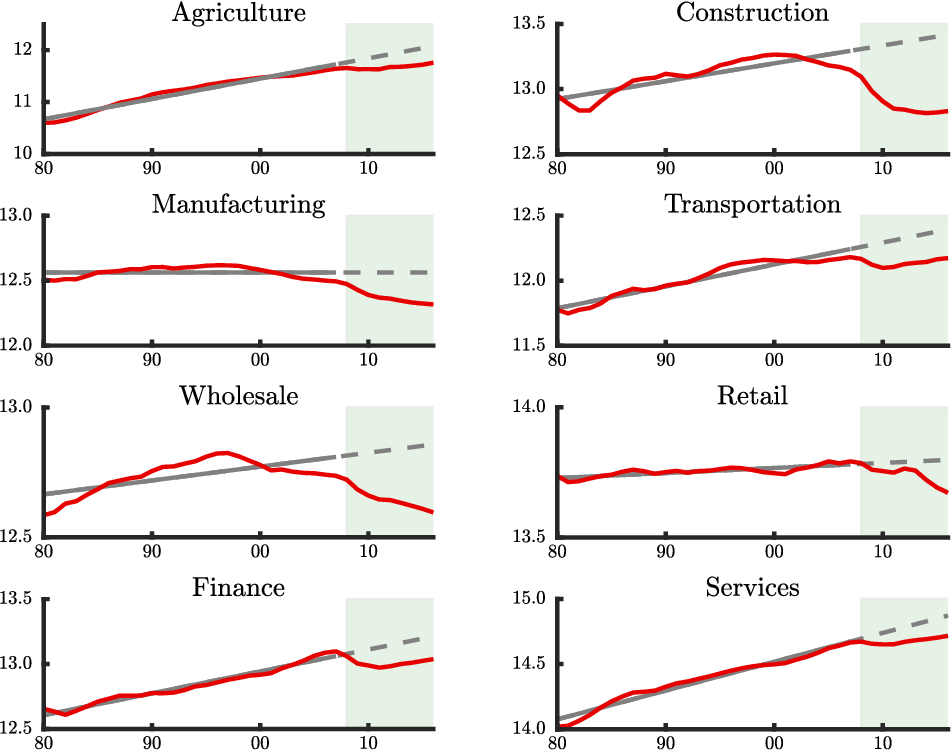} 
\par\end{centering}
\caption{\textbf{Number of Firms per Sector: 1980-2018}\protect \\
{\small{}Each panel shows the number of firms with at least one employee in each sector (in logs). For
each series, the dashed grey line shows a linear trend computed over the 1980-2007 period. Data is from the US Business Dynamics Statistics\label{fig:nfirms_sector}}}
\end{figure}

\section{Alternative Source of Fluctuations}

\begin{minipage}{\textwidth}
\begin{minipage}[b]{.49\linewidth}
\centering{}
\captionsetup{type=figure}
\includegraphics*[scale=0.675]{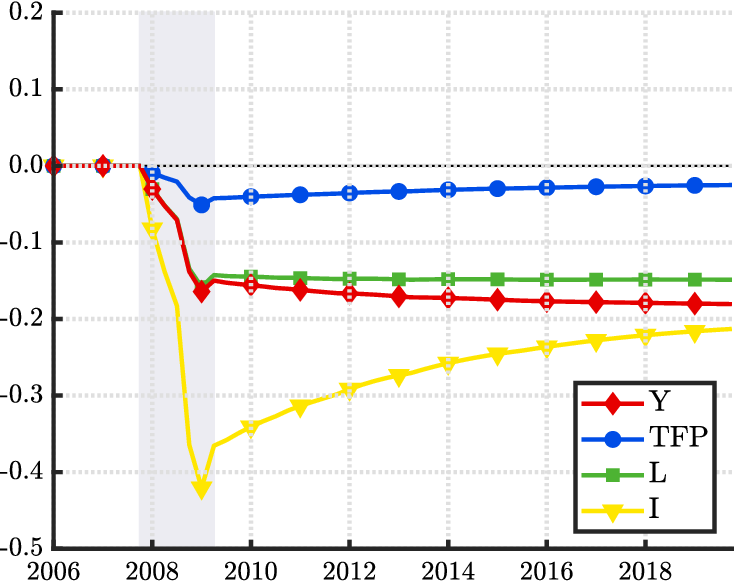}
\captionof{figure}{The \textit{great recession} in the 2007 model} \label{fig:GR_model_fshock}
\end{minipage}
\hfill
\begin{minipage}[b]{.49\linewidth}
\centering
\captionsetup{type=table}
\resizebox{0.9\textwidth}{!} {
\begin{tabular}{c c c c c } \thickhline
& & & &  \\[-1.5ex]
& \multicolumn{4}{c}{\Large 2007 Model}  \\[1ex]
& {\large 2009Q4} & {\large 2015Q1}&{\large 2019Q1} & {\large 2040Q1} \\[1.5ex]
\thickhline
& & & &  \\[-1.25ex]
{\large Output} &-0.154& -0.174&	-0.18&	-0.182  \\[1ex]
{\large TFP} & -0.041&	-0.030&	-0.025&	-0.022 \\[1ex]
{\large Hours} & -0.144&	-0.148&	-0.149	&-0.147\\[1ex]
{\large Investment} & -0.350&-0.250&	-0.212&	-0.191\\[1ex]
\thickhline 
\end{tabular}}
\captionof{table}{Deviation from the high steady-state \label{tab:great-recession-counterfactual-fshock}}
\end{minipage}
{\small Note: This figure shows the response of the 2007 model to a sequence of shocks to the number of markets with positive fixed costs (see the main text for details).}
\end{minipage}

\vspace{0.75cm}

\section{Evaluating Different Channels}
\label{sec:different-channels}

\subsection*{Fixed Number of Firms}

\begin{figure}[H]
\begin{minipage}[b]{.285\linewidth}
\centering\includegraphics*[scale=0.335]{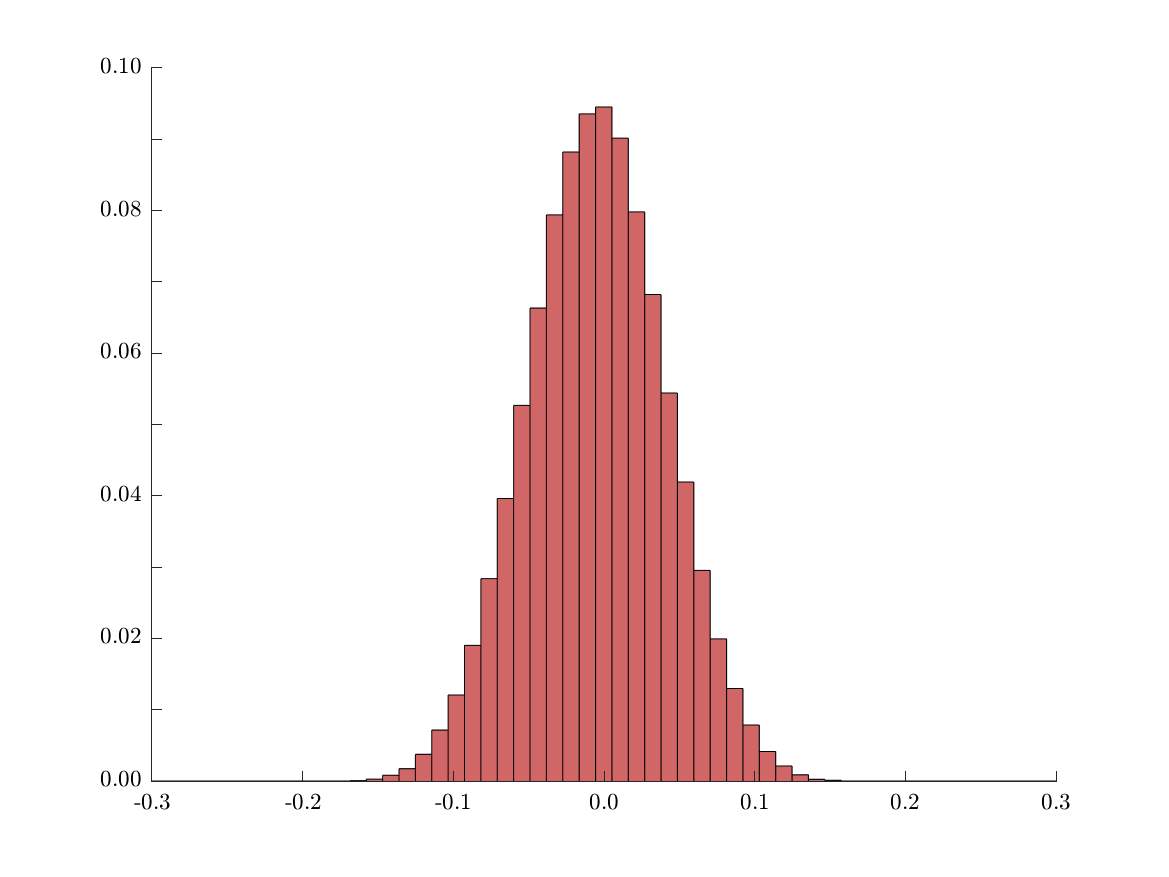}\subcaption{1975} \label{fig:y_dist_1975_fixed_N}
\end{minipage}%
\hspace{0.25cm}
\begin{minipage}[b]{.285\linewidth}
\centering\includegraphics*[scale=0.335]{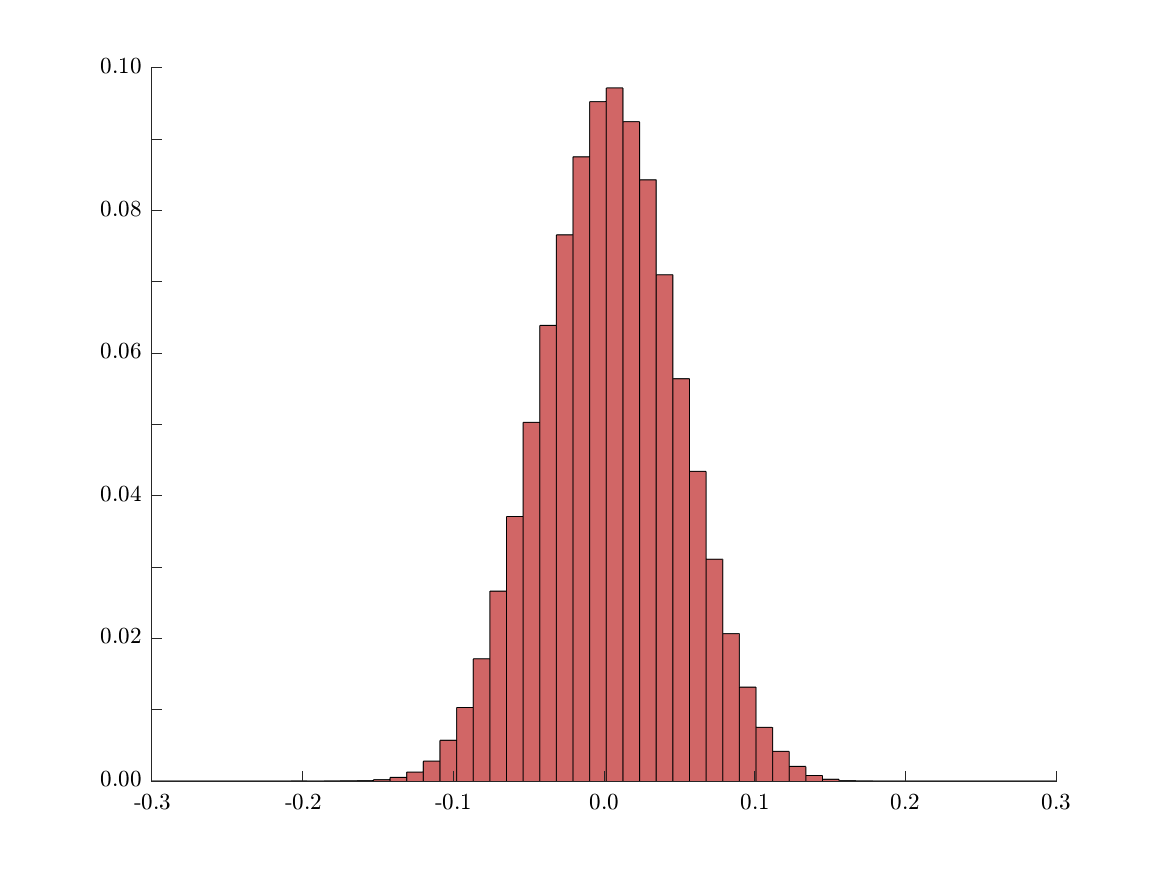}\subcaption{1990} \label{fig:y_dist_1990_fixed_N}
\end{minipage}
\hspace{0.25cm}
\begin{minipage}[b]{.285\linewidth}
\centering\includegraphics*[scale=0.335]{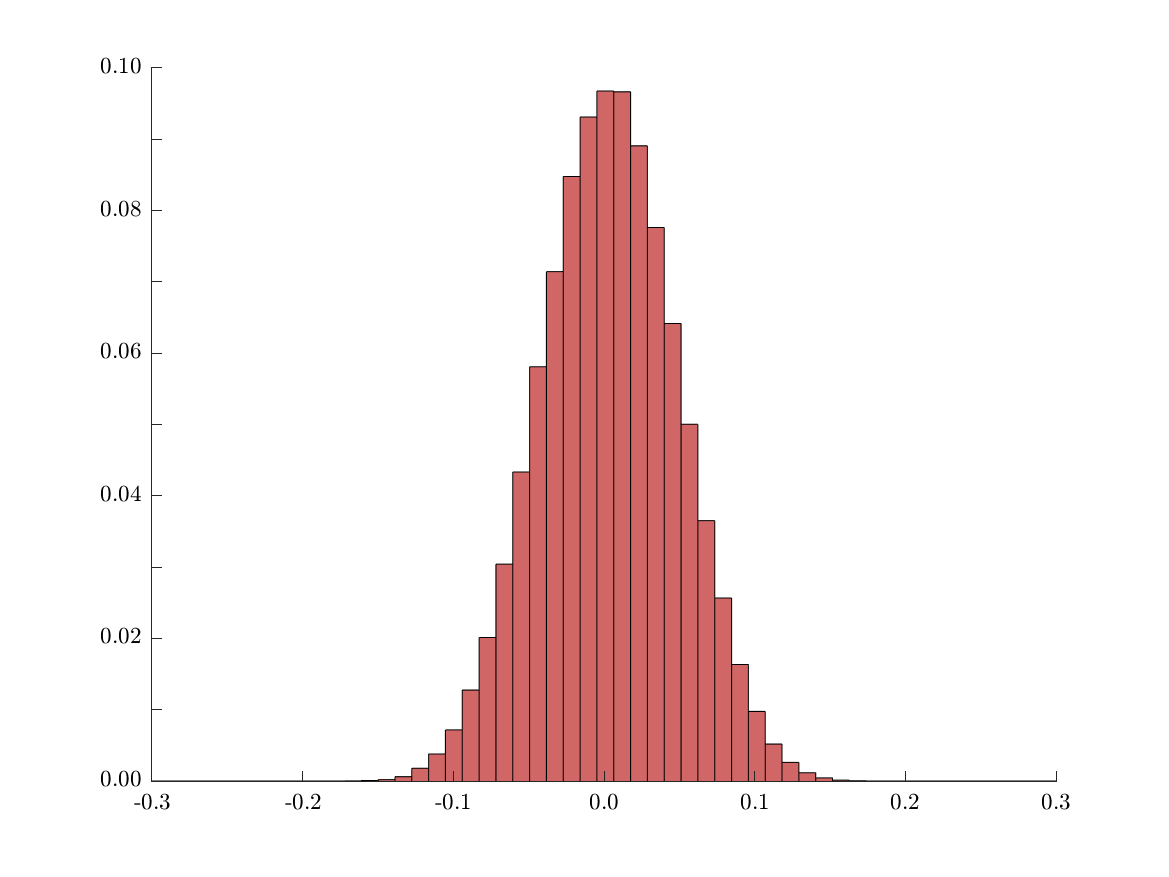}\subcaption{2007} \label{fig:y_dist_2007_fixed_N}
\end{minipage}
\caption{Ergodic distribution of output (fixed set of firms) \protect \\ {Note: \small{}\small{} This figure shows the distribution of log output for the 1975, 1990, and 2007 economies. We simulate each economy for 1,000,000 periods and plot log output in deviation from the steady state.}} \label{fig:hist_fixed_N}
\end{figure}

First, we consider a version of the model without the extensive margin of firms. We do so by taking our calibrated model with the extensive margin and i) setting fixed costs to zero\footnote{We do this so that the fixed cost is not paid every period. Note that without setting the fixed cost to zero, the economy would look identical other than that some units of the final good being subtracted from profits.} and ii) fixing the set of active firms $N_j$. All the model parameter values are the same as our baseline models (with the exception of fixed costs, which are set to zero). We also use the same values for the exogenous TFP process. In this way, by comparing the volatility of this model to our baseline, we can quantify the role of endogenous entry and exit in contributing to aggregate fluctuations.
We then subject the three economies (1975, 1990 and 2007) to a large history of draws and plot the ergodic distributions in Figure \ref{fig:hist_fixed_N}. 

All three economies have an unimodal distribution of output, which indicates a unique (stochastic) steady state. To better understand this result, note that aggregate TFP is effectively fixed in this economy, up to the changes in the exogenous component.

Note that without entry and exit, the markup distribution is fixed since market shares do not move over the business cycle, and, in our model, they uniquely pin down markups. Since the markup distribution is constant over the cycle, the only remaining endogenous component of TFP is love for variety but this is driven by $N$, which is naturally kept fixed in this special case. As a consequence, the only endogenous response to changes in the exogenous component of TFP is coming from variable factor supply, as in a standard RBC model. This force alone cannot generate multiplicity since, absent the endogenous market structure, the rental rate is monotonically decreasing in the capital. There is just one value of capital such that $R_{t} = R^{*} \coloneqq \beta^{-1} - \left(1-\delta\right) $. The economy will only move around the unique steady state.

The three different economies (1975, 1990, and 2007) have very similar business cycle patterns. This is not surprising since they only differ in the endogenous (but fixed) component of TFP. Consequently, output changes are only driven by the exogenous TFP shocks, and by the factor supply responses (which are identically parametrized across economies). Hence, the three differ in levels but not in changes. From equation (\ref{eq:agg_output_capital}), we know that
 \begin{align*}
     Y_{t} \; = & \; A_{t}\,\Phi\left(\mathbf{\Gamma},\mathbf{N}_{t}\right)\:\left[\left(1-\alpha\right)\Theta\left(\mathbf{\Gamma},\mathbf{N}_{t}\right)\right]^{\frac{1-\alpha}{\nu+\alpha}}\:K_{t}^{\alpha\frac{1+\nu}{\nu+\alpha}}.
 \end{align*}
We can simplify this expression by taking logs, and defining $\zeta$ and $\xi$ appropriately: 
\[\log Y_t= \log A_t +\log \Phi_t +\zeta \log (1-\alpha)\Omega_t+\xi \log K_t.
\]
As a consequence, log deviations from the steady state will be\begin{align*}
    \log Y_t-\log\bar Y= \log A_t- \log \bar A+\log \Phi_t-\log \bar\Phi+\zeta(\log \Omega_t-\log \bar{\Omega})+\xi (\log K_t-\log \bar K),
\end{align*}
where upper bars denote steady-state variables. Since the market structure is constant, $\log \Phi_t=\log \bar \Phi,\,\forall t$ and $\log \Omega_t=\log \bar \Omega,\,\forall t$, hence the only difference across economies in log deviations from the steady state is driven by differences in the response of the endogenous factor supply, the last term in the equation above.

To further compare these alternative models we compute the degree of amplification and persistence in each configuration. We use the stochastic process calibrated to our baseline economy and simulate a large time series for each model. In each time series, we then compute moments of log GDP deviations. These are reported in Table \ref{tab:submodels}.

\subsection*{Monopolistic Competition}

We consider a version of the model where firms operate under monopolistic competition, as in \cite{dixit_stiglitz}. Specifically, we look at a special case of our model in which there is at most one firm in each market.
Aggregate output is
\[
Y_{t}=\left(\sum_{i=1}^{I}{\vphantom{\Pi^P}y}_{it}^{\rho}\right)^{1/\rho}, \]
and each firm produces with the production function
\[
y_{it}=A_t\;\gamma_{i}\left(k_{it}\right)^{\alpha}\left(l_{it}\right)^{1-\alpha}.
\]
Each firm charges the monopoly markup $1/\rho$. This version of the model implies that the markup distribution is time-invariant and degenerate at $1/\rho$. We then have a free entry condition across markets. 

In this special case, there is no heterogeneity in markups, and the factor share is equal to $\rho$ (the inverse of the aggregate markup). We can write the aggregate production function as 
\begin{align*}
    Y\;=\;\dfrac{\left(\sum\limits _{i=1}^{N}\:\gamma_{i}^{\rho/\left(1-\rho\right)}\right)^{1/\rho}}{\sum\limits _{i=1}^{N}\gamma_{i}^{\rho/\left(1-\rho\right)}}AK^{\alpha}L^{1-\alpha},
\end{align*}
where $\dfrac{\left(\sum\limits _{i=1}^{N}\:\gamma_{i}^{\rho/\left(1-\rho\right)}\right)^{1/\rho}}{\sum\limits _{i=1}^{N}\gamma_{i}^{\rho/\left(1-\rho\right)}}A$ is measured TFP. Absent heterogeneity, this boils down to the familiar love-of-variety in production term equal to $N^{(1-\rho)/\rho}$.

To calibrate the model, we use the average sales-weighted markup in the data to obtain $\rho=1/\hat\mu$. Hence, we set $\rho=1/1.46=0.68$, so that the elasticity $\sigma_I=1/(1-\rho)=3.17$. 
As in our baseline model, we calibrate $\lambda$ to match the standard deviation in log revenues. Under monopolistic competition, the market share of firm $i$ is given by
\begin{align*}
    s_i=\frac{p_i q_i}{PQ}=\frac{\gamma_i^{\epsilon-1}}{\sum_j \gamma_j^{\epsilon-1}}.
\end{align*}
Therefore, the standard deviation of log market shares is simply given by
\begin{align*}
    \text{SD} \log\left(s\right) =(\sigma_{I}-1)\text{SD} \log\left(\gamma\right),
\end{align*}
where $\text{SD} \log\left(\gamma\right)$ is the standard deviation of log productivity. 
To make this model more directly comparable to our baseline model, we assume that only a fraction $f_u$ of firms is subject to productivity differences. Among these, a fraction $x_c$ is subject to positive fixed costs. These parameters are again calibrated to match the employment share in COMPUSTAT and the employment share in highly concentrated markets.
This economy has a free entry condition (across industries), which holds exactly. 
To calibrate the fixed cost, we follow the steps used in the oligopoly model and target the fixed-to-total cost ratio in steady-state. 
The drawback of this approach is that the elasticity of substitution puts an upperbound $\overline{c/(vc+c)}=1-\rho$ on the value of the fixed cost as a fraction of total costs.\footnote{The net profit margin is equal to $vc_{i}\left(1/\rho-1-c/vc_{i}\right)$. This is positive provided that $c/vc_{i}<1/\rho-1$,which is equivalent to
\[
\dfrac{c}{vc_{i}+c}<1-\rho.
\]
} This upperbound is binding for our model. Given our value of $\rho=1/1.46=0.685$ the bound is 0.315. This is below the fixed to total cost ratio that we find in the data for 2007, which is equal to 0.369.
Therefore, if the fixed-to-total cost ratio would take its data value, no firm would be active. To overcome this, we target a ratio of fixed to total costs equal to $0.8\times\overline{c/(vc+c)}$.

Once again, we use the same values for the exogenous TFP process. By comparing this model's volatility to our baseline, we can get a sense of how variable markups contribute to aggregate fluctuations.


\begin{table}[H]
\setlength{\tabcolsep}{0.15cm} \begin{center} \resizebox{0.8\textwidth}{!} {  \begin{tabular}{lccccl}
\thickhline
\\[-1ex]
{\large Description} & {\large Parameter} & \multicolumn{3}{c}{{\large Value}} & {\large Source/Target} \\[1ex] \thickhline
\\[-1ex]
[B.1] Calibrated Parameters: Fixed \\ \\[-1ex]
\hline
\\[-1ex]
Between product markets ES & $\sigma_{I}$ & \multicolumn{3}{c}{3.17} & Sales-weighted average markup\\
\\[-1ex]
Share of \textit{uncompetitive} sector & $f_u$ & \multicolumn{3}{c}{0.31} &  Emp share COMPUSTAT \\
\\[-1ex]
Share of \textit{uncompetitive} markets with $c_{i}>0$ & $x_c$ & \multicolumn{3}{c}{0.60} &  Emp share concentrated industries \\
\\[-1ex] \hline
\\[-1ex]
[B.2] Calibrated Parameters: Variable & & 1975 & 1990 & 2007 & \\ \\[-1ex] \hline
\\[-1ex]
Standard deviation of $\gamma_{ij}$ & $\lambda$ & 0.66 & 0.85 & 0.93 & Std log revenues \\
\\[-1ex]
Fixed cost ($\times 10^{-3}$) & $c$ & 1.80 & 50.46 & 55.83 & Average ratio fixed/total costs \\
\\[-1ex]
\thickhline
\end{tabular}  
} \end{center}
\caption{Parameter Values\label{tab:parameter_values_mc}}
\end{table}

\begin{figure}[H]
\hspace{-1cm}
\begin{minipage}[b]{.315\linewidth}
\centering\includegraphics*[scale=0.365]{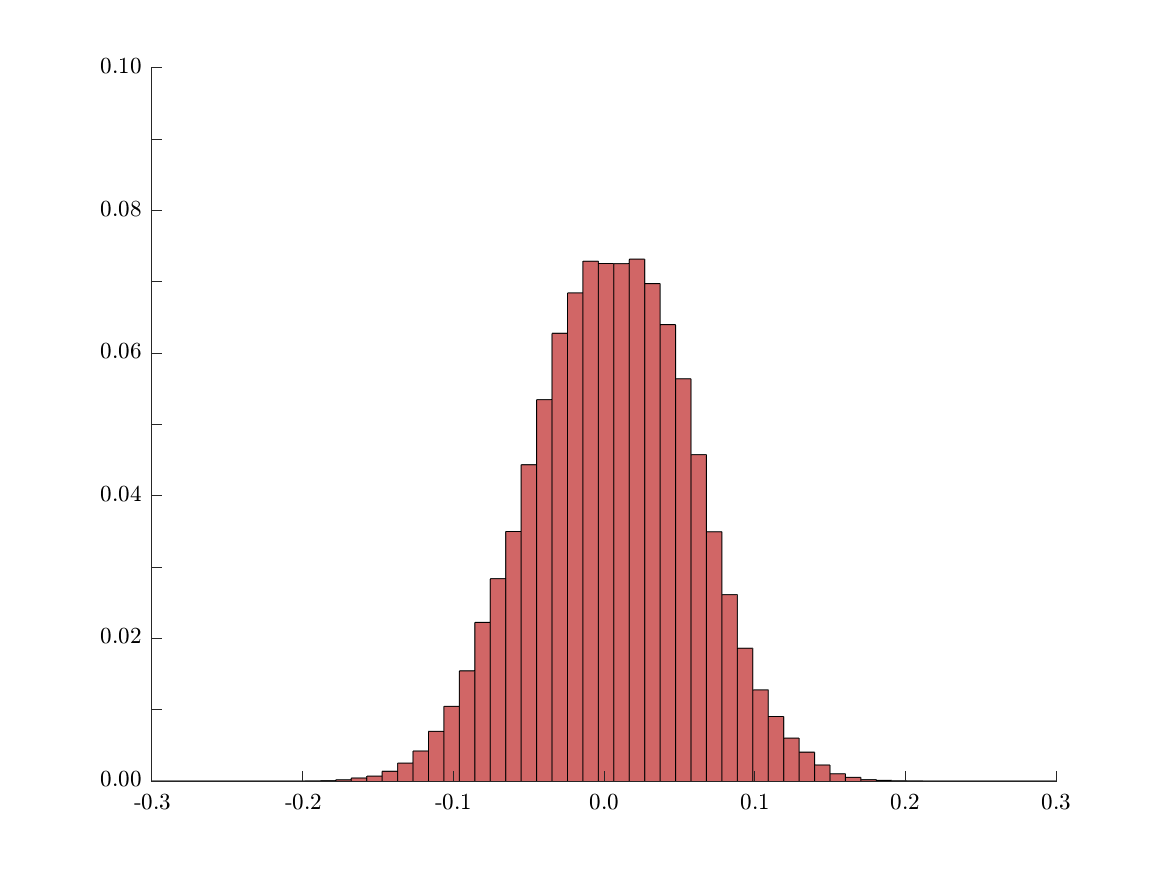}\subcaption{1975} \label{fig:y_dist_mc_1975}
\end{minipage}%
\hspace{0.25cm}
\begin{minipage}[b]{.315\linewidth}
\centering\includegraphics*[scale=0.365]{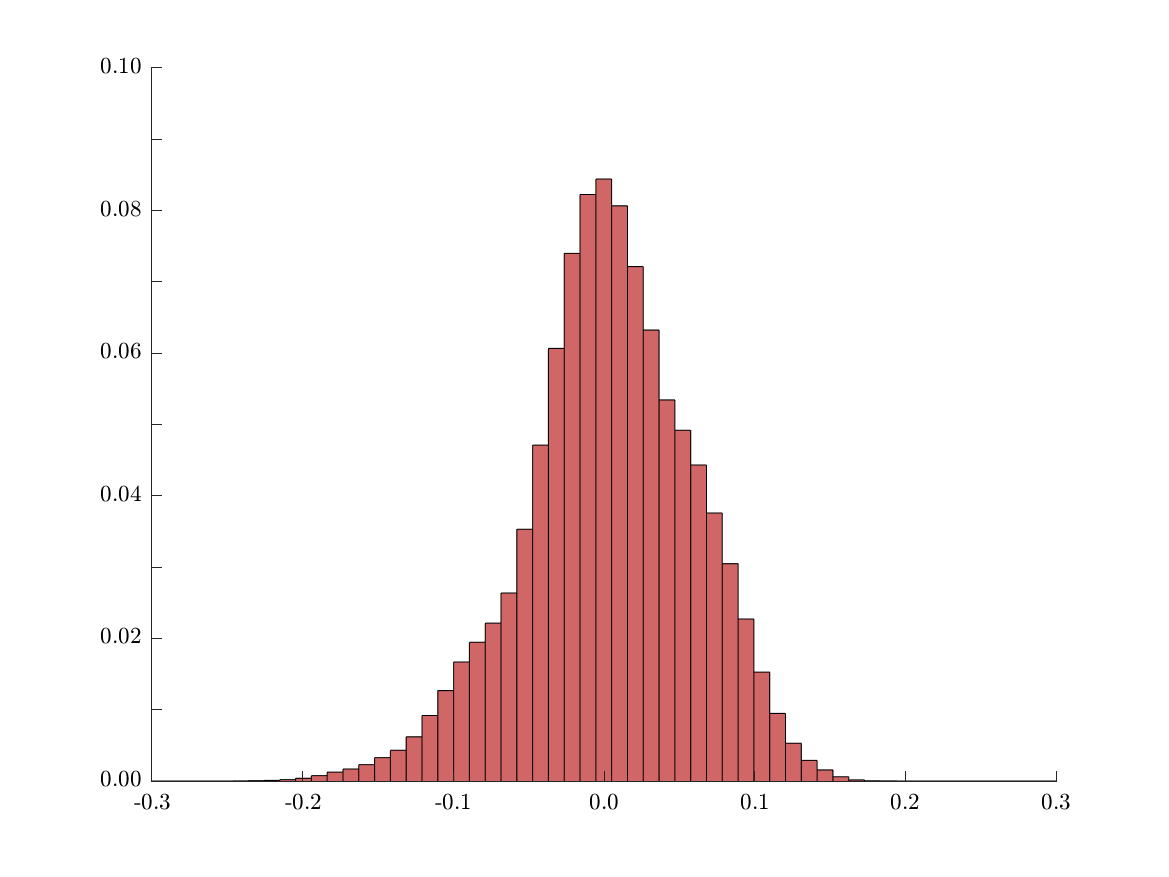}\subcaption{1990} \label{fig:y_dist_mc_1990}
\end{minipage}
\hspace{0.25cm}
\begin{minipage}[b]{.315\linewidth}
\centering\includegraphics*[scale=0.365]{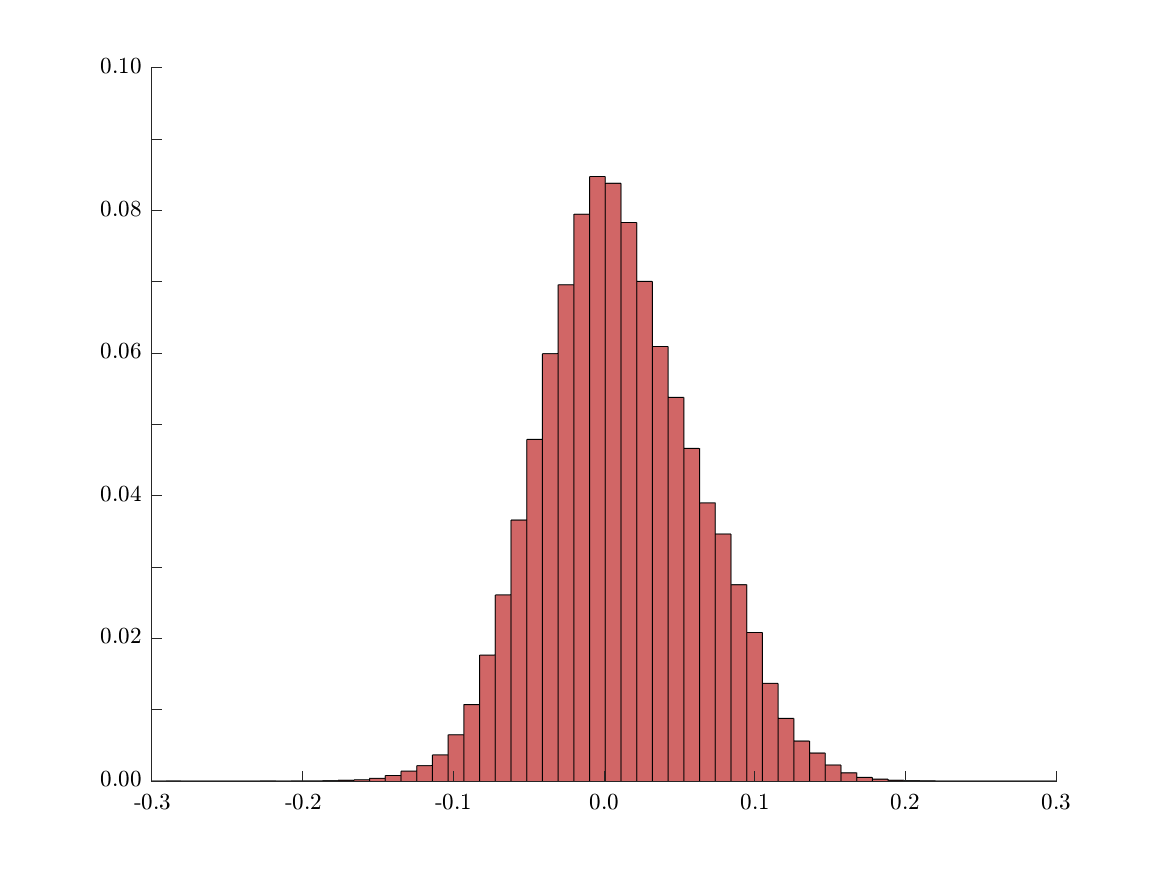}\subcaption{2007} \label{fig:y_dist_mc_2007}
\end{minipage}
\caption{Ergodic distribution of output (monopolistic competition) \protect \\ {Note: \small{}\small{} This figure shows the distribution of log output for the 1975, 1990, and 2007 economies. We simulate each economy for 1,000,000 periods and plot log output in deviation from the high steady-state.}} \label{fig:mc_histograms}
\end{figure}

We then subject the three economies to a large sequence of draws and plot the ergodic distributions in Figure \ref{fig:mc_histograms}. 
All three economies have an unimodal distribution of output, which indicates a unique (stochastic) steady state. 

\subsection*{Fixed Labor Supply}

The special case where all factors are in fixed supply is of somewhat limited interest \textendash{} if capital cannot be accumulated, transitions across steady-states are not possible. For this reason, we consider the case where labor is in fixed supply and equal to $L_t = 1$, while capital is supplied elastically. 

Households have an intertemporal utility
\[
U_{0}=\mathbb{E}\sum_{t=0}^{\infty}\:\beta^{t}\:\log\left(C_{t}\right),
\]
and are subject to the budget constraint
\[
K_{t+1}=\left[R_{t}+\left(1-\delta\right)\right]K_{t}+W_{t}+\Pi^{N}_{t}-C_{t}.
\]
The aggregate production function becomes
\[
Y_{t}\;=\;A_{t}\,\Phi\left(\mathbf{\Gamma},\mathbf{N}_{t}\right)\:K_{t}^{\alpha},
\]
with the understanding that endogenous productivity $\Phi\left(\cdot\right)$ depends on the number of active firms and, hence, on the aggregate capital stock $K_t$. 

We follow the same steps as in the main model to calibrate our parameters of interest and report the model fit and the key experiment. Regarding the process of exogenous TFP, we use the same values for $\phi_A = 0.95$ and $\sigma_{\varepsilon} = 0.003$ as in the baseline model. In this way, by comparing the volatility of this model to our baseline, we can isolate the role of endogenous labor supply in contributing to aggregate fluctuations.

\begin{table}[ht!]
\setlength{\tabcolsep}{0.15cm} \begin{center} \resizebox{0.8\textwidth}{!} {  \begin{tabular}{lccccl}
\thickhline
\\[-1ex]
{\large Description} & {\large Parameter} & \multicolumn{3}{c}{{\large Value}} & {\large Source/Target} \\[1ex] \thickhline
\\[-1ex]
[B.1] Calibrated Parameters: Fixed \\ \\[-1ex]
\hline
\\[-1ex]
Between product markets ES & $\sigma_{I}$ & \multicolumn{3}{c}{1.48} & Sales-weighted average markup\\
\\[-1ex]
Within product market ES & $\sigma_{G}$ & \multicolumn{3}{c}{12.0} & Sales-weighted average markup\\
\\[-1ex]
Share of \textit{uncompetitive} sector & $f_u$ & \multicolumn{3}{c}{0.42} &  Emp share COMPUSTAT \\
\\[-1ex]
Share of \textit{uncompetitive} markets with $c_{i}>0$ & $x_c$ & \multicolumn{3}{c}{0.28} &  Emp share concentrated industries \\
\\[-1ex] \hline
\\[-1ex]
[B.2] Calibrated Parameters: Variable & & 1975 & 1990 & 2007 & \\ \\[-1ex] \hline
\\[-1ex]
Standard deviation of $\gamma_{ij}$ & $\lambda$ & 0.175 & 0.210 & 0.220 & Std log revenues \\
\\[-1ex]
Fixed cost ($\times 10^{-4}$) & $c$ & 0.240 & 0.392 & 0.550 & Average ratio fixed/total costs \\
\\[-1ex]
\thickhline
\end{tabular}  
} \end{center}
\caption{Parameter Values\label{tab:parameter_values_fixed_L}}
\end{table}

\begin{figure}[htbp]
\hspace{-1cm}
\begin{minipage}[b]{.315\linewidth}
\centering\includegraphics*[scale=0.335]{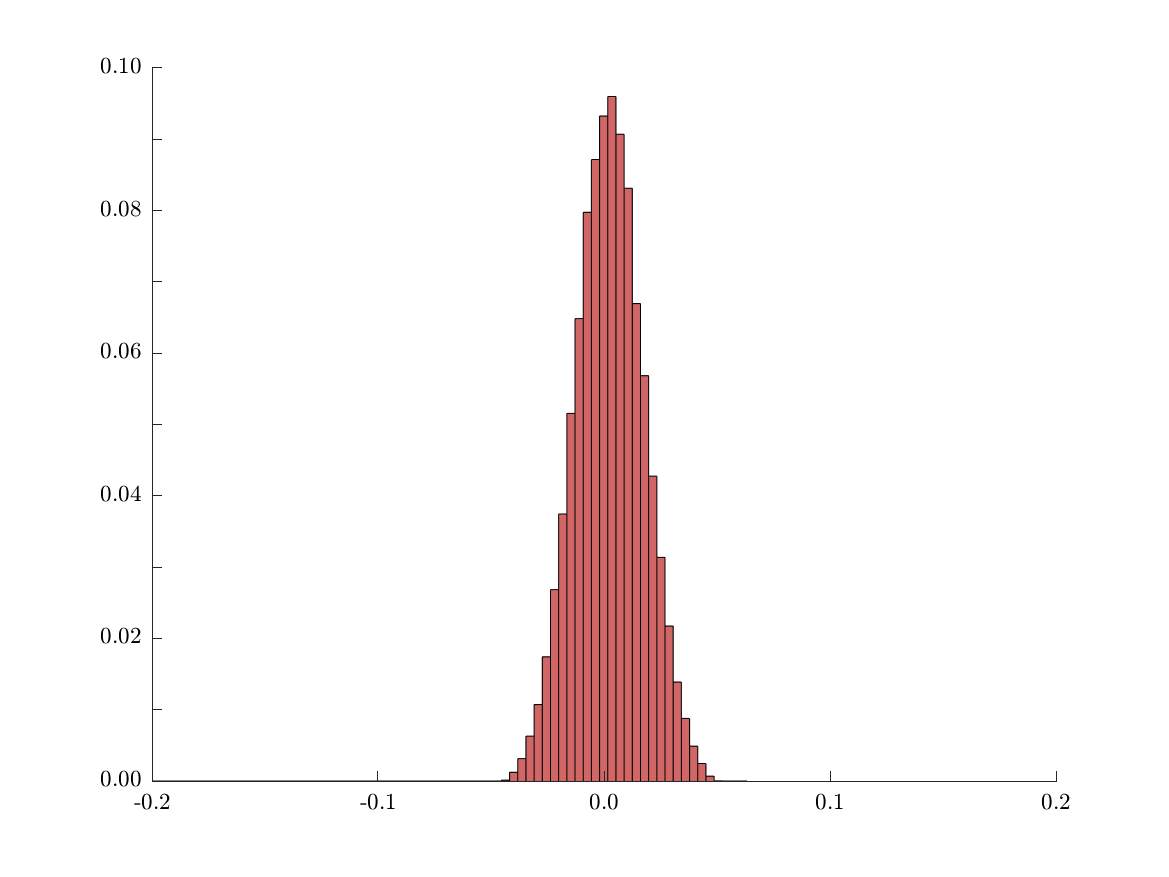}\subcaption{1975} \label{fig:y_fls_dist_1975}
\end{minipage}
\hspace{0.15cm}
\begin{minipage}[b]{.315\linewidth}
\centering\includegraphics*[scale=0.335]{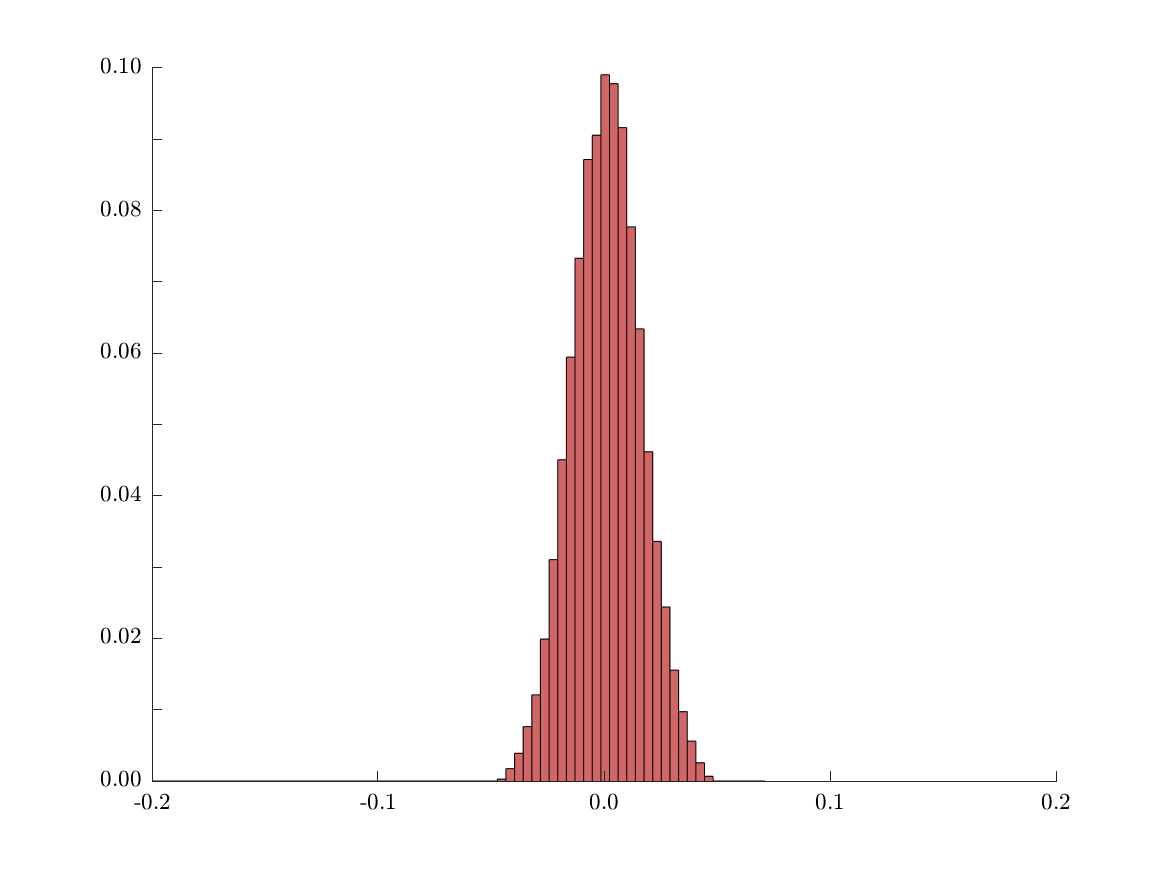}\subcaption{1990} \label{fig:y_fls_dist_1990}
\end{minipage}
\hspace{0.15cm}
\begin{minipage}[b]{.315\linewidth}
\centering\includegraphics*[scale=0.335]{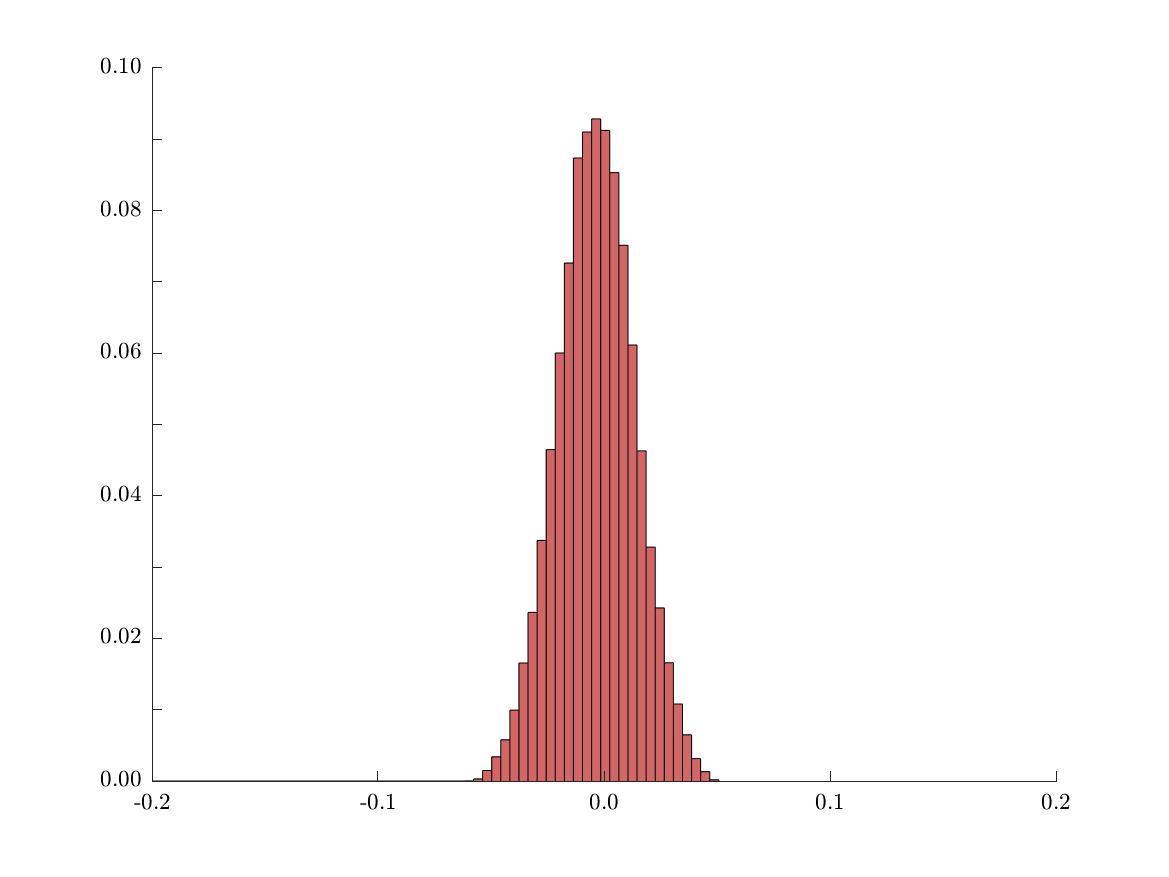}\subcaption{2007} \label{fig:y_fls_dist_2007}
\end{minipage}
\caption{Ergodic distribution of output (fixed labor supply) \protect \\ {Note: \small{}\small{} This figure shows the distribution of log output for the 1975, 1990, and 2007 economies. We simulate each economy for 1,000,000 periods and plot log output in deviation from the high steady-state.}} \label{fig:histograms_fls}

\end{figure}

We simulate each economy 1,000,000 times and plot the distribution of log output (in deviation from its mode). The three distributions are unimodal, which indicates that the three economies feature a unique (stochastic) steady state.

\section{Aggregate Productivity \label{subsec:agg_TFP}}

\subsubsection*{Average Firm Level TFP}

\begin{figure}[H]
\centering
\includegraphics*[scale=0.8]{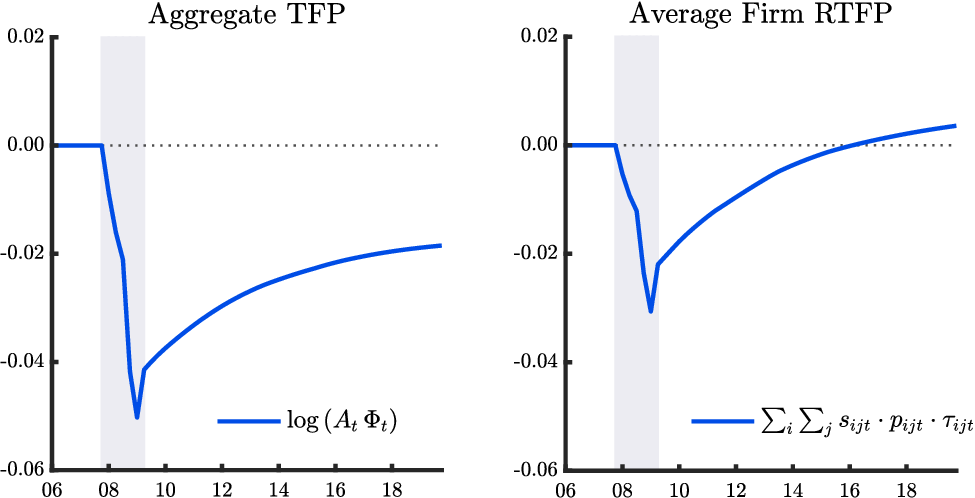}\caption{Aggregate TFP versus Average Firm Level TFP \protect \\
{Note: \small{}The left panel shows aggregate TFP. The right panel shows a sales-weighted average of firm-level revenue TFP $p_{ijt}\cdot\tau_{ijt}$.
\label{fig:TFP_Model_transition}}}
\end{figure}

Figure \ref{fig:TFP_Model_transition} reports a sales-weighted average of firm-level revenue TFP. A similar pattern emerges if one uses physical TFP instead.

\subsubsection*{Dispersion in Industry Output}
\begin{figure}[H]
\centering
\includegraphics*[scale=0.5]{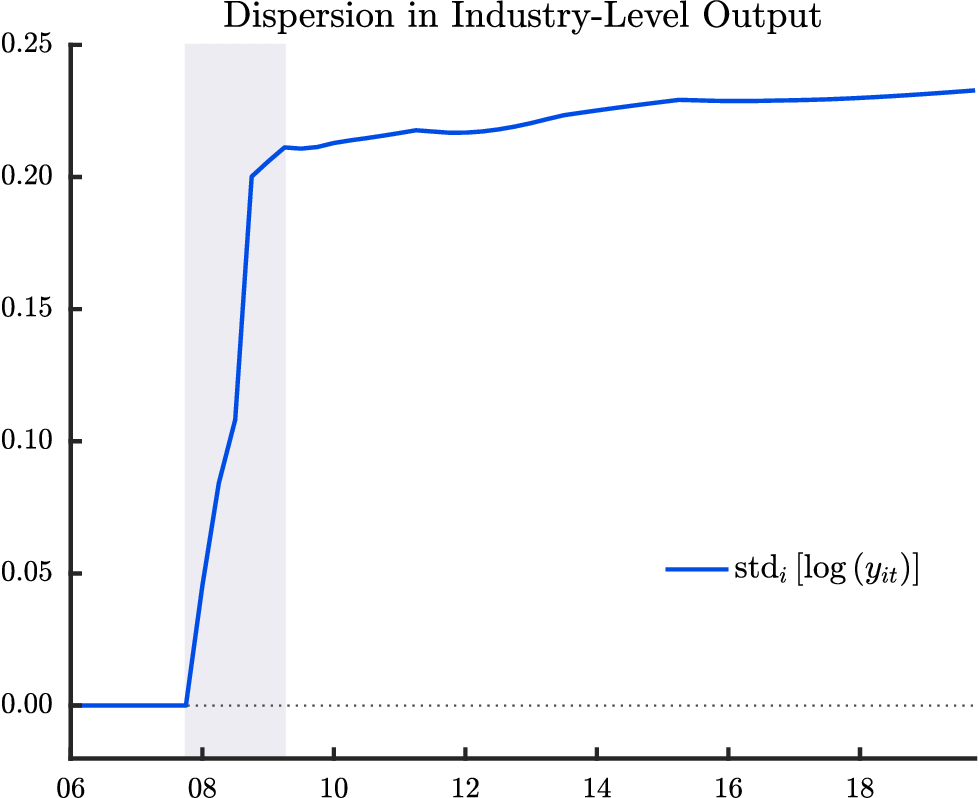}\caption{Dispersion in $\log\left(y_{it}\right)${\small{} \label{fig:output_dispersion_transition}}}
\end{figure}

\section{Regression Tables \label{sec:reg-tables}}

\begin{table}[H]
\setlength{\tabcolsep}{0.1cm}  \begin{center}
\scalebox{.7}{\begin{tabular}{lcccc} 				\thickhline
			\\[-2ex]
			 & (1) & (2) & (3) & (4) \\
			 \\[-1.5ex]
			VARIABLES & $\Delta \log \textrm{emp}_{07-16}$ & $\Delta \log \textrm{emp}_{07-16}$ & $\Delta \log \textrm{emp}_{07-16}$ & $\Delta \log \textrm{emp}_{07-16}$ \\
			\\[-1.5ex]
            \thickhline
			&  &  &  \\
			$ \textrm{concentration}_{07} $ & -0.0223*** & -0.0160** & -0.0177*** & -0.0178** \\
			& (0.00667) & (0.00688) & (0.00682) & (0.00732) \\
			&  &  &  & \\
			$ \log \textrm{firms}_{07} $ &  & 0.00239*** & 0.00193*** & 0.00151 \\
			&  & (0.000705) & (0.000706) & (0.000983) \\
			&  &  &  & \\
			$\Delta \log \textrm{emp}_{03-07}$ &  &  & 0.0984*** & 0.0901*** \\
			&  &  & (0.0241) & (0.0247) \\
			&  &  &  &  \\
			Observations & 770 & 770 & 769 & 761 \\
			R-squared & 0.014 & 0.029 & 0.050 & 0.064 \\
			Sector FE & NO & NO & NO & YES \\ 
			\\[-2ex] \thickhline
			\\[-2ex]
			\multicolumn{5}{c}{ Standard errors in parentheses} \\
			\multicolumn{5}{c}{ *** p$<$0.01, ** p$<$0.05, * p$<$0.1} \\
\end{tabular} }
\caption{Change in Employment: 2007-2016\protect \\ {Note: \small{} the table shows the results of regressing the growth rate of sectoral employment between 2007 and 2016 on the measure of concentration in 2007. The table presents the results of progressively adding controls and, in the last column, sector fixed effects.}}
\label{tab:employment}
\end{center}
\end{table}

\begin{table}[H]
\setlength{\tabcolsep}{0.1cm}  \begin{center} 
\scalebox{.7}{\begin{tabular}{lcccc} 				\thickhline
			\\[-2ex]
			& (1) & (2) & (3) & (4) \\
			\\[-1.5ex]
			VARIABLES & $\Delta \log \textrm{payroll}_{07-16}$ & $\Delta \log \textrm{payroll}_{07-16}$ & $\Delta \log \textrm{payroll}_{07-16}$ & $\Delta \log \textrm{payroll}_{07-16}$ \\
			\\[-1.5ex]
            \thickhline
			&  &  &  \\
			$ \textrm{concentration}_{07} $ & -0.0231*** & -0.0177** & -0.0189*** & -0.0194*** \\
		    & (0.00679) & (0.00702) & (0.00697) & (0.00749) \\
		    &  &  &  &  \\
			$ \log \textrm{firms}_{07} $ &  & 0.00203*** & 0.00164** & 0.000991 \\
			&  & (0.000724) & (0.000725) & (0.00101) \\
			&  &  &  &  \\
			$\Delta \log \textrm{payroll}_{03-07}$ &  &  & 0.0823*** & 0.0697*** \\
			&  &  & (0.0219) & (0.0225) \\
			&  &  &  &  \\
			Observations & 774 & 774 & 773 & 765 \\
			R-squared & 0.015 & 0.025 & 0.043 & 0.054 \\
			Sector FE & NO & NO & NO & YES \\ 
			\\[-2ex] \thickhline
			\\[-2ex]
			\multicolumn{5}{c}{ Standard errors in parentheses} \\
			\multicolumn{5}{c}{ *** p$<$0.01, ** p$<$0.05, * p$<$0.1} \\
\end{tabular} }
\end{center}
\caption{Change in Total Payroll: 2007-2016\protect \\ {Note: \small{} The table shows the results of regressing the growth rate of sectoral total payroll between 2007 and 2016 on the measure of concentration in 2007. The table presents the results of progressively adding controls and, in the last column, sector fixed effects.}}
\label{tab:payroll}
\end{table}

\begin{table}[H]
\setlength{\tabcolsep}{0.2cm}  \begin{center} 
\scalebox{.7}{\begin{tabular}{lcccc} 				\thickhline
			\\[-2ex]
			& (1) & (2) & (3) & (4) \\
			\\[-1.5ex]
			VARIABLES & $\Delta \log \textrm{firms}_{07-16}$ & $\Delta \log \textrm{firms}_{07-16}$ & $\Delta \log \textrm{firms}_{07-16}$ & $\Delta \log \textrm{firms}_{07-16}$ \\
			\\[-1.5ex]
            \thickhline
			&  &  &  \\
			$ \textrm{concentration}_{07} $ & -0.0432*** & -0.0391*** & -0.0406*** & -0.0231*** \\
            & (0.00608) & (0.00637) & (0.00635) & (0.00666) \\
            &  &  &  &  \\
            $ \log \textrm{firms}_{07} $ &  & 0.00137** & 0.00119* & 0.00449*** \\
            &  & (0.000663) & (0.000661) & (0.000897) \\
            &  &  &  &  \\
            $\Delta \log \textrm{firms}_{03-07}$ &  &  & 0.0881*** & 0.0808*** \\
            &  &  & (0.0270) & (0.0273) \\
            &  &  &  &  \\
            Observations & 791 & 791 & 791 & 782 \\
            R-squared & 0.060 & 0.065 & 0.078 & 0.151 \\
			Sector FE & NO & NO & NO & YES \\ 
			\\[-2ex] \thickhline
			\\[-2ex]
			\multicolumn{5}{c}{ Standard errors in parentheses} \\
			\multicolumn{5}{c}{ *** p$<$0.01, ** p$<$0.05, * p$<$0.1} \\
\end{tabular} }
\end{center}
\caption{Change in Number of Firms: 2007-2016\protect \\ {Note: \small{} the table shows the results of regressing the growth rate of the industry number of firms between 2007 and 2016 on the measure of concentration in 2007. The table presents the results of progressively adding controls and, in the last column, sector fixed effects.}}
\label{tab:nfirms}
\end{table}

\begin{table}[H]
\setlength{\tabcolsep}{0.2cm}  \begin{center} 
\scalebox{.7}{\begin{tabular}{lcccc} 				\thickhline
			\\[-2ex]
			& (1) & (2) & (3) & (4) \\
			\\[-1.5ex]
			VARIABLES & $\Delta \textrm{labor share}_{07-16}$ & $\Delta \textrm{labor share}_{07-16}$ & $\Delta \textrm{labor share}_{07-16}$ & $\Delta \textrm{labor share}_{07-16}$ \\
			\\[-1.5ex]
            \thickhline
			&  &  &  \\
			$ \textrm{concentration}_{07} $ & -0.0314* & -0.0319* & -0.0314* & -0.0301 \\
            & (0.0167) & (0.0168) & (0.0167) & (0.0196) \\
            &  &  &  &  \\
            $ \log \textrm{firms}_{07} $ &  & -0.00111 & -0.00120 & -0.00255 \\
            &  & (0.00240) & (0.00240) & (0.00335) \\
            &  &  &  &  \\
            $\Delta \textrm{labor share}_{03-07}$ &  &  & 0.169* & 0.146* \\
            &  &  & (0.0867) & (0.0871) \\
            &  &  &  &  \\
            Observations & 99 & 99 & 98 & 97 \\
            R-squared & 0.035 & 0.037 & 0.075 & 0.111 \\
			Sector FE & NO & NO & NO & YES \\ 
			\\[-2ex] \thickhline
			\\[-2ex]
			\multicolumn{5}{c}{ Standard errors in parentheses} \\
			\multicolumn{5}{c}{ *** p$<$0.01, ** p$<$0.05, * p$<$0.1} \\
\end{tabular} }
\end{center}
\caption{Change in Labor Share: 2007-2016\protect \\ {Note: \small{} The table shows the results of regressing the growth rate of sectoral labor share between 2007 and 2016 on the measure of concentration in 2007. The table presents the results of progressively adding controls and, in the last column, sector fixed effects.}}
\label{tab:labor_share}
\end{table}

\end{document}